\documentclass[10pt,aps,prx,twocolumn,amsmath,amssymb,superscriptaddress,nobibnotes]{revtex4-2}
\usepackage[T1]{fontenc}
\usepackage{enumitem}
\usepackage[latin9]{inputenc}
\usepackage{amstext}
\usepackage{amsmath}
\usepackage{todonotes}
\usepackage{verbatim}
\usepackage{mathtools}
\usepackage{bbold}
\newcommand{\angstrom}{\mbox{\normalfont\AA}}

\usepackage{graphicx}
\usepackage{lipsum}

\usepackage{esint}
\usepackage{babel}

\begin{document}
\title{Field-Theoretic Simulation of Dean--Kawasaki Dynamics for Interacting Particles}
\author{Jaehyeok Jin}
\email{jj3296@columbia.edu}
\affiliation{Department of Chemistry, Columbia University, 3000 Broadway, New York, NY 10027, USA}
\author{Chen Liu}
\altaffiliation{Current Address: Innovation and Research Division, Ge-Room, Inc., 93160 Noisy le Grand, France}
\affiliation{Department of Chemistry, Columbia University, 3000 Broadway, New York, NY 10027, USA}
\author{David R. Reichman}
\email{drr2103@columbia.edu}
\affiliation{Department of Chemistry, Columbia University, 3000 Broadway, New York, NY 10027, USA}
\date{\today}
\begin{abstract}
The formulation of a fluctuating hydrodynamic theory for interacting particles is a crucial step in the theoretical description of liquids. The microscopic mappings proposed decades ago by Dean and Kawasaki have played a central role in the analytical treatment of such problems. However, the singular mathematical nature of the density distributions used in these derivations raises concerns about the validity and practical utility of the resulting stochastic partial differential equations, particularly for direct numerical simulations. Recent efforts have centered on establishing a rigorous coarse-graining procedure to regularize the effective Dean--Kawasaki equation. Building on this foundation, we numerically investigate weakly interacting fluids within such a regularized framework for the first time. Our work reveals, at the level of structural correlations, the effects of regularization on the Dean--Kawasaki formalism and paves the way for improved numerical approaches to simulate fluctuating hydrodynamics in liquids.
\end{abstract}
\maketitle

\nopagebreak

\section{Introduction} 
The formulation of a fluctuating hydrodynamic theory that connects microscopic dynamics to mesoscopic fields remains a centerpiece of modern liquid state theory \cite{landau2013fluid,bixon1969boltzmann,fox1970contributions,kelly1971hydrodynamic,ladd1993short, groot1997dissipative, ahlrichs1999simulation}. The exact mappings proposed independently by Dean and Kawasaki \cite{dean1996langevin,kawasaki1994stochastic} provide a formal framework in which the microscopic density of interacting particles follows a stochastic partial differential equation (SPDE) with multiplicative noise. This mapping offers a direct route from particle-level dynamics to field-level descriptions, opening the possibility of interpreting and extending established approaches to complex fluids, such as polymer field theories \cite{fredrickson2006equilibrium} and lattice models \cite{baxter2016exactly}, within a unified microscopic framework. In principle, such a connection could also enable simulations over time and length scales far beyond those accessible to particle-based methods \cite{frenkel2001understanding,allen2017computer}, providing new insights into slow collective processes including glassy dynamics \cite{berthier2011theoretical,berthier2023modern}.

Despite its theoretical rigor, the Dean--Kawasaki equation is ill-suited for direct simulation. The density field is expressed as a sum of singular delta functions, giving rise to both mathematical ambiguities and severe computational obstacles \cite{cornalba2019modelling}. To address this, Cornalba and co-workers have introduced a mathematically rigorous regularization procedure that smooths the microscopic density into well-defined mesoscopic fields, and have established the existence and uniqueness of the resulting regularized dynamics \cite{cornalba2019regularized,cornalba2020weakly,cornalba2021well}. However, their treatment was primarily a formal mathematical analysis and did not explicitly incorporate microscopic interactions between particles. Furthermore, the systematic coarse-graining of such particle-level interactions at the field-theoretic level has so far remained unaddressed and requires further consideration. As a result, a practical route to numerically simulating bottom-up density field dynamics for interacting molecules remains unresolved.

In this work, we build on the essence of this regularization approach and extend it to systems of interacting particles by developing a mesoscopic coarse-graining framework for microscopic densities. This approach allows us to examine how correlations, densities, and interactions are coarse-grained at the field level. We then perform the first numerical simulations of regularized Dean--Kawasaki dynamics for interacting fluids, thereby expanding its applicability beyond purely formal treatments or ideal gases. In doing so, we take an initial step toward developing Dean--Kawasaki dynamics into a practical simulation framework for fluids, with potential applications to long-time collective behavior in simple liquids, polymers, and glasses.

\section{Dean's Equation} \label{sec:dean}
\subsection{Dean's Equation as a Fluctuating Hydrodynamic Framework}
To derive fluctuating hydrodynamics from first principles, Dean formally derived an SPDE that governs the time evolution of the instantaneous \textit{microscopic density field}, $\hat{\rho}(x,t)$ \cite{dean1996langevin} from a system of $N$ Brownian particles interacting via a pair potential $V(r)$. Specifically, the position $q_i$ of particle $i$ evolves according to the overdamped Langevin equation of $dq_i(t)/dt = - \sum_{j(\neq i)}  \nabla V(q_i-q_j)  + \xi_i(t)$, where $\xi_i(t)$ is a thermal noise obeying the fluctuation-dissipation relation. Starting from this particle-based description, a closed-form stochastic equation, commonly referred to as \textit{Dean's equation}, is derived: 
\begin{align} \label{eq:dean}
        \frac{\partial \hat{\rho}(x,t)}{\partial t} =& \nabla \cdot \left( \xi(x,t)\hat{\rho}^{\frac{1}{2}}(x,t) \right) \nonumber 
        + T\nabla^2 \hat{\rho}(x,t)  \nonumber
        \\ & + \nabla \cdot \Biggl(\hat{\rho}(x,t) \int dy \hat{\rho}(y,t) \nabla V(x-y) \Biggr),
\end{align}
where $\hat{\rho}(x,t) = \sum_i^N \delta(x-q_i(t))$ is a sum of Dirac delta functions centered at particle positions $q_i(t)$. The three terms in Eq. \eqref{eq:dean} correspond to thermal noise, diffusion, and interaction.
The thermal noise is modeled as uncorrelated white noise in space and time, $\xi$: $\langle \xi_i (x,t)\xi_j (y,t')\rangle = 2T \delta(t-t')\delta_{ij} \delta(x-y)$
Notably, the noise is multiplicative, which is physically consistent as noise should be absent in regions devoid of particles. The temperature-dependent diffusion term reads as $T\nabla^2 \hat{\rho}(x,t) := T \Delta \hat{\rho}(x,t)$, while the interaction term is written as a convolution over densities: $\nabla \cdot \left(\hat{\rho}(x,t) \int dy \hat{\rho}(y,t) \nabla V(x-y) \right)$, which can, in principle, be linked to a coarse-grained free energy functional $\mathcal{F}$ via $\nabla \cdot \left( \hat{\rho}(x,t) \nabla \frac{\delta \mathcal{F}}{\delta \hat{\rho}(x)}|_{\hat{\rho}(x,t)}\right)$ \cite{godreche1992solids}. 

A similar equation was independently derived by Kawasaki \cite{kawasaki1994stochastic}. Despite their minor differences \cite{das2011statistical,gupta2011time}, both formulations share the same general SPDE structure:
\begin{equation} \label{eq:dk}
    \frac{\partial \hat{\rho}}{\partial t} = \nabla \cdot (\hat{\rho} \nabla V * \hat{\rho})+\Delta \hat{\rho} +\nabla \cdot (\sigma \sqrt{\hat{\rho}} \xi),
\end{equation}
which is commonly referred to as Dean--Kawasaki dynamics \cite{illien2024dean}. Mathematically, Eq. (\ref{eq:dk}) can be interpreted as a stochastic perturbation of a Wasserstein gradient flow, with the noise term conforming to Otto's formal Riemannian structure for optimal transport \cite{otto2001geometry}. As this mathematical structure is closely linked to non-equilibrium statistical mechanics through large deviation theory \cite{benamou2000computational, ellis2007entropy} and macroscopic fluctuation theory \cite{bertini2015macroscopic}, the Dean--Kawasaki equation provides a key framework for describing the non-equilibrium dynamics of fluctuating fields.  

While Dean's equation offers a powerful \textit{formal} route to incorporating microscopic physics into a field-level description and enables theoretical analysis via the use of field-theoretic tools (e.g. the Martin--Siggia--Rose--De Dominicis--Janssen formalism \cite{zinn2021quantum, andreanov2006field}), we note that Dean's original formulation [Eq. \eqref{eq:dean}] is not directly suitable for simulating dynamics because (1) it provides an approximate rather than exact representation of dynamics in the general interacting case, and (2) it is mathematically ill-posed without regularization.

\subsection{Approximation and Generalized Dean's Equation}
To demonstrate that Dean's equation is only an approximate physical description, we derive a generalized form of Dean's equation by closely following the original steps outlined in Ref. \onlinecite{dean1996langevin}, with particular care in handling interactions and reviewing the underlying assumptions.

Using Ito's formula for a regular function (e.g. functions that at least twice continuously differentiable) $f(q)$ \cite{gardiner1985handbook}, we obtain:
\begin{widetext}
\begin{align}\label{eq:ito}
    \frac{d}{dt}f(q_i(t)) &= \nabla f(q_i(t))\cdot \bigg[-\sum_{j(\neq i)}  \nabla V(q_i-q_j)  + \xi_i(t)\bigg] + T \nabla^2 f(q_i(t))  \\
    &= \int dx \delta(x-q_i(t))\bigg\{ \nabla f(x) \cdot \bigg[-\int dy \bigg(\sum_{j(\neq i)}\delta(y-q_j)\bigg)  \nabla V(x-y) \bigg]  + \nabla f(x) \cdot \xi_i(t) + T \nabla^2 f(x) \bigg\}. \nonumber
\end{align}
\end{widetext}
As $\int dx \partial_t\hat{\rho}(x,t) f(x) \equiv \sum_i^N df(q_i(t))/dt$ holds for the one-particle microscopic density operator $\hat{\rho}(x,t)$, by performing an integration by parts and eliminating the integration over $x$, we obtain 
\begin{align}\label{eq:dean_general}
    \partial_t \hat{\rho}(x,t) =&  \nabla\cdot \int dy \hat{\rho}^{(2)}(x,y,t) \nabla V (x-y) + T\nabla^2 \hat{\rho}(x,t) \nonumber \\ & +  \nabla \cdot \bigg( \hat{\rho}(x,t)^{1/2} \xi(x,t) \bigg), 
\end{align}
where the multiplicative noise, $\hat{\rho}^{1/2}\xi$, originates from $\sum_{i}\delta(x-q_i)\xi_i$, as defined in Ref. \onlinecite{dean1996langevin}, and the two-particle density is defined as $\hat{\rho}^{(2)}(x,y,t)\equiv \sum_i^N \sum_{j(\neq i)}^N \delta (x-q_i(t))  \delta (y-q_j(t)).$

Equation (\ref{eq:dean_general}) generalizes Dean's original formulation by retaining the two-particle density $\hat{\rho}^{(2)}$ rather than replacing it with the product $\hat{\rho}(x,t)\hat{\rho}(y,t)$. This substitution is only valid under the assumptions that $\int dx \delta(x)\nabla V(x) = \nabla V(0)$ and $\int d\vec{k}V(k)<\infty$ hold. However, the identity $\int dx \delta(x) g(x) = g(0)$ requires that $g(x)$ be continuous and compactly supported, or more generally, ``well-behaved'' \cite{arfken2011mathematical}, which is not satisfied by many realistic interaction potentials, such as the Lennard-Jones potential. 

The discussion above illustrates that in a formal sense, the exact microscopic dynamics of a liquid in the general interacting case is not closed at the level of the one-particle density alone, and the resulting theory obeys a hierarchy in which the stochastic dynamics of the $n$-particle density depends on the $(n+1)$-particle density. This hierarchy also carries over to correlation functions, e.g., the intermediate scattering function $F(k,t)$ (see Appendix A). In summary, Dean's original equation is formally exact only for interaction potentials that are finite and well-behaved at all distances. In contrast, Eq. \eqref{eq:dean_general} provides a more general expression applicable to realistic systems with hard-core divergences \cite{widom1967intermolecular}. 
\begin{figure}
    \includegraphics[width=0.5\textwidth]{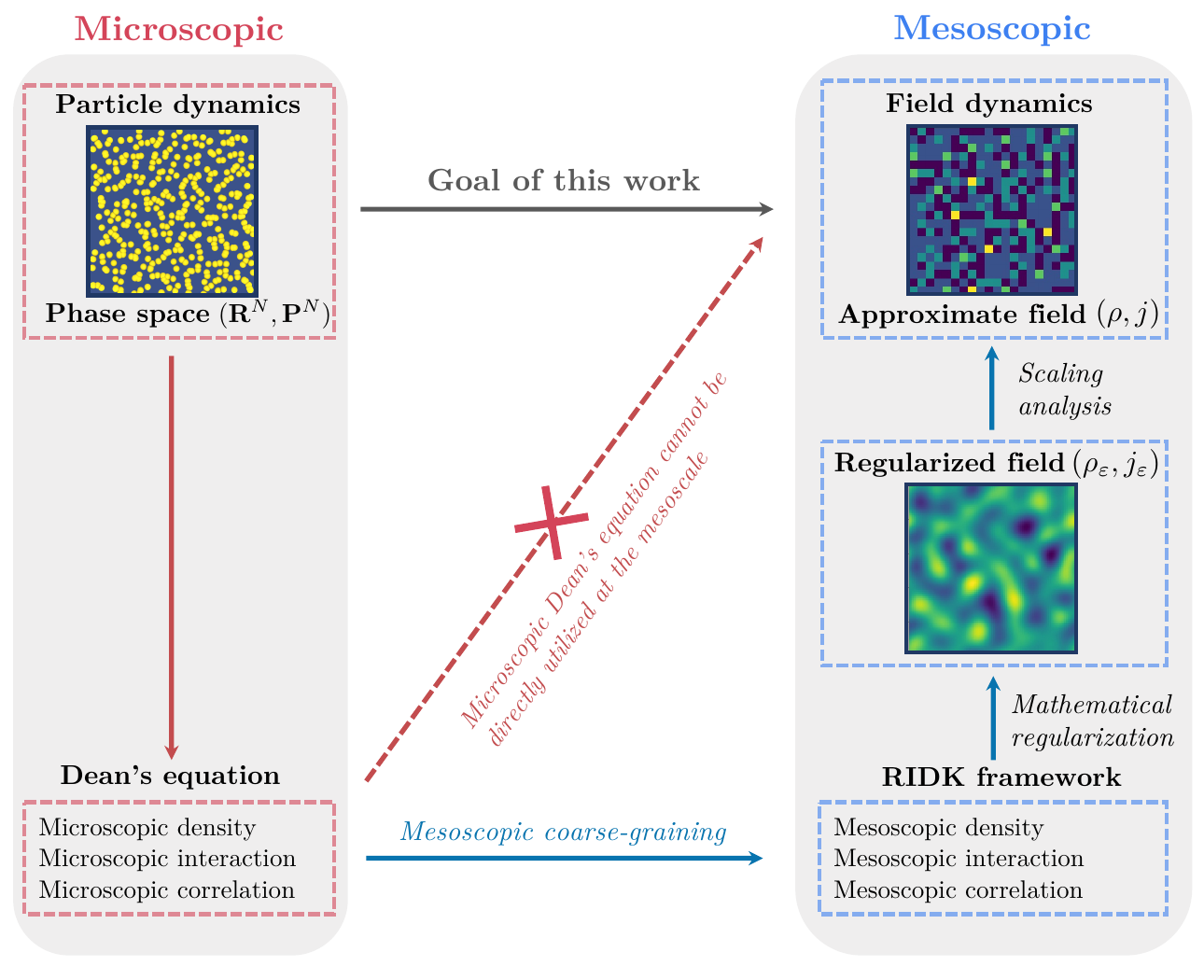}
    \caption{\label{fig:figure1} Schematic diagram illustrating the path taken in this work: bridging microscopic particle dynamics and mesoscopic field dynamics via Dean's equation. Since the original Dean's equation, involving microscopic density, interactions, and correlations (red), cannot be directly applied at the mesoscopic field level, mesoscopic coarse-graining is required under the RIDK framework (blue). From the mathematically regularized fields $(\rho_\varepsilon, j_\varepsilon)$, we perform a scaling analysis to approximate these fields as $(\rho,j)$ under specific conditions, allowing us to efficiently propagate the field dynamics of $(\rho,j)$ using the finite element method.}
\end{figure}
\subsection{Mathematical Problems with Dean's Equation}
The critical caveat with Dean's equation is that the density field employed in the derivation of Dean's equation is a singular distribution. Hence, relying solely on Dean's original formulation does not offer a route for direct simulation by conventional numerical techniques, e.g., finite element methods, because the finite element method discretizes space and time into finite elements under the assumption that fields are continuous and differentiable \cite{logan2011first}. More importantly, a mathematical issue arises from the stochastic multiplicative noise term, as traditional methods employ the conventional definition of the divergence operator for continuous fields, which does not apply to the microscopic density operator. 

From a mathematical perspective, solutions of SPDEs of the form in Eq. (\ref{eq:dk}) are generally not suitable for numerical implementation due to the low regularity of stochastic noise. In particular, Dean's equation poses several difficulties: (1) the regularity of the divergence operator is unknown, and (2) even if the noise is sufficiently regular, the square root function still remains irregular and lacks Lipschitz continuity (see Ref. \onlinecite{cornalba2019modelling} for a detailed mathematical discussion of these issues). These characteristics often prevent finding strong or mild solutions \cite{prevot2007concise, da2014stochastic}, and thus pose a significant challenge in determining the existence and uniqueness of solutions to Dean's equation, which is a mathematically ill-posed conservative supercritical singular SPDE \cite{hairer2014theory, gubinelli2015paracontrolled}.

Since Dean's equation contains the same information as the original Langevin equations, the trivial solution is the empirical microscopic density field. Importantly, von Renesse and coworkers further proved that no smooth solutions other than this trivial atomic measure exist; they demonstrated this argument for a one-dimensional (1D) purely diffusive system \cite{konarovskyi2019dean} and later extended it to more general smooth drift potentials \cite{konarovskyi2020dean}. 

Altogether, while Eqs. (\ref{eq:dean}) and (\ref{eq:dean_general}) formally provide the correct foundation for a field-theoretic description, the mathematically ill-defined nature of Dean's equation prevents its direct numerical implementation for mesoscopic field dynamics. 

\section{Mathematically Regularized Dean's Equation} 
\subsection{Regularization as Coarse-Graining}
The ill-posed nature of Dean's equation has led to several mathematical regularization approaches. Early work by von Renesse and coworkers corrected the drift term to ensure the existence of solutions \cite{von2009entropic, andres2010particle, konarovskyi2019modified, konarovskyi2017reversible}, but left the noise term unregularized and formulated only in terms of probabilistic arguments without a link to the underlying microscopic description \cite{cornalba2019modelling}. Here, we instead adopt the regularization approach by Zimmer, Shardlow, and Cornalba to regularize the smooth noise coefficient and the colored noise \cite{cornalba2019regularized,cornalba2020weakly,cornalba2021well}. This approach replaces the singular Dirac delta representation of particles with a smooth kernel by defining the $\varepsilon$-regularized density with $\varepsilon$ as the kernel length scale as
\begin{equation} \label{eq:rho}
    \rho_\varepsilon(x,t) := \frac{1}{N} \sum_{i=1}^N \omega_\varepsilon(x-q_i(t)),
\end{equation}
where $\omega_\varepsilon(x)$ regularizes the microscopic Dirac delta distribution \cite{cornalba2019regularized} and ensures that the noise field is smooth and its divergence is well-defined. To further ensure numerical stability and preserve physical meaning, we follow the underdamped extension by Cornalba \textit{et al.} and introduce the regularized momentum density \cite{cornalba2021well}:
\begin{equation} \label{eq:j}
    j_\varepsilon(x,t) := \frac{1}{N} \sum_{i=1}^N p_i(t) \omega_\varepsilon(x-q_i(t)),
\end{equation}
which avoids the need to directly compute the divergence of the microscopic density operator for the noise and offers a more comprehensive representation of the target system compared to the overdamped dynamics \cite{von2009entropic, andres2010particle, konarovskyi2019modified, konarovskyi2017reversible}, while maintaining the same interpretability as the original Dean--Kawasaki dynamics with a multiplicative noise structure in divergence form \cite{cornalba2021well}. As in Refs. \onlinecite{von2009entropic, andres2010particle, konarovskyi2019modified, konarovskyi2017reversible}, we rescale the field variables in Eqs. \eqref{eq:rho} and \eqref{eq:j} by the number of particles, which allows $\rho_\varepsilon(x,t)$ to be well-defined as $N\rightarrow \infty$.

The smoothing kernel $\omega_\varepsilon$ is chosen based on the nature of interactions. For non-interacting systems (i.e. ideal gas), we adopt the Gaussian kernel \cite{cornalba2019regularized}
\begin{equation}
    \omega_\varepsilon(x) = \frac{1}{(2\pi \varepsilon^2)^{1/2}} \exp \left( -\frac{x^2}{2\varepsilon^2} \right),
\end{equation}
whereas for weakly interacting systems on the flat torus $\mathbb{T} := [0, 2\pi]$, we use its toroidal analog, i.e., the periodic von Mises distribution \cite{forbes2011statistical}:
\begin{eqnarray}
    w_{\varepsilon}(x)
:=Z_{\varepsilon}^{-1}e^{-\frac{\sin^2(x/2)}{\varepsilon^2/2}},\quad Z_{\varepsilon}:=\int_{\mathbb{T}}{e^{-\frac{\sin^2(x/2)}{\varepsilon^2/2}} dx},
\end{eqnarray}
which follows the same scaling as a Gaussian with variance $\varepsilon^2$, see Sec. II A of the Supplemental Material (SM). In the limit $\varepsilon \to 0$, both kernels recover the Dirac delta distribution $\omega_\varepsilon \rightarrow \delta$. Since the von Mises kernel is the periodic counterpart of the Gaussian, we proceed with Gaussian regularization for simplicity and without loss of generality. Physically, this regularization represents a mesoscopic coarse-graining procedure: $\omega_\varepsilon(x)$ acts as a coarse-grained Dirac delta function with a characteristic length scale of $\varepsilon$, bridging the microscopic (particle) and mesoscopic (field) levels. Namely, Eqs. \eqref{eq:rho} and \eqref{eq:j} operate at the microscopic level of Dean's equation to derive the mesoscopic level dynamics.

With this mathematical regularization, our overarching aim is to build a quantitative link between microscopic particle-level and mesoscopic field-level physics, as illustrated in Fig. \ref{fig:figure1}. In the following, we first review the dynamical equations for the regularized fields $(\rho_\varepsilon,\,j_\varepsilon)$ from Ref. \onlinecite{cornalba2019regularized}, then examine how mesoscopic coarse-graining affects the microscopic observables, e.g., density, momentum, and the interaction. This analysis provides the basis for a consistent statistical-mechanical correspondence between particle-based and field-based descriptions. Finally, we numerically implement the regularized mesoscopic Dean--Kawasaki dynamics using finite element methods for interacting fluids to gauge its feasibility as a system-specific simulation method.

\subsection{Regularized Inertial Dean--Kawasaki (RIDK) Framework}

Under the $\varepsilon-$regularization, exact stochastic dynamic equations for the regularized (coarse-grained) field variables differ from the microscopic equation and should be derived from the particle dynamics. Following Ref. \onlinecite{cornalba2021well}, we consider the underdamped Langevin dynamics interacting with 
nontrivial potential $V$: 
\begin{equation}
  \left\{
    \begin{array}{l}
      \displaystyle \dot{q}_i= p_i,  \vspace{0.2 pc}\\
      \displaystyle \dot{p}_i= -\gamma p_i-\frac{1}{N}\sum_{j=1}^{N}{V'(q_i-q_j)}+\sigma\,\dot{\beta}_i,\qquad i=1,\dots,N,
    \end{array}
  \right.
\end{equation}
where $\beta_i$ is a family of independent Brownian motions and the friction $\gamma$ and diffusion $\sigma$ determine the temperature via the fluctuation-dissipation relation, $k_B T = \sigma^2/(2\gamma).$ 

From density conservation and Ito's formula \cite{oksendal2013stochastic}, the evolution equations for $(\rho_\varepsilon, j_\varepsilon)$ are found to be
\begin{widetext}
    \begin{align}
    \frac{\partial \rho_\varepsilon}{\partial t}(x,t) &= -\frac{\partial j_\varepsilon}{\partial x} (x,t), \label{eq:rhoepsilon} \\
    \frac{\partial j_\varepsilon}{\partial t}(x,t) &= -\gamma j_\varepsilon(x,t) - j_{2,\varepsilon}(x,t) - \frac{1}{N} \sum_{i=1}^N \left( \frac{1}{N} \sum_{j=1}^N V'(q_i(t)-q_j(t)) \right)\omega_\varepsilon(x-q_i(t))+\frac{\sigma}{N} \sum_{i=1}^N \omega_\varepsilon(x-q_i(t))\dot{\beta}_i ,\label{eq:jepsilon}
\end{align}
\end{widetext}
where the auxiliary variable $j_{2,\varepsilon}$ is defined as
\begin{align}
    j_{2,\varepsilon}(x,t) = \frac{1}{N}\sum_{i=1}^N p_i^2 (t)\omega_\varepsilon'(x-q_i(t)),  
\end{align}
which introduces a hierarchy involving higher-order moments of the regularized density operator, e.g., $j_{2,\varepsilon},\ j_{3,\varepsilon},\cdots$, similar to that found in the direct consideration of Dean's equation in Sec. II. Therefore, a closure relationship is needed. Another issue arises from the stochastic term, $\sigma \sum_{i=1}^N \omega_\varepsilon(x-q_i(t))\dot{\beta}/N$, which cannot be expressed solely in terms of $\rho_\varepsilon$ or $j_\varepsilon$, and thus requires approximation to obtain a closed form. 

To this end, we adopt the Regularized Inertial Dean--Kawasaki (RIDK) model developed by Cornalba, Shardlow, and Zimmer through key approximations that close this stochastic dynamics \cite{cornalba2021well, cornalba2023regularised}:
\begin{align}
\label{eq:finalridk}
\frac{\partial \rho_{\varepsilon}}{\partial t}(x,t) =& -\frac{\partial j_{\varepsilon}}{\partial x}(x,t), \\
\frac{\partial  j_{\varepsilon}}{\partial t}(x,t)  =& 
  -\gamma  j_{\varepsilon}(x,t)-\left(\frac{\sigma^2}{2\gamma}\right)\frac{\partial  \rho_{\varepsilon}}{\partial x}(x,t) \nonumber \\ \label{eq:finalridk2}
&-\{V'\ast \rho_{\varepsilon}(\cdot,t)\} \rho_{\varepsilon}(\cdot,t)
  +\frac{\sigma}{\sqrt{N}}\sqrt{ \rho_{\varepsilon}(x,t)}.
\end{align}

\subsection{Key Approximations for Numerical Implementation}
Equations (\ref{eq:finalridk}) and (\ref{eq:finalridk2}) are closed by introducing four approximations. While detailed derivations are provided in SM Secs. I and II based on Refs. \onlinecite{cornalba2019regularized,cornalba2020weakly,cornalba2021well}, we briefly summarize them here, as they are crucial for the numerical implementation discussed in later sections. The first is the \textit{kinetic approximation}, which assumes the low-temperature limit by replacing $v_i$ with their mean values $ v_i \approx \bar{v}$. In this regime, the variance $\sigma^2/2\gamma$ is controllably small, so the resulting error is negligible, and the kinetic term closes as
\begin{equation} \label{eq:approx1}
    j_{2,\varepsilon} \approx \frac{\sigma^2}{2\gamma} \frac{\partial \rho_\varepsilon}{\partial x},
\end{equation}
which should remain stationary when local equilibrium is maintained. Next, the \textit{regularized stochastic noise}, $\dot{\mathcal{Z}}_N$, is approximated by the closed form $\dot{\mathcal{Y}}_N$, which has the same spatial covariance: 
\begin{equation} \label{eq:approx2}
    \dot{\mathcal{Z}}_N:=\frac{\sigma}{N} \sum_{i=1}^N \omega_\varepsilon(x-q_i(t))\dot{\beta}_i \approx \frac{\sigma}{\sqrt{N}} \sqrt{\rho_{\varepsilon/\sqrt{2}}}Q_{\sqrt{2}\varepsilon}^{1/2} \xi,
\end{equation} 
where $Q_\varepsilon$ is the convolution operator defined with the von Mises kernel $\omega_\varepsilon$. For the \textit{pair interaction terms}, the original interaction term in Eq. (\ref{eq:jepsilon}) is replaced with a nonlocal convolution term in Eq. (\ref{eq:finalridk2}):
\begin{align} \label{eq:approx3}
    &\frac{1}{N}\sum_{i=1}^N \left( \frac{1}{N} \sum_{j=1}^N \nabla V(q_i(t)-q_j(t)) \right) \omega_\varepsilon(x-q_i(t)) \nonumber \\ \approx & (\nabla V \ast \rho_\varepsilon)\rho_\varepsilon(x).
\end{align}

Finally, to mitigate numerical instabilities, the \textit{finite element approximation} explicitly tracks the mesoscopic field $(\rho,j)$ by approximating 
\begin{equation} \label{eq:approx4}
    (\rho,j)\approx (\rho_\varepsilon,j_\varepsilon).
\end{equation}

\section{Scaling Analysis of the RIDK Model}
\subsection{Scaling Analysis and Coarse-Graining}
Cornalba and coworkers have demonstrated that, under these approximations [Eqs.~\eqref{eq:approx1}-\eqref{eq:approx4}], the RIDK model is well-posed and possesses a nontrivial solution, unlike the original microscopic case \cite{cornalba2019regularized,cornalba2020weakly,cornalba2021well}. Nevertheless, this framework has not yet been numerically tested on systems with nontrivial pair interactions. For the framework to be applicable to interacting systems, one should carefully select $\varepsilon$ to ensure that the error from closing the equations remains bounded and small. Since $\varepsilon$ determines the size of the regularization, it can be physically interpreted as the \textit{level of coarse-graining}, where $\varepsilon$ determines the characteristic length scale. 

In the RIDK framework, the mathematical conditions on $\varepsilon$ that bound these errors are governed by the scaling relationship \cite{cornalba2019modelling}:
\begin{eqnarray}
    N\varepsilon^\theta = 1,
\end{eqnarray}
where $\theta$ is a positive exponent that depends on the specific setting of the SPDE. The optimal $\theta=\theta_0$ is determined by deriving the particular condition $\theta\ge\theta_0$ that can bound the errors from the four approximations [Eqs. \eqref{eq:approx1}--\eqref{eq:approx4}]. Since a larger value of $\theta$ implies a larger particle size with greater $\varepsilon$ values, $\theta_0$ defines the optimal level of coarse-graining while retaining possible details beyond a certain scale $\varepsilon_0$. 

We highlight that this scaling idea serves as a building block for numerically implementing the mesoscopic coarse-graining of molecular systems onto a real-space grid. As previous mathematical work on the RIDK focused mainly on ideal gas systems, it did not explicitly involve systematic coarse-graining. Here, we integrate coarse-graining with the scaling analysis provided by the RIDK model to implement this approach for mesoscopic field-level dynamics with the goal of systematically bridging the microscopic and mesoscopic regimes. In doing so, determining optimal $\theta$ values is essential for minimizing error bounds in numerical RIDK simulations of interacting systems. 

\subsection{Stochastic Noise Approximation}
The stochastic noise terms $\dot{\mathcal{Z}}_N$ and $\dot{\mathcal{Y}}_N$ are not identical, and hence to faithfully apply the final RIDK model to molecular systems, the difference $\mathcal{R}_N :=\mathcal{Z}_N - \mathcal{Y}_N$ should be analytically identified to examine the error bound under coarse-graining. For non-interacting systems, the error bound for the covariance between $\mathcal{Z}_N$ and $\mathcal{Y}_N$ can be sharply formulated (see Ref. \onlinecite{cornalba2019regularized} and SM Sec. I). For interacting systems, despite a less sharp and complex form of the error bound, Ref. \onlinecite{cornalba2020weakly} showed that this bound can be expressed as
\begin{widetext}
    \begin{align} \label{eq:inequal1}
\left|\mathbb{E}\left[\mathcal{Z}_N(x_1,t)\mathcal{Z}_N(x_2,t)\right]-\mathbb{E}\left[{\mathcal{Y}_N(x_1,t)\mathcal{Y}_N(x_2,t)}\right]\right|  & \leq  \frac{C\sigma^2}{N}w_{\sqrt{2}\varepsilon}(x_1-x_2)\left\{|x_1-x_2|+\varepsilon^{c_1(\theta)}+\varepsilon^{\alpha}+\varepsilon^{c_2(\theta)}|x_1-x_2|^{\frac{1}{2}}\right\} \nonumber \\ &+ \frac{C\sigma^2}{N}\varepsilon^{\alpha},
\end{align}
\end{widetext}
where $C,\,c_1(\theta)$, and $c_2(\theta)$ are positive constants for sufficiently large $\theta$ following the scaling $N\varepsilon^\theta = 1$. Here, $\mathbb{E}[\cdot]$ denotes the expectation value of a random variable over its defined probability space. This stochastic error can then be approximately bounded as
\begin{align} \label{eq:inequal2}
&\left|\mathbb{E}\left[\mathcal{Z}_N(x_1,t)\mathcal{Z}_N(x_2,t)\right]-\mathbb{E}\left[{\mathcal{Y}_N(x_1,t)\mathcal{Y}_N(x_2,t)}\right]\right| \nonumber \\
\lesssim & \, C\sigma^2 N^{-1} w_{\sqrt{2}\varepsilon}(x_1-x_2) |x_1-x_2|^\alpha,
\end{align}
where the exponent in the error bound $\alpha \approx 1$ ($\alpha=2$ for non-interacting independent particles, see SM Sec. I). Additionally, the relationship $\left| \mathbb{E}[\mathcal{Z}_N(x_1,t)\mathcal{Z}_N(x_2,t)]\right| \lesssim N^{-1} \omega_{\sqrt{2}\varepsilon}(x_1-x_2)$ is derived for interacting systems \cite{cornalba2020weakly}. 

Equations (\ref{eq:inequal1}) and (\ref{eq:inequal2}) form the basis for estimating the error incurred when substituting noise terms. We first consider the relative error, defined as $\left(\mathrm{size}\mathcal{Z}_N-\mathrm{size}\mathcal{Y}_N\right)/\mathrm{size}\mathcal{Z}_N$, where the \textit{size} denotes the magnitude of the spatial covariance of a given noise term. For coordinates $x_1$ and $x_2$ separated by less than $\varepsilon$, i.e., $|x_1-x_2|\lesssim\varepsilon$, we have
\begin{equation} \label{eq:relerror}
    \frac{\left|\mathbb{E}\left[\mathcal{Z}_N(x_1,t)\mathcal{Z}_N(x_2,t)\right]-\mathbb{E}\left[{\mathcal{Y}_N(x_1,t)\mathcal{Y}_N(x_2,t)}\right]\right| }{\left| \mathbb{E}[\mathcal{Z}_N(x_1,t)\mathcal{Z}_N(x_2,t)]\right|} \propto \varepsilon^\alpha,
\end{equation}
indicating that the relative error scales as $\varepsilon^\alpha$. For distant points ($|x_1-x_2|>1$), $N^{-1}\omega_{\sqrt{2}\varepsilon}(x_1-x_2)$ decays exponentially, making both $\mathcal{Z}_N$ and $\mathcal{Y}_N$ negligibly small. For intermediate distances ($\varepsilon \lesssim |x_1 - x_2| \lesssim 1$), we can write $|x_1 - x_2| \propto a\varepsilon$ for some $a \in {1, 2, \ldots, \varepsilon^{-1}}$, and the relative error becomes $a^{-\alpha} \varepsilon^\alpha < 1$ \cite{cornalba2019regularized}.

Another way to assess this error is by computing the maximum possible value of the noise difference. Optimizing over $|x|$ yields the maximum noise difference for $|x|\propto \varepsilon$, leading again to a relative error scaling as $\varepsilon^\alpha$, consistent with Eq. (\ref{eq:relerror}).

\subsection{Kinetic Approximation}
The low-temperature limit [Eq. (\ref{eq:approx1})] is suitable for typical simulations of interacting fluids, as nontrivial structural correlations generally emerge at relatively low temperatures. To estimate the kinetic contribution, we consider
\begin{align} \label{eq:kinetic2}
    \mathbb{E}\left[\left\Vert N^{-1} \sum_{i=1}^N (v_i^2 -\bar{v}^2)\omega_\varepsilon'(x-q_i(t))\right\Vert^2\right].
\end{align}
Expanding the square and taking expectation values yields $N^{-2}$ multiplied by a double sum whose cross terms vanish, leaving a diagonal contribution proportional to $\textrm{Var}(v^2) N^{-1} \Vert \omega_\varepsilon'\Vert^2$. Since at low temperature the velocity fluctuations satisfy $\textrm{Var}(v^2)=\mathcal{O}(T^2)$, and $\Vert \omega_\varepsilon'\Vert^2 \approx \varepsilon^{-d-2}$ due to $\omega_\varepsilon(x)=\varepsilon^{-d}\omega(x/\varepsilon)$, the bound of Eq. \eqref{eq:kinetic2} becomes $\mathcal{O}(T^2 N^{-1}\varepsilon^{-d-2}).$ Under $N\varepsilon^\theta =1$ scaling, this reduces to $\mathcal{O}(T^2\varepsilon^{\theta-d-2})$. Hence, for $\theta>d+2$, the kinetic error vanishes as $\varepsilon \rightarrow 0$ and is bounded in terms of $\varepsilon$. 

\subsection{Convolution Approximation}
According to Ref. \onlinecite{cornalba2020weakly}, the stochastic remainders derived from the difference between the pair interactions $V(q_i(t)-q_j(t))$ and the convolution term $\left\{V'\ast \rho_\varepsilon(\cdot,t) \right\}(x)\rho_\varepsilon(x,t)$ can be further reduced and bounded as (see SM Sec. II):
\begin{widetext}
\begin{align} \label{eq:conv_error}
    \left| \frac{1}{N}\sum_{i=1}^N\left(\frac{1}{N}\sum_{j=1}^N V'(q_i(t)-q_j(t)) \right) \omega _\varepsilon (x-q_i(t)) - \left\{V'\ast \rho_\varepsilon(\cdot,t) \right\}(x)\rho_\varepsilon(x,t)\right| \lesssim  r_{1,\varepsilon}\rho_\varepsilon(x,t)+r_{2,\varepsilon},
\end{align}
\end{widetext}
where, for sufficiently smooth regularization, this error is correctly bounded, and one can roughly estimate bounds using a strong Sobolev norm. Namely, under $N\varepsilon^\theta=1$, the error can be estimated as
\begin{equation} \label{eq:sobolev}
    r_{1,\varepsilon}\lesssim \left\lVert V\right\rVert_{V^{1,\infty}} \sqrt{\varepsilon} ,\qquad r_{2,\varepsilon} \lesssim \left\Vert V\right\Vert_{V^{2,\infty}} \sqrt{\varepsilon},
\end{equation}
where $\left\Vert \cdot \right\Vert_{V^{1,\infty}}$ and $\left\Vert \cdot \right\Vert_{V^{2,\infty}}$ denote the strong Sobolev norms. These norms provide a quantitative estimate of the convolution error's order of magnitude. We will later assess this approximation by evaluating these norms for realistic pair interactions derived from mesoscopic coarse-graining combined with numerical implementation in Appendix B.

\subsection{Optimal $\varepsilon$ for Scaling Analysis}
\subsubsection{$N\varepsilon^\theta=1$ Analysis: Strong Metrics}
Given the error magnitudes as a function of $\varepsilon$, a key challenge in implementing the RIDK model is determining an appropriate bound for $\varepsilon$ or $\theta$ via $N\varepsilon^\theta=1$. While larger $\varepsilon$ reduces approximation errors, overly large $\varepsilon$ leads to excessively regularized fields and negatively impacts the underlying reference correlations (see Sec. V). A lower bound, $\varepsilon_0$, can be determined from the minimum admissible $\theta_0$ from the derived error bound: $\varepsilon \ge \varepsilon_0 = (1/N)^{1/\theta_0}.$

We note that the original RIDK framework for non-interacting particles in one dimension using the Gaussian kernel was developed under several different regularization conditions. To ensure tightness of $\{\rho_\varepsilon\}_\varepsilon$ and $\{j_\varepsilon\}_\varepsilon$ (i.e., to prevent divergence), $\theta_0 \ge 3$ is required; for $\{j_{2,\varepsilon}\}_\varepsilon$, a stronger condition $\theta_0 \ge 5$ is needed. Replacing the noise terms with bounded error requires $\theta_0 \ge 7/2$, and the high-probability existence and uniqueness of the solution requires $\theta_0 > 7$ \cite{cornalba2019regularized}. However, these thresholds were not optimized in Ref. \onlinecite{cornalba2019regularized}, and lower $\theta_0$ could be beneficial for practical implementations while maintaining key physical correlations.

Subsequent work extended this framework to weakly interacting systems using auxiliary Langevin dynamics \cite{cornalba2020weakly}, while sufficiently large $\theta$ values were assumed to ensure the existence of a well-defined mild solution. A more refined condition on $\theta_0$ was later introduced in Ref. \onlinecite{cornalba2021well}, which showed that for a $d$-dimensional torus $\mathbb{T}^d$, the RIDK equation is well-posed under:
\begin{equation} \label{eq:scaling1}
    N\varepsilon^\theta = 1, \qquad \theta=\theta_0 >2d,
\end{equation}
which reduces the earlier 1D requirement from $\theta_0=7$ to slightly larger than 2. For general dimensions, Eq. (\ref{eq:scaling1}) can be interpreted as a condition on the effective particle volume $v$, namely $Nv^2\approx1$. 

Altogether, different $\theta_0$ values have been derived to provide sufficient regularity for a Sobolev space analysis, and we refer these analyses as providing a \textit{strong metric}. While strong Sobolev norms are useful for bounding and controlling residuals, they also imply large $\varepsilon$, which can lead to excessive particle overlap and diminish the ability to describe structural features. In our case, implementing RIDK in two-dimensional (2D) systems implies a typical threshold of $\theta \approx 5$. As we will show in Sec. V, this requirement can be severely limiting, since large $\varepsilon$ values tend to smooth out essential correlations.

\subsubsection{Weaker Metric: $\theta_0\approx2$}
To reduce $\varepsilon$ in the RIDK implementation, it is desirable to consider a \textit{weaker metric}. While choosing $\varepsilon$ smaller than $\varepsilon_0$ implied by the strong Sobolev analysis may yield less smooth fields $\rho_\varepsilon$ and $j_\varepsilon$, this situation can be improved in practice by evaluating these fields against smooth test functions ${\varphi} = (\varphi_1, \varphi_2)$ via $\int_{\mathbb{T}^d} \rho_\varepsilon(x,t)\varphi_1(x) dx$ and $\int_{\mathbb{T}^d} j_\varepsilon(x,t) \varphi_2(x) dx$, respectively. The authors of Ref. \onlinecite{cornalba2023dean} adopted this approach and employed a finite difference scheme to solve the Dean--Kawasaki dynamics under the \textit{high-density condition}: 
\begin{equation} \label{eq:scaling2}
    N \ge h^{-d},
\end{equation}
where the grid spacing $h$, instead of $\varepsilon$, is introduced by the finite difference scheme. Equation (\ref{eq:scaling2}) implies that the number of particles should exceed the number of grid points in a $d$-dimensional domain. Assuming $h$ is of a similar order to the coarse-graining scale $\varepsilon$, this weaker norm relaxes the requirement from $\theta_0>2d$ to $\theta_0\approx d$. Using this weaker metric, SM Sec. IV demonstrates and validates a proof-of-concept implementation of particle-level interactions on mesoscopic fields. However, the precise relationship between $\varepsilon$ and $h$ remains unclear, and we will address this in Sec. V by establishing a systematic connection between these two coarse-graining lengths. We note that this weaker norm condition ($\theta_0\approx d$) is consistent with the analytical results of Ref. \onlinecite{fehrman2023non}, which established well-posedness under a noise truncation procedure using a finite number of Fourier modes, a method numerically analogous to imposing the high-density condition in Eq. (\ref{eq:scaling2}). 

In summary, recent analytical and numerical studies suggest that weaker norms, rather than strict Sobolev norms under a high-density setting, can be employed for RIDK implementations. Nonetheless, two key questions remain: (1) What is the correct correspondence between the regularization scale $\varepsilon$ and the numerical grid spacing $h$? and (2) what does the high-density setting physically mean in the context of molecular systems? As this condition has thus far been imposed only in the mathematical context, its physical and practical implications remain ambiguous. In the following sections, we focus on field-level correlations and use the aforementioned findings to answer the two questions above.

\section{Mesoscopic Coarse-Graining on A Real-Space Grid}
\subsection{Mesoscopic Coarse-Graining and Correlations}
We start by analyzing the effects of coarse-graining on the density field and structural correlations using a Gaussian kernel form for simplicity (extendable to toroidal domains via the von Mises kernel). For a 2D system with particle positions $\mathbf{r}_i$,  the regularized density is given by 
\begin{align} \label{eq:gaussdist}
     \rho_\varepsilon(\mathbf{r})=\sum_{i=1}^N \frac{1}{2\pi\varepsilon^2} \exp \left[ -\frac{1}{2}\frac{(\mathbf{r}-\mathbf{r}_i(t))^2}{\varepsilon^2} \right].
\end{align}
As shown in Fig. 3, increasing the coarse-graining length $\varepsilon$ smooths the density field by blurring the underlying microscopic configurations and correlations.

\begin{figure*}
    \includegraphics[width=0.85\textwidth]{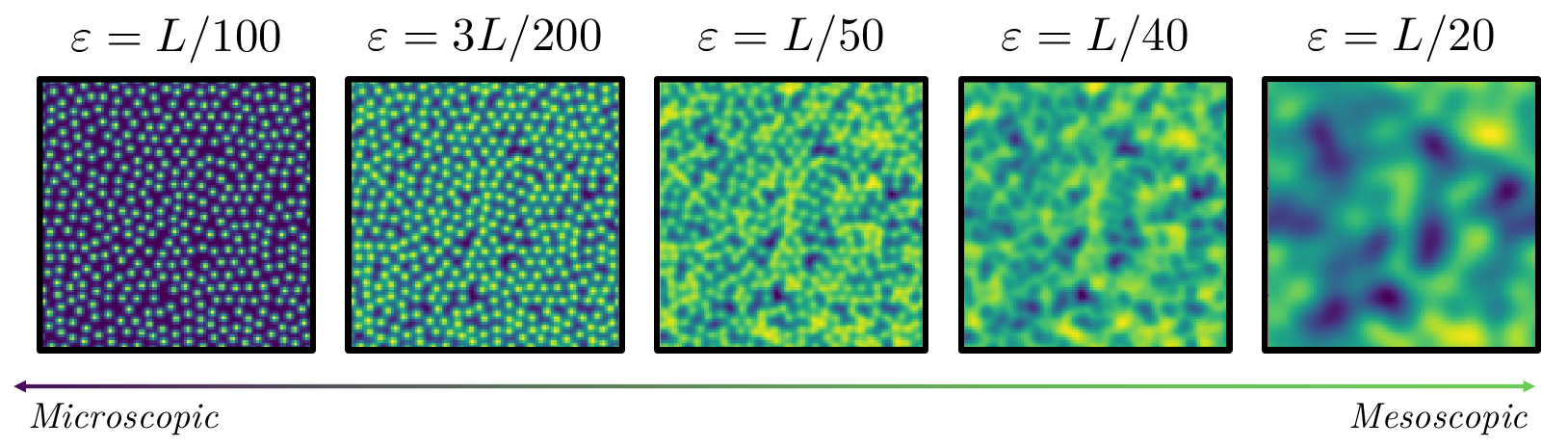}
    \caption{\label{fig:figure3} Effect of mesoscopic coarse-graining on molecular configuration. Here, we compare various levels of mesoscopic coarse-graining applied to the MD simulation snapshot of the Gaussian core model (box size: $L=2\pi$) by imposing Gaussian kernels with variances $\varepsilon^2$ from $\varepsilon=L/100$ to $L/20$ at each particle position. For small $\varepsilon$ (left), the coarse-grained configuration resembles the microscopic trajectory, while larger $\varepsilon$ (right) smooths out microscopic details, consistent with $\rho_\varepsilon(\mathbf{k},t)=e^{-\tfrac{1}{2}\mathbf{k}^2\varepsilon^2}\hat{\rho}(\mathbf{k},t)$.}
\end{figure*}

The effect of mesoscopic coarse-graining on structural and dynamical correlations can be quantified by examining the dynamic structure factor $F_\varepsilon(\mathbf{k},t):=\langle \rho_\varepsilon(\mathbf{k},t)\rho_\varepsilon(-\mathbf{k},t)\rangle$. We focus here on its static component, $S_\varepsilon(\mathbf{k})=F_\varepsilon(\mathbf{k},0)$. For the Gaussian kernel [Eq. (\ref{eq:gaussdist})], the coarse-grained density is $\rho_\varepsilon(\mathbf{k},t) = e^{-\frac{1}{2} \mathbf{k}^2 \varepsilon^2} \hat{\rho}(\mathbf{k},t)$, where $\hat{\rho}(\mathbf{k}) = \sum_i e^{i\mathbf{k}\cdot\mathbf{r}_i(t)}$ denotes the microscopic density mode, and $S_\varepsilon(\mathbf{k})$ is expressed as
\begin{align} \label{eq:gauss}
    S_\varepsilon(\mathbf{k})=\langle \rho_\varepsilon(\mathbf{k},0) \rho_\varepsilon(\mathbf{-k},0)\rangle = e^{-\mathbf{k}^2 \varepsilon^2} S(\mathbf{k}),
\end{align}
with the microscopic static structure factor is defined as $S(\mathbf{k})=\langle \hat{\rho}(\mathbf{k})\hat{\rho}(\mathbf{-k})\rangle/N$. The same relation holds for $F_\varepsilon(\mathbf{k},t)$. Thus, Eq. (\ref{eq:gauss}) shows that mesoscopic coarse-graining with $\varepsilon$-regularization effectively filters $S(\mathbf{k},t)$ through the Gaussian kernel.

Since the scale of the coarse-graining, $\varepsilon$, is constrained by $N\varepsilon^\theta=1$ to bound the error residuals, this finding also suggests that numerically discretizing the mesoscopic field through coarse-graining Dean's equation cannot fully capture microscopic (particle-level) structural or dynamical correlations. At best, it reproduces the Gaussian-smeared correlations set by the chosen coarse-graining length, implying that the RIDK model inevitably loses structural and dynamical detail by design.

\subsection{Grid Size--Coarse-Graining Length Correspondence}
In practice, RIDK simulations are implemented on a numerical grid with spacing $h=2\pi/n_g$ by approximating $(\rho_\varepsilon,j_\varepsilon)\approx(\rho,j)$ under the scaling relation $N\varepsilon^\theta=1$. Since $\varepsilon$ does not appear explicitly in the numerical scheme, the grid spacing $h$ acts as the effective coarse-graining scale. Thus, understanding the relationship between $\varepsilon$ and $h$ is essential for numerically implementing the $\varepsilon$-based RIDK framework. 

Since the grid density $\rho_h$ represents the coarse-grained density after discretization, it does not inherently carry the coarse-graining kernel $\omega_\varepsilon$ or $\varepsilon$, and the direct relationship between $\varepsilon$ and $h$ remains unclear. Nevertheless, inspired by the observation that one can discretize the configuration with spacing $h$ by histogramming particle positions to construct $\rho_h$ and evaluate density correlations $\langle \rho_h(\mathbf{k},t)\rho_h(-\mathbf{k},0)\rangle$, we establish an indirect link between $\varepsilon$ and $h$ by matching $S(\textbf{k})$. Specifically, we determine $\varepsilon$ that yields the structure factor $S_\varepsilon(\textbf{k})$ computed from the $\rho_\varepsilon(\textbf{k})$ that most closely matches $S_h(\textbf{k})$ from the grid-based density for a given discretization $h$.

In practice, we first select a desired level of mesoscopic coarse-graining by choosing the specific number of grid points $n_g^\ast$. This determines the grid spacing $h^\ast=L/n_g^\ast$, which corresponds to $2\pi/n_g^\ast$ in $\mathbb{T}$ for the RIDK simulation. From the microscopic trajectories, we then construct a density histogram on the $n_g^\ast \times n_g^\ast$ grid to obtain $\rho_{h^\ast}(\mathbf{r})$. A numerical Fourier transform gives $\rho_{h^\ast}(\mathbf{k})$, from which we evaluate the structure factor $\langle \rho_{h^\ast}(\mathbf{k}) \rho_{h^\ast}(-\mathbf{k}) \rangle$. To determine the optimal coarse-graining length $\varepsilon^\ast$ (or $n_g^\ast$), we generate a set $\varepsilon = L/n,2L/n,\cdots,L$ for sufficiently large $n$. For each $\varepsilon$, we construct a Gaussian-kernel coarse-grained density $\rho_\varepsilon$ of width $\varepsilon$ and compute its structure factor $\langle \rho_\varepsilon(\mathbf{k}) \rho_\varepsilon(-\mathbf{k}) \rangle$. The optimal $\varepsilon^\ast$ is identified by minimizing the difference between the $h^\ast$- and $\varepsilon$-dependent radial distribution functions (RDFs) obtained from these structure factors.

\subsection{Grid Size--Coarse-Graining Length Correspondence: Results}
\begin{figure*}
    \includegraphics[width=1\textwidth]{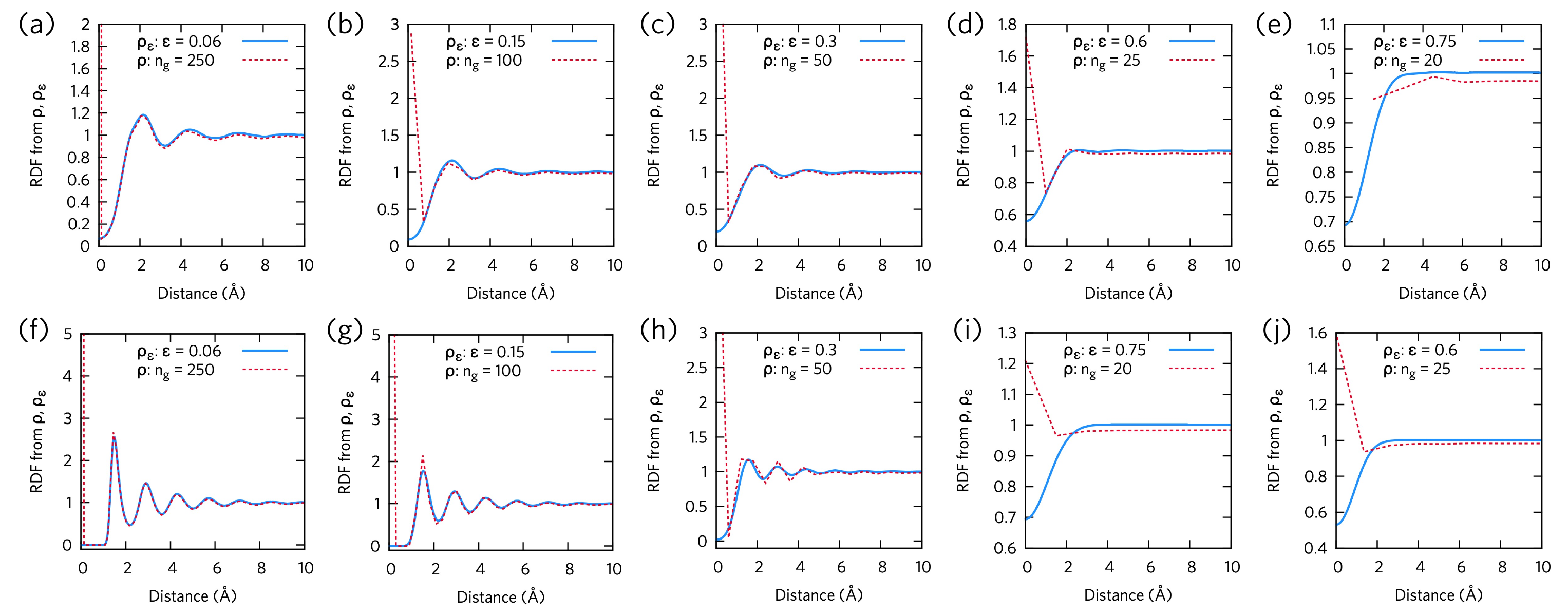}
    \caption{\label{fig:figure4} Structural correspondence between the number of numerical grid points $n_g$ (or grid size $h$) used in the finite element method and the mesoscopic coarse-graining length $\varepsilon$ for interacting particles. This correspondence was established by examining the coarse-grained RDF, equivalent to $g(r)$ defined as $g(\textbf{r}):=\langle \sum_i \delta(\mathbf{r}-\mathbf{r}_i)\rangle/\rho$, at varying $n_g$ and $\varepsilon$ values. We find that two differently coarse-grained RDFs are nearly identical when $h=2\varepsilon$ across different coarse-graining levels: (a, f) $\varepsilon=0.06\,\textrm{\AA}$ with $n_g=250$, (b, g) $\varepsilon=0.15\,\textrm{\AA}$ with $n_g=100$, (c, h) $\varepsilon=0.3\,\textrm{\AA}$ with $n_g=50$, (d, i) $\varepsilon=0.6\,\textrm{\AA}$ with $n_g=25$, and (e, j) $\varepsilon=0.75\,\textrm{\AA}$ with $n_g=20$. Notably, this finding also holds for different interaction types: (a-e) Gaussian core model and (f-j) bare Lennard-Jones model, where the RDFs are indistinguishable from those of the regularized potentials (see Sec.~IX)}.
\end{figure*}

We first examine the effect of Gaussian coarse-graining on the RDF of the Gaussian core model using MD simulations (for computational details, see SM Sec. V). At small $\varepsilon$ ($\varepsilon=0.06\,\textrm{\AA}$), the coarse-grained RDF reproduces the microscopic reference, while increasing $\varepsilon$ smooths out structural correlations until they disappear beyond $\varepsilon \gtrsim 0.3\,\textrm{\AA}$. To compare with grid-based discretization, we computed RDFs using different $n_g^\ast$ values and compared them with the $\varepsilon$-smoothed results [Fig. \ref{fig:figure4}(a)-(e)], omitting the artificial zero-distance peak introduced by discretization. As expected, decreasing $n_g^\ast$ produces the same trend as increasing $\varepsilon$, with correlations washed out at coarse resolution. Interestingly, the Gaussian kernel with $\varepsilon^\ast = h^\ast/2 = L/(2n_g^\ast)$ yields an RDF nearly identical to that from an $n_g^\ast \times n_g^\ast$ grid, establishing the correspondence $h \approx 2\varepsilon$ for Gaussian core models.

We further validated this correspondence in a bare Lennard-Jones liquid. While the atomistic RDF exhibits pronounced structure, mesoscopic correlations decay from the reference ($n_g^\ast\approx250$) toward ideal-gas-like behavior ($n_g^\ast\approx20$) as $h^\ast=L/n_g^\ast$ increases, mirroring the effect of larger $\varepsilon$. Again, we find that $h \approx 2\varepsilon$ holds, with $n_g^\ast=250,100,50,25,$ and $20$ corresponding to $\varepsilon^\ast=0.06,0.15,0.3,0.6,$ and $0.75\,\textrm{\AA}$, respectively [Fig. \ref{fig:figure4}(f)-(j)].

In summary, our analysis indicates that discretizing molecular systems onto a finite numerical grid effectively corresponds to imposing a Gaussian kernel with width $\varepsilon\approx0.5h$. The observed correspondence implies that the open ball of radius $\varepsilon$ should fit within the grid spacing $h$, which is physically reasonable, suggesting that this relationship may generalize to other systems as well. Moreover, establishing the correspondence $h = 2\varepsilon$ allows these two distinct coarse-graining length scales to be used interchangeably and facilitates the simulation of the RIDK equations. Specifically, the high-density condition from Eq. (\ref{eq:scaling2}) \cite{cornalba2023dean} can be further reduced to $N\ge h^{-2} \approx \left( {\varepsilon}/{2} \right)^{-2}$. Compared to Eq. (\ref{eq:scaling1}), this reduces the required scaling exponent from 5 to 2, making the weaker metric condition readily applicable to the RIDK formalism. Accordingly, we adopt $\theta_0 = 2$ throughout the remainder of the manuscript to determine the minimum admissible coarse-graining length.

\section{Grid-based Renormalized Interaction Design by Coarse-Graining (GRID CG)}
\subsection{GRID CG Approach: Principles}
Having established the role of coarse-graining in structural correlations, we now turn to its effect on the interactions. Since mesoscopic coarse-grained configurations exhibit simplified correlations relative to the atomistic reference, the effective interaction between mesoscopic density fields must differ accordingly \cite{kadanoff1966scaling,wilson1971renormalization,wilson1974renormalization,cardy1996scaling}. 
In molecular coarse-graining, this concept is well established: coarse-grained particles interact through simplified yet renormalized potentials compared to their atomistic counterparts \cite{noid2013perspective,saunders2013coarse,jin2022bottom,noid2023perspective, noid2024rigorous}. Numerous ``bottom-up'' coarse-graining methodologies have been developed to systematically derive these interactions from microscopic statistics \cite{reith2003deriving, izvekov2005multiscale1,izvekov2005multiscale2,noid2008multiscale1,noid2008multiscale2,shell2008relative,chaimovich2010relative,chaimovich2011coarse,shell2016coarse}. However, most existing approaches focus on particle-level coarse-grained models, making it difficult to directly derive field-level interactions \cite{jin2024first,jin2025field}. 

\begin{figure*}
    \includegraphics[width=0.85\textwidth]{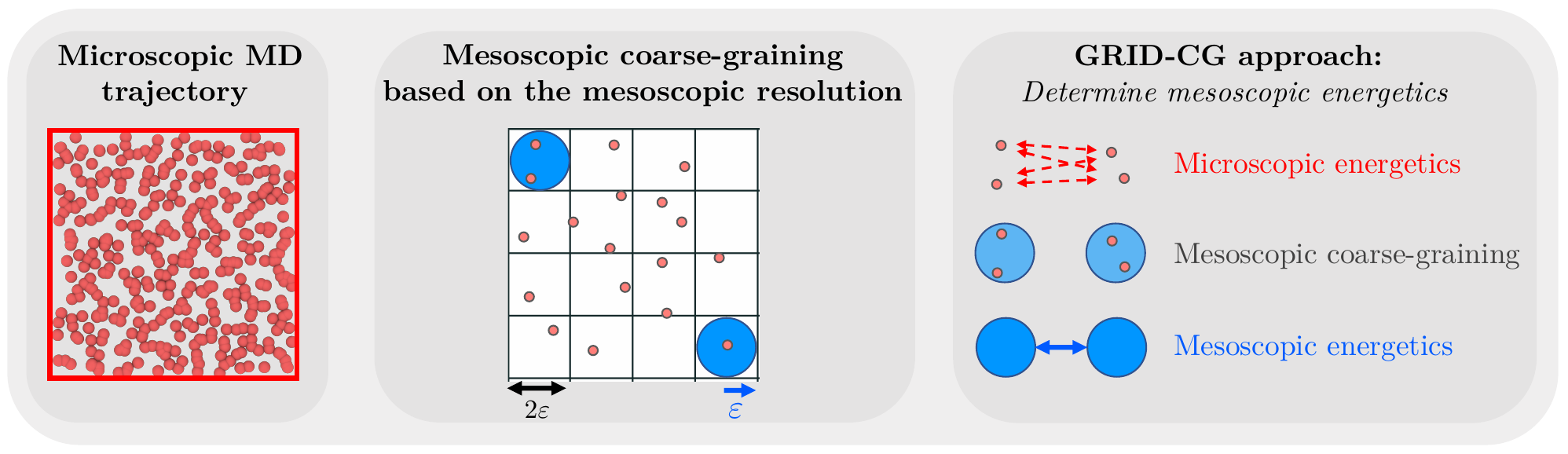}
    \caption{\label{fig:figure5} Essential steps underlying the GRID CG method for determining mesoscopic coarse-grained energetics. (Left): Initially prepared microscopic MD trajectory of the system of interest due to the bottom-up nature of our approach. (Middle): Construction of an approximated mesoscopic density histogram at the specified coarse-grained resolution ($\varepsilon$), where the grid size $h$ corresponds to $2\varepsilon$ based on the established correspondence. (Right): Determination of mesoscopic interaction $V_\mathrm{GRID}$ between grid elements by matching it to the microscopic energetics between particles within each grid element using Eq. (\ref{eq:ECG}).}
\end{figure*}

To address this gap, we propose a parametrization scheme inspired by energy-matching techniques in molecular coarse-graining \cite{wang2009effective,lebold2019dual1,lebold2019dual2}, which we call Grid-based Renormalized Interaction Design by Coarse-Graining (GRID CG). As illustrated in Fig. \ref{fig:figure5}, the statistical mechanical principle underlying GRID CG is to match the effective energetics of the mesoscopic system to those of the reference atomistic system. We assume that interactions between coarse-grained density fields can be represented as interactions between densities at discretized grid points. By matching the effective energetics at the field level to those at the reference molecular level, GRID CG aims to capture the microscopic energetics and associated correlations. 

Unlike particle-level interactions, which are functions of interparticle distances [e.g., $V(q_i(t)-q_j(t))$], field-level interactions are encoded through convolutions with the coarse-grained density field $\rho_\varepsilon(\mathbf{r},t)$. Thus, GRID CG matches the interaction energy between grid points to the overall energy among particles within those grid regions (Fig. \ref{fig:figure5}, middle panel). In essence, GRID CG renormalizes particle-level energetics into discretized grid-level interactions over a finite set of inter-grid distances, $\mathcal{D} = \{h,\,\sqrt{2}h,\,\sqrt{3}h,\,2h,\cdots\}$, where $\mathcal{D}$ depends on the total number of grid points $n_g$. In this context, matching the thermodynamic forces, and thus energies, is known to capture both two-body \cite{chaimovich2010relative} and three-body \cite{noid2007multiscale} correlations in liquid-state systems \cite{hansen2013theory}. While this principle is established for particle-level coarse-graining, capturing configuration-dependent energetics at the grid level is necessary to ensure structural fidelity in field-theoretic numerical simulations.

\subsection{GRID CG Approach: Algorithms}
As illustrated in Fig. \ref{fig:figure5}, the GRID CG method provides a systematic way to determine renormalized interactions at a given grid discretization $h$ (or equivalently $n_g$). Starting from microscopic MD trajectories (Fig. \ref{fig:figure5}, left panel), we first construct the coarse-grained density $\rho_h$. Our objective is to determine the effective mesoscopic interaction as a function of inter-grid distances $\mathcal{D}=\{d_i\}_i$. For each $d\in\mathcal{D}$, the GRID CG energy $V_\mathrm{GRID}(d)$ is obtained by matching the coarse-grained interaction energy to its microscopic counterpart at that separation:
\begin{equation} \label{eq:ECG}
    \left[ \sum_{n_I,n_J,|n_I-n_J|=d} \rho_{n_I} \rho_{n_J} \right] \frac{V_\mathrm{GRID}(d)}{2} = E_d(\mathbf{r}^N),
\end{equation}
where $\rho_{n_I}$ and $\rho_{n_J}$ are obtained from histogramming over the $n_g \times n_g$ grid points, and $
    E_d (\mathbf{r}^n) = \frac{1}{2}\sum_{i \in n_I ,\,j \in n_J,\,|r_{ij}|\approx d} U(r_{ij})$ represents the total microscopic energy between particles $i$ and $j$, whose grid positions are separated by $d$. Here, $r_{ij}$ is only approximately equal to $d$ ($|r_{ij}|\approx d$), as it includes microscopic fluctuations of particle configurations mapped onto the grid. For all sampled distances where $\rho_{n_I} \neq 0$ and $\rho_{n_J} \neq 0$, we solve Eq. (\ref{eq:ECG}) for $V_\mathrm{GRID}(d)$:
\begin{equation} \label{eq:ECG3}
    V_\mathrm{GRID}(d) = \frac{\displaystyle 
 \sum_{i \in n_I ,\,j \in n_J,\,|r_{ij}|\approx d} U(r_{ij})}{\displaystyle  \sum_{n_I,n_J,|n_I-n_J|=d} \rho_{n_I} \rho_{n_J}},\qquad \forall d\in\mathcal{D}.
\end{equation}
By repeating this for all $d \in \mathcal{D}$, we obtain the full discrete interaction $V_\mathrm{GRID}(d)$, which can then be fit to a continuous functional form for use in the RIDK simulations. In turn, Eqs. (\ref{eq:ECG})--(\ref{eq:ECG3}) represent a field-level renormalization of microscopic energetics, generalizing force-matching and energy-matching methods in molecular coarse-graining \cite{noid2008multiscale1,noid2008multiscale2,wang2009effective}.

The GRID CG approach is designed to integrate directly into the RIDK framework. Although the RIDK equations are formulated using the coarse-graining length $\varepsilon$ rather than $h$, setting $\varepsilon = h/2$ allows each particle to be convolved with a Gaussian of variance $\varepsilon^2 = (h/2)^2$, such that each grid point effectively represents the Gaussian density. The resulting $V_\mathrm{GRID}(d)$ can thus be consistently employed in RIDK simulations with $\varepsilon = h/2$. Furthermore, GRID CG can, in principle, directly regularize divergent hard-core interactions at zero distance ($d=0$) by assigning a finite, well-behaved interaction at $d=0$ reflecting self-interactions among particles within the same grid cell. While such divergent interactions cannot be directly incorporated into Dean's equation (as discussed in Sec. II), the renormalized values of $V_\mathrm{GRID}(d=0)$ remain finite and vary systematically with the coarse-graining level, providing a practical route to regularize divergent atomistic interactions for a broader class of grid-based field-theoretical models. 

We conclude this section by noting that this matching strategy is not necessarily unique. For example, in effective interaction models for liquids, matching the virial in repulsive and attractive potentials has been highly successful in reproducing structural and dynamical properties \cite{pedersen2010repulsive}. While spatial discretization may introduce additional complexity, similar strategies may offer promising directions for future research. Our approach also differs from that of Ref. \onlinecite{jin2024perturbative}, which applies classical perturbation theory to mesoscopic interactions in $k$-space via the Hubbard-Stratonovich transformation \cite{hubbard1959calculation}. That work defines field interactions in reciprocal space and introduces perturbations to mitigate the divergence at $d \rightarrow 0$, without involving real-space coarse-graining. In contrast, GRID CG renormalizes particle-level interactions through real-space density fields, enabling direct simulation of mesoscopic dynamics in configuration space. GRID CG thus aims to capture the essential energetics of molecular systems within the mesoscopic field representation. While the difference may be modest for systems with soft, convergent interactions, GRID CG offers advantages for treating atomistic potentials with short-ranged repulsions and multiple characteristic length scales.

\section{Particle-Field Correspondence}
\subsection{Equivalence Mapping between RIDK and MD} 
To faithfully implement RIDK simulations for molecular systems, the underlying physics of RIDK (field-level) must be equivalent to that of molecular dynamics (atomistic-level). MD simulations typically use physical units and realistic length scales, whereas RIDK simulations are performed on a different domain $\mathbb{T}^2=[0,\,2\pi]\times[0,\,2\pi]$ using von Mises kernels and operating in dimensionless units. Hence, establishing a correspondence between MD and RIDK is necessary. 

We propose an \textit{equivalence mapping} by matching reduced temperature and density to ensure that the underlying physics remains consistent. Consider scaling up an atomistic MD simulation of the Gaussian core model (see Sec. VIII) with the interaction form
\begin{equation}
    V_\mathrm{MD}(r) = \epsilon_\mathrm{MD} \exp \left[ -\frac{1}{2} \left(\frac{r}{\mathbb{L}_\mathrm{MD}}\right)^2 \right],
\end{equation}
where $\epsilon_\mathrm{MD}$ is the interaction strength and $\mathbb{L}_\mathrm{MD}$ is the characteristic length. We assume that the atomistic system consists of $N_\mathrm{MD}$ particles in a 2D box of size $L_\mathrm{MD}\times L_\mathrm{MD}$ at temperature $T_\mathrm{MD}$. Our goal is to determine the corresponding RIDK parameters: temperature $T_\mathrm{RIDK}$ [given by the fluctuation-dissipation relation $T_\mathrm{RIDK} = \sigma^2 / (2\gamma)$], particle number $N_\mathrm{RIDK}$, and the domain length of $2\pi$. Since both RIDK and MD simulations employ periodic boundary conditions, we set $N_\mathrm{RIDK}=N_\mathrm{MD}$ and assume that the mesoscopic coarse-grained interaction retains the Gaussian form, with adjusted parameters $\epsilon_\mathrm{RIDK}$ and $\mathbb{L}_\mathrm{RIDK}$:
\begin{equation} \label{eq:gausscg}
    V_\mathrm{RIDK}(r) = \epsilon_\mathrm{RIDK} \exp \left[ -\frac{1}{2} \left(\frac{r}{\mathbb{L}_\mathrm{RIDK}}\right) ^2 \right].
\end{equation}

To ensure temperature equivalence, we match the dimensionless temperature ($\tilde{\beta}^{-1}$ or $\tilde{T}$):
\begin{equation} \label{eq:Tmatching}
    \tilde{\beta}^{-1} = \frac{T_\mathrm{MD}}{\epsilon_\mathrm{MD}} = \frac{T_\mathrm{RIDK}}{\epsilon_\mathrm{RIDK}}.
\end{equation}
Given a chosen pair of $\sigma$ and $\gamma$ that defines $T_\mathrm{RIDK}$, Eq. (\ref{eq:Tmatching}) determines the rescaled RIDK interaction strength as  $\epsilon_\mathrm{RIDK}=\epsilon_\mathrm{MD}\times T_\mathrm{RIDK}/T_\mathrm{MD}.$

Next, we match the dimensionless density $\tilde{\rho}$ to determine the rescaled interaction length $\sigma_\mathrm{RIDK}$ in 2D:
\begin{equation} \label{eq:rhomatching}
    \tilde{\rho} = \frac{N_\mathrm{RIDK}}{(2\pi)^2}\sigma_\mathrm{RIDK}^2 = \frac{N_\mathrm{MD}}{L_\mathrm{MD}^2}\sigma_\mathrm{MD}^2.
\end{equation}
Assuming $N_\mathrm{RIDK}=N_\mathrm{MD}$, this reduces to $\sigma_\mathrm{RIDK} = 2\pi \times \sigma_\mathrm{MD}/L_\mathrm{MD}$, effectively rescaling the length from the molecular box to the $\mathbb{T}$ domain. Although the GRID CG process is not explicitly applied here, this mapping is fully compatible with the GRID CG interaction $V_\mathrm{GRID}(r)$ derived from atomistic MD trajectories
\begin{equation} \label{eq:gausscg2}
    V_\mathrm{GRID}(r) = \epsilon_\mathrm{GRID} \exp \left[ -\frac{1}{2} \left(\frac{r}{\mathbb{L}_\mathrm{GRID}}\right)^2 \right],
\end{equation}
where $\epsilon_\mathrm{GRID}$ and $\mathbb{L}_\mathrm{GRID}$ are obtained by fitting $V_\mathrm{GRID}(r)$ at various grid distances to the functional form above. In Secs. VIII and IX, we will apply this equivalence mapping in combination with GRID CG to atomistic interacting systems.

\subsection{Practical Requirements}
While Eqs. (\ref{eq:Tmatching}) and (\ref{eq:rhomatching}) determine the conditions for thermodynamic consistency between MD and RIDK, additional practical constraints must be addressed for numerical implementation on a grid. First, the RIDK interaction length, $\sigma_\mathrm{RIDK}$, must exceed the grid spacing $h$, i.e., $\sigma_\mathrm{RIDK} \ge h$. For some constant $\alpha\ge1$, this condition can be written as 
\begin{align} \label{eq:require1}
    \sigma_\mathrm{RIDK} = \alpha \frac{2\pi}{n_g}.
\end{align}

Another critical issue in RIDK simulations is the possible occurrence of unphysical \textit{negative density}. Since the RIDK equation resembles a damped wave equation, it lacks a maximum principle to enforce positivity of the density field \cite{bessemoulin2012finite, carrillo2015finite}. To prevent negative densities, the number of particles should exceed the number of grid cells, ensuring that, on average, no grid element is empty. For some constant $\beta\ge1$, this requirement can be expressed as
\begin{equation} \label{eq:require2}
    N_\mathrm{RIDK} = \beta n_g^2.
\end{equation} 
Notably, Eq. (\ref{eq:require2}) is consistent with the scaling analysis under a weak metric to determine the optimal coarse-graining size
\begin{equation} \label{eq:37}
    N_\mathrm{RIDK} \ge n_g^2 = \left(\frac{2\pi}{h}\right)^2,
\end{equation}
which implies $N_\mathrm{RIDK}h^2 \ge 4\pi^2\gg 1$.

Equations (\ref{eq:require1}) and (\ref{eq:require2}), can be combined in terms of the \textit{reduced density} $\tilde{\rho}= \sigma^2_\mathrm{RIDK}{N_\mathrm{RIDK}}/{(2\pi)^2}= \alpha^2 \beta \gg 1,$ indicating that \textit{RIDK simulations are numerically stable on average only under high-density conditions}. This high-density requirement can limit practical applications. For example, if the Gaussian interaction decays over two grid cells and one particle is assigned per grid point, then $\tilde{\rho}=4$, corresponding to an extremely dense system. 

We note that several positivity-preserving algorithms, such as the Brownian bridge technique \cite{sotiropoulos2008adaptive} and complex averaging schemes for field variables \cite{gupta2011time,kim2017stochastic}, may reduce the occurrence of negative densities, potentially allowing for smaller $\beta$. These techniques could be explored in future work. However, Eq. (\ref{eq:require2}) naturally arises from scaling arguments that bound errors in the RIDK model, so high-density conditions are still fundamentally required, even with advanced numerical techniques. In this study, our numerical parameters for finite difference simulations follow the Courant--Friedrichs--Lewy condition \cite{courant1928partiellen} to prevent negative density sampling
 (see SM Sec. V for computational details). 

Finally, since $N_\mathrm{RIDK}$ is matched to $N_\mathrm{MD}$, Eq. (\ref{eq:require2}) also sets a upper bound on coarse-graining by limiting the maximum grid resolution $n_g \le \sqrt{N_\mathrm{RIDK}}$. As the mesoscopic correlations from RIDK simulations are coarse-grained via the Gaussian kernel, such high-density constraints with fewer grid points imply lower spatial resolution, potentially leading to a loss of important correlations. The remainder of this paper will apply the aforementioned scaling analysis to coarse-grained microscopic systems at resolutions that ensure positivity and evaluate the effect of coarse-graining.  

\section{Gaussian Core Model}
\subsection{Microscopic Setting and Simulation}
\begin{figure}
    \includegraphics[width=0.5\textwidth]{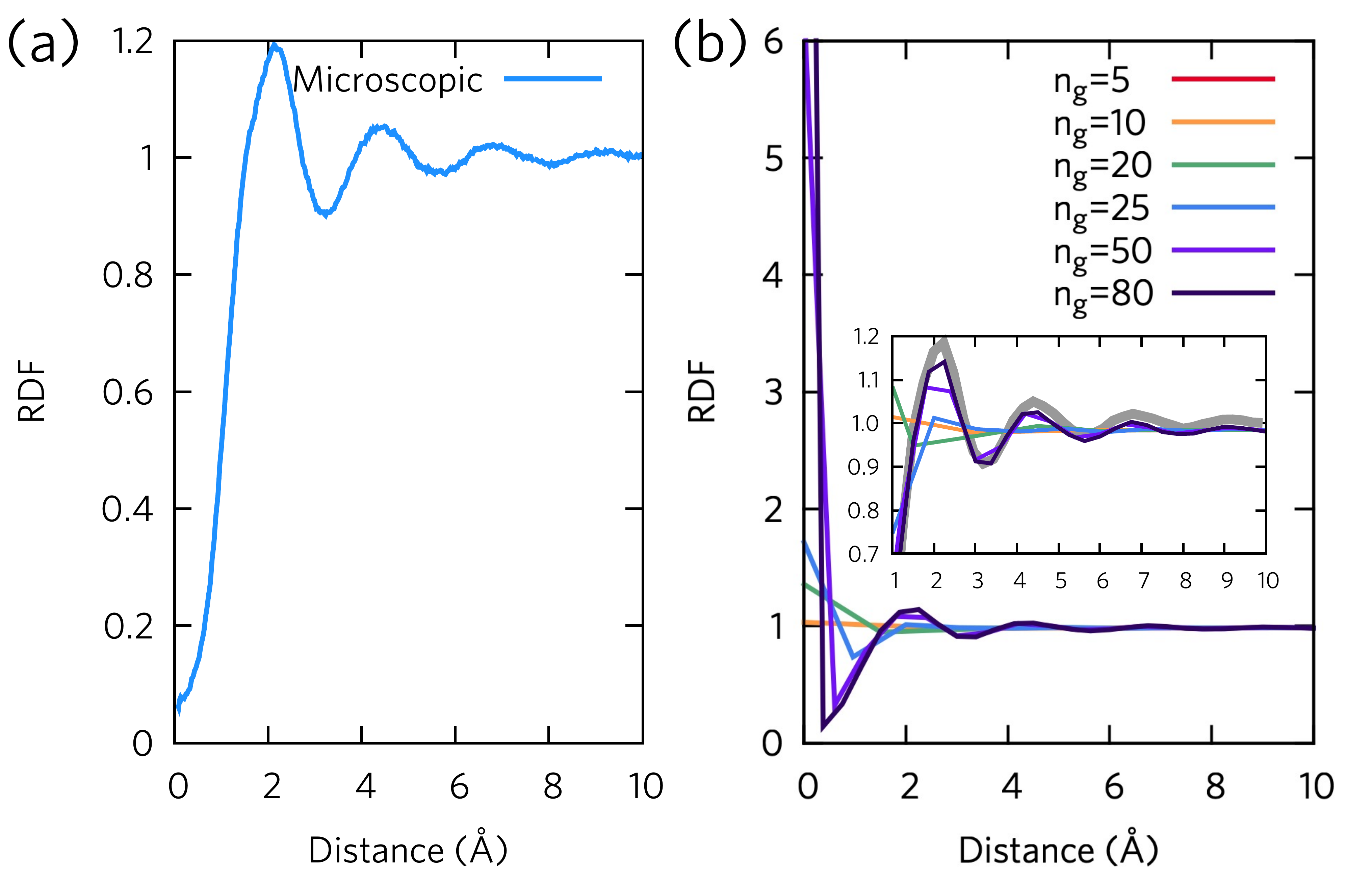}
    \caption{\label{fig:figure6} Structural correlations (RDF) of the Gaussian core model at different scales: (a) microscopic and (b) mesoscopic. In panel (b), using the microscopic RDF (gray) with a weakly structured profile, the mesoscopic coarse-grained RDFs are shown for various grid sizes $n_g$: 5 (red), 10 (orange), 20 (green), 25 (blue), 50 (purple), and 80 (navy). As $n_g$ decreases (or $\varepsilon$ increases), the structural correlations gradually diminish,  consistent with Fig. \ref{fig:figure4}.}
\end{figure}

\begin{figure*}
    \includegraphics[width=1\textwidth]{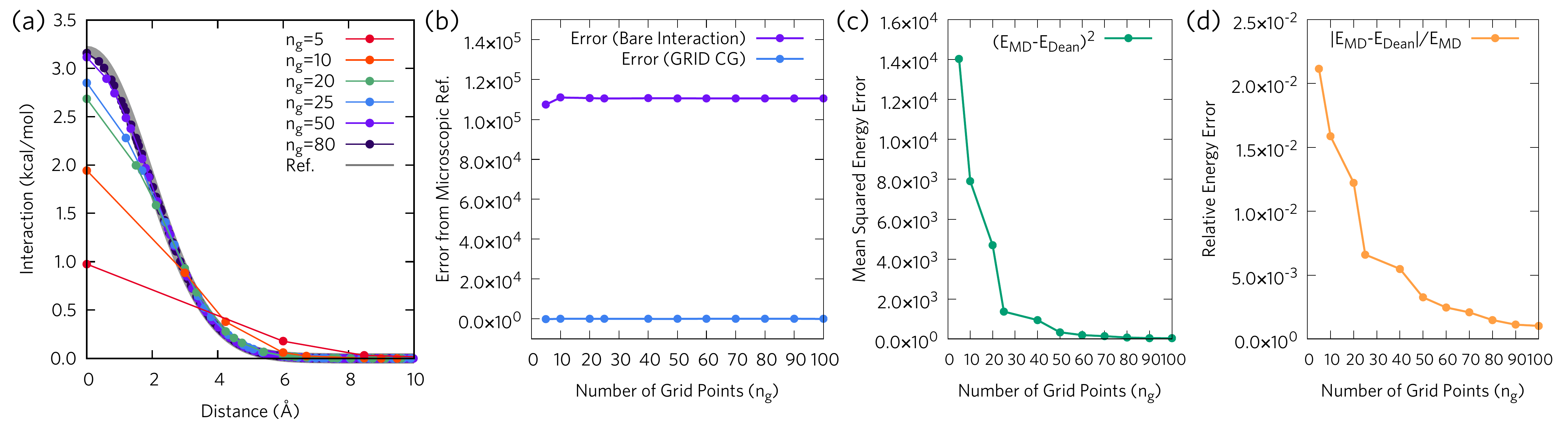}
    \caption{\label{fig:figure7} Microscopic and mesoscopic energetics of the Gaussian core model under mesoscopic coarse-graining. (a) Coarse-grained interactions at different grid sizes [color-coded as in Fig. \ref{fig:figure6}(b)] by applying Eq. (\ref{eq:ECG}) to the microscopic potential [Eq. (\ref{eq:gcmref})] (gray). (b) Fidelity of the GRID CG approach, shown by the energy error relative to the microscopic reference energetics (units in kcal/system): bare interaction (purple) and GRID CG interaction (blue). (c-d) Energy mismatch upon coarse-graining. While the GRID CG method aims to capture underlying microscopic energetics, excessive coarse-graining (low $n_g$ values) inevitably deviate from the reference energetics, quantified by (b) mean squared errors and (c) relative energy errors, where the energy mismatch decreases as $n_g$ increases.}
\end{figure*}
We begin by applying our methodology to implement the \textit{Gaussian core model} \cite{stillinger1976phase}, described by
\begin{align} \label{eq:gcmaa}
    V^\mathrm{GCM}(r)=\epsilon \exp \left(-\frac{r^2}{\sigma^2} \right),
\end{align}
where $\epsilon$ sets the interaction strength and $\sigma$ defines the characteristic interaction length. Even though the Gaussian core model may oversimplify atomistic interactions, it still captures essential microscopic features relevant to various soft matter systems, such as polymers and glasses \cite{lang2000fluid,louis2000can,louis2000mean}. 

Notably, the monodisperse Gaussian core model exhibits glassy behavior at high densities \cite{ikeda2011glass}, making it a compelling target for field-theoretic numerical simulation, as simulating such glassy behavior over long time scales at the particle level is often computationally prohibitive \cite{berthier2011theoretical,berthier2023modern}. Its non-divergent potential at $r=0$ and absence of singularities make it an ideal prototype for RIDK or auxiliary field methods \cite{fredrickson2002field, fredrickson2006equilibrium, fredrickson2023field}. Thus, the Gaussian core model serves as a practical baseline for demonstrating the fidelity of RIDK. 

However, the non-divergent nature of the Gaussian core model also presents challenges for constructing high-density conditions ($\tilde{\rho}=\alpha^2\beta>1$) that exhibit strong, nontrivial pair correlations. This issue is particularly pronounced in 2D systems, where the system is more prone to crystallization at high density \cite{prestipino2005phase,prestipino2011hexatic}. While most prior studies focus on relatively dilute regimes ($\tilde{\rho}\lesssim1$), our implementation requires much denser systems. For example, consider a system of 400 particles interacting via Eq. \eqref{eq:gcmaa} that does not decay beyond the grid spacing ($\sigma = 1.25 h$) on a $10\,\times\,10$ grid (approximately four atoms per cell). This setup yields $\tilde{\rho}=6.25$, an extremely dense regime where microscopic correlations are already weak and are further diminished upon coarse-graining (SM Sec. VI). Hence, we consider a moderately dense Gaussian system of 400 particles in a $30\,\angstrom\,\times\,30\,\angstrom$ domain with $\epsilon = 3.178$ kcal/mol and $\sigma=1.875$ $\angstrom$ at 0.1 K:
\begin{equation} \label{eq:gcmref}
    V^\mathrm{GCM}_\mathrm{MD}(r)=3.178 \cdot \exp \left( -\frac{1}{2}\frac{r^2}{1.875^2}\right),
\end{equation}
which corresponds to $T^\ast=1/16000$ and $\tilde{\rho}=1.5625$. While this density is still markedly higher than values typically studied \cite{prestipino2005phase,prestipino2011hexatic}, atomistic MD simulations exhibit nontrivial structural correlations, as seen in the RDF [Fig. \ref{fig:figure6}(a)] while crystallization can be suppressed. Notably, at this high density the RDF retains a finite value at zero separation, suggesting that some correlation features may be lost upon mesoscopic coarse-graining.

\subsection{Mesoscopic Coarse-Graining}
\subsubsection{Mesoscopic Interaction}
We next perform coarse-graining of the 400 Gaussian particles in a 2D system into a field-level representation using grid sizes ranging from $5\times5$ to $80\times80$ [Fig. \ref{fig:figure7}(a)]. The coarsest grid ($5\times5$) averages about 16 particles per cell, too coarse to capture microscopic detail, while the finest grid ($80\times80$) approaches the microscopic limit with only 0.0625 particles per cell. As the grid becomes finer, the effective interaction from GRID CG converges toward the bare microscopic potential. Grid cells finer than $25\times25$ recover the microscopic interaction well at longer distances, while coarser grid cells with multiple particles per cell produce smoother interactions that lose short-range detail. A crossover appears near 3 \angstrom: short-range repulsion is reduced, whereas long-range interactions become more repulsive, consistent with the behavior observed under dense conditions (see Fig. S4 in SM).

Figure \ref{fig:figure7}(b) highlights the importance of preserving energetics during coarse-graining by comparing field-level energetics computed with both the bare interaction and the GRID CG interaction to the microscopic reference. Note that the total potential energy of the system, $V(\mathbf{r}^N)$, can be expressed as
\begin{align} \label{eq:energyeval}
        V(\mathbf{r}^N) = \frac{1}{2}\sum_{i\neq j} V(r_{ij}).
\end{align}
By introducing the self-energy $V_\mathrm{self}:= \frac{N}{2} V(r=0)$, $ V(\mathbf{r}^N) $ can be rewritten as
\begin{align} \label{eq:energyeval2}
    V(\mathbf{r}^N) = \frac{1}{2}\sum_{i,j} V(r_{ij}) -V_\mathrm{self}.
\end{align}
While evaluating the microscopic energy via Eq. \eqref{eq:energyeval} is straightforward at the particle level, the corresponding field-level expression with the renormalized potential can be efficiently computed in Fourier space, where $\hat{\rho}(\textbf{k})=\sum_i \mathrm{exp}[-i\textbf{k}\cdot \textbf{r}_i]$, transforming the renormalized pair interaction sum over $V^\mathrm{GRID}(r_{ij})$ into  
\begin{align} \label{eq:transformpair}
    V^\mathrm{GRID}(\mathbf{r}^N) &= \frac{1}{2}\int d\mathbf{r} d\mathbf{r}' \hat{\rho}(\mathbf{r})V^\mathrm{GRID}(\mathbf{r}-\mathbf{r}')\hat{\rho}(\mathbf{r}') - V^\mathrm{GRID}_\mathrm{self} \nonumber \\ &= \frac{1}{2} \sum_\mathbf{k} \hat{\rho}^*(\mathbf{k}) V^\mathrm{GRID}(\mathbf{k})\hat{\rho}(\mathbf{k}) - V^\mathrm{GRID}_\mathrm{self},
\end{align}
which we approximate as $\frac{1}{2} \sum_\mathbf{k} \rho_{h}^*(\mathbf{k}) V^\mathrm{GRID}(\mathbf{k})\rho_{h}(\mathbf{k}) - V^\mathrm{GRID}_\mathrm{self}.$ Remarkably, Fig. \ref{fig:figure7}(b) shows that na\"{i}vely applying the bare Gaussian interaction in the field-level representation, as commonly done in the literature, results in errors spanning up to seven orders of magnitude, thereby producing a significant deviation from the true microscopic reference. In contrast, the GRID CG interaction yields overall energetics that are quantitatively similar to the reference. Figures \ref{fig:figure7}(c) and (d) further quantify the relative deviation of GRID CG energetics by estimating the variance and relative error compared to the microscopic reference. These errors remain minor and decrease with increasing grid resolution. Since the GRID CG method yields the coarse-grained potential only as a set of discrete values from Eq. \eqref{eq:ECG3}, we fit these to a continuous Gaussian form for simulation purposes (see Subsection C and Table SI in SM).

\subsubsection{Mesoscopic Correlation} 
Similar to the renormalized interaction, density correlations also change upon mesoscopic coarse-graining. Prior to coarse-graining, the reference pair correlation function exhibits a structured profile with a peak around 1.2, despite the fact that the zero-distance value is nonzero due to the finite interaction at $r=0$.

Figure \ref{fig:figure6}(b) shows the effect of mesoscopic coarse-graining on the RDF. This dependence can also be interpreted through $\varepsilon$ using $\varepsilon\approx0.5h$. In Sec. V, we observed that density correlations become smoother as coarse-graining progresses. While grid resolutions finer than $40\,\times\,40$ still retain the key features of the microscopic RDF, coarser grid cells significantly suppress these features. At a resolution of $20\,\times\,20$, much of the structural information is lost, making it difficult to resolve nontrivial correlations, in part because the Gaussian core model itself has inherently weak structural correlations at high density. Therefore, we adopt a $25\,\times\,25$ grid for the Gaussian system, which preserves the first RDF peak even after mesoscopic coarse-graining. Although this grid choice slightly violates $n_g \le \sqrt{N_\mathrm{RIDK}}$, it still satisfies the weaker metric condition, $N\varepsilon^2\ge1$, ensuring that RIDK simulations can be carried out without numerical artifacts. Our goal is to assess whether the density correlations produced by the RIDK simulation under this grid setting are consistent with the expected coarse-grained correlations. 

\subsection{Mesoscopic RIDK Simulation}
For a $25\,\times\,25$ grid, we obtain a mesoscopic CG interaction potential $V^\mathrm{GCM}_\mathrm{GRID}(r)=2.820 \times \exp \left[ -\frac{1}{2} \left(\frac{r}{1.981}\right) ^2 \right]$, expressed in microscopic units. Using the equivalence mapping (See SM Sec. VIII for details), the rescaled interaction for the RIDK simulation is
\begin{equation} \label{eq:ridk_gauss}
    V^\mathrm{GCM}_\mathrm{RIDK}(r) = 7.100 \times \exp \left[ -\frac{1}{2} \left(\frac{r}{0.4149}\right) ^2 \right].
\end{equation}
The initial condition for RIDK was obtained by applying $\varepsilon$-smoothing to the final snapshot of the microscopic MD simulation. The full computational pipeline, including the finite element implementation, is described in SM Secs. III and V.

\begin{figure}
    \includegraphics[width=0.3\textwidth]{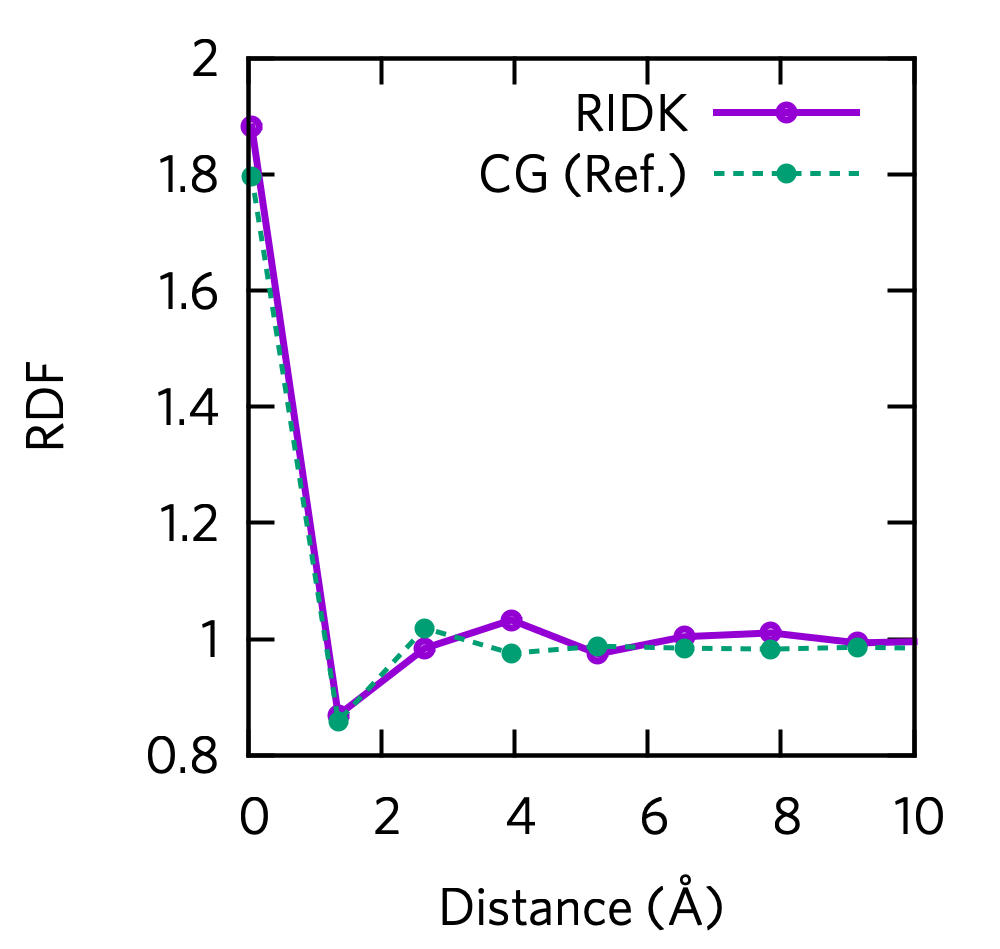}
    \caption{\label{fig:figure8} Mesoscopic correlations estimated from the numerical RIDK simulation using the $25\,\times\,25$ grid setting on the Gaussian core model (purple lines), compared to the Gaussian filtered reference correlation (green dots) from Fig. \ref{fig:figure6}(b) with $n_g=25$ (blue).}
\end{figure}

From the RIDK simulation, we numerically estimated the renormalized static density correlations by computing the density histogram discretized on the grid. Figure \ref{fig:figure8} compares the RDF obtained from RIDK to the manually coarse-grained RDF of the microscopic MD simulations under the same grid resolution. Although the coarse grid setting used to bound the residuals in RIDK simplifies correlations significantly, the RDF from the RIDK simulation reproduces the key structural features, including the correlation hole, with nearly exact values.

In summary, for the Gaussian core model, the structural correlations are well reproduced, capturing the reference behavior observed in atomistic simulations performed with the smeared interaction derived through GRID CG. To further evaluate the effectiveness of this bottom-up framework, we next apply the RIDK approach combined with GRID CG to the Lennard-Jones interaction. 

\section{Lennard-Jones Liquids}
\subsection{Microscopic Setting and Simulation}
\begin{figure}
    \includegraphics[width=0.5\textwidth]{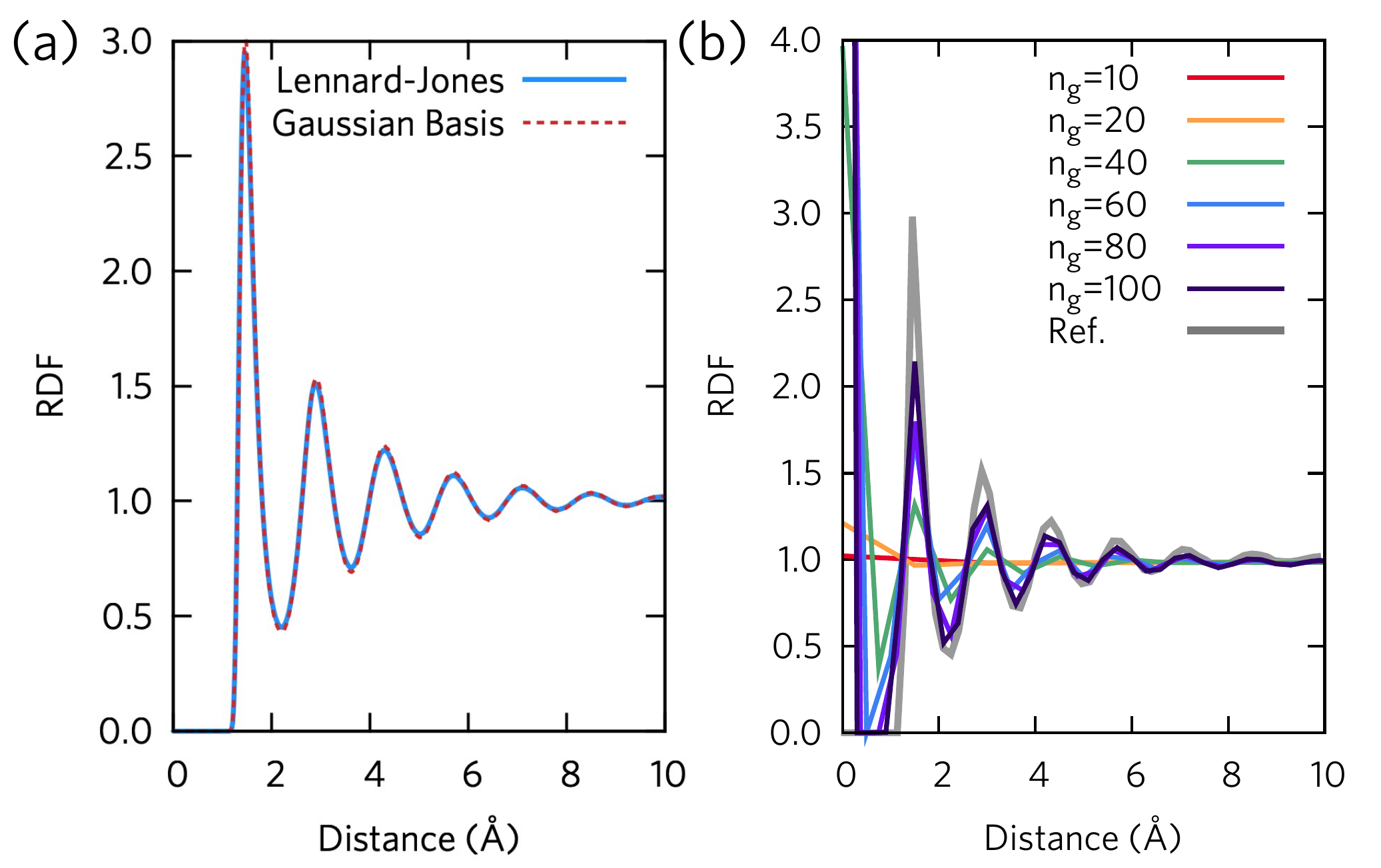}
    \caption{\label{fig:figure9} Structural correlations (RDF) of the Lennard-Jones liquid at different scales: (a) microscopic, showing the divergent bare interaction (solid) and its Gaussian-regularized form (dashed line), and (b) mesoscopic. In panel (b), using the microscopic RDF (gray) with a weakly structured profile, the mesoscopic coarse-grained RDFs are shown for various grid sizes $n_g$: 10 (red), 20 (orange), 40 (green), 60 (blue), 80 (purple), and 100 (navy).}
\end{figure}

At the atomistic level, the Lennard-Jones interaction $V^\mathrm{LJ}_\mathrm{MD}(r)$ is defined as
\begin{align} \label{eq:lj}
    V^\mathrm{LJ}_\mathrm{MD}(r)=4\epsilon_{LJ} \left[\left(\frac{\sigma}{r}\right)^{12}-\left(\frac{\sigma}{r}\right)^{6}\right],
\end{align}
where the short-range $r^{-12}$ divergence, unlike the Gaussian core model, stabilizes the liquid phase even in two dimensions. We used the same number density as in the Gaussian core system, with parameters $\epsilon=3.9745\times 10^{-3}$ kcal/mol and $\sigma=2.121$ \textrm{\AA} at $1000$ K, corresponding to $\tilde{\rho}=2$ and $\tilde{\beta}^{-1}=T/\epsilon=500$. To further mitigate issues associated with the divergence of Lennard-Jones interactions at short distances, we followed Refs. \onlinecite{singer1958use} and \onlinecite{ma1993approximate} and represented the Lennard-Jones interaction as a sum of two Gaussian basis functions. The first Gaussian, centered at the origin, captures the short-range repulsion, while the second, with negative magnitude, models the attractive well. This regularized interaction reproduces the RDF of the bare Lennard-Jones potential, which at this state point exhibits strong structural features with a first peak intensity of about 3 [Fig. \ref{fig:figure9}(a)], while the system remains in the diffusive liquid state regime, as indicated by the mean square displacement.

\subsection{Mesoscopic Coarse-Graining}
\subsubsection{Mesoscopic Interaction}
\begin{figure*}
    \includegraphics[width=0.8\textwidth]{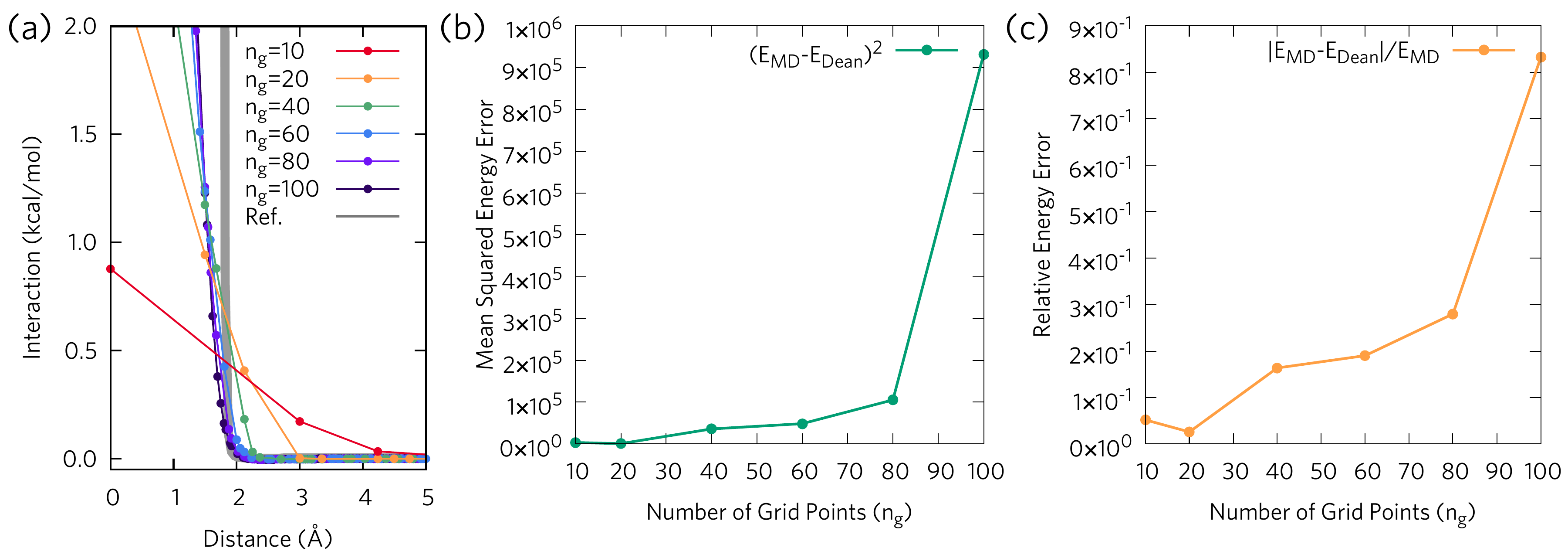}
    \caption{\label{fig:figure10} 
    Microscopic and mesoscopic energetics of the Lennard-Jones model under mesoscopic coarse-graining. (a) Coarse-grained interactions at different grid sizes [color-coded as in Fig. \ref{fig:figure9}(b)] by applying Eq. (\ref{eq:ECG}) to the microscopic potential [Eq. (\ref{eq:lj})] (gray). (b-c) Energy mismatch during coarse-graining. Unlike the Gaussian core case (Fig. \ref{fig:figure7}), where energy mismatch decreases as $n_g$ increases, the divergent hard-core repulsion of the Lennard-Jones model at the microscopic level results in a completely different trend. We find that $n_g=20$ minimizes the energy mismatch in (b) and (c), while also satisfying $N\ge n_g^2$.}
    \end{figure*}

We applied mesoscopic coarse-graining to the regularized Lennard-Jones interaction using the GRID CG approach across various grid sizes, from $10\times10$ to $100\times100$, derived from the microscopic trajectory. Unlike the Gaussian core model, the coarse-grained Lennard-Jones interactions show significant variation across resolutions, even at relatively coarse grid cells, as shown in Fig. \ref{fig:figure10}(a). At finer resolutions, the interactions become weakly repulsive, with a crossover to the microscopic interactions also observed (Fig. S5 in SM).

Despite some similarities to the Gaussian core case, the GRID CG Lennard-Jones interaction exhibits distinct features. The GRID CG approach renormalizes  $V^\mathrm{LJ}_\mathrm{MD}(r\rightarrow0)$  into a significantly less steep repulsive profile. This results from multiple particles being grouped into the same grid cell, which softens the self-interaction at zero distance as the coarse-graining length increases. Even at the finest resolution ($100\,\times\,100$), where the effective RDF remains comparable to the microscopic RDF [Fig. \ref{fig:figure9}(b)], the short-range repulsion is noticeably softened. 

Similar to the microscopic case, we fitted the numerically obtained GRID CG interactions to a sum of two Gaussian basis functions (see Table SII in SM). To more accurately describe such complex interaction profiles, future work may consider generalized interaction forms that combine hard-core repulsion with multiple Gaussian components \cite{jin2023gaussian}. As shown in Fig. \ref{fig:figure10}, the repulsive magnitude decreases monotonically from 55.4575 ($n_g=100$) to 0.8783 kcal/mol ($n_g=10$), while the corresponding length scales increase from 0.5141 to 1.6922 $\textrm{\AA}$. A similar trend is observed for the weaker attractive term, where its magnitude decreases and its characteristic range increases, reflecting a smooth crossover between coarse-grained attraction and repulsion (see Fig. S6). While here we report the mesoscopic coarse-graining of the regularized Lennard-Jones interaction, the GRID CG approach can also be applied to the divergent bare Lennard-Jones potential. As expected from the agreement of microscopic RDFs, the GRID CG interaction obtained from the bare potential reproduces the same interaction profiles as the regularized case, demonstrating that grid-based coarse-graining can in principle renormalize divergences in hard-core interactions at the mesoscopic level.

\subsubsection{Mesoscopic Correlation}
We now examine the effect of coarse-graining on the RDF by varying the grid size. As with the Gaussian core model, the strongly structured Lennard-Jones RDF must be convolved with the Gaussian kernel corresponding to the coarse-graining length to make a consistent comparison with the output of the RIDK simulations. Figure \ref{fig:figure9}(b) illustrates that the sharp RDF peak around 3 decays to 2.1 at $n_g=100$, 1.8 at $n_g=80$, 1.3 at $n_g=40$, and further to 0.96 at $n_g=20$ and below upon coarse-graining. However, weak nontrivial pair correlations near the first coordination shell persist even at $n_g=20$, suggesting that certain microscopic structural correlations survive under high-density conditions.

\label{fig:Schematic}
\subsubsection{Choice of the Optimal Grid: Mesoscopic Energetics}
To construct an RIDK model, selecting an appropriate grid resolution is essential. For the Gaussian core system, the finest grid satisfying the weaker metric condition was sufficient. However, because Lennard-Jones interactions involve two characteristic length scales, the situation is more complicated. Even though the GRID CG scheme matches interactions as closely as possible at a given resolution, it may still fail to recover the correct microscopic energetics. Thus, one must choose a grid size that best captures the underlying energetics.

With this in mind, we apply Eqs. (\ref{eq:energyeval}) and (\ref{eq:energyeval2}), originally derived for the Gaussian core model to guide this choice. Although the Lennard-Jones potential does not have a well-defined self-interaction $V_\mathrm{self}$ at $r=0$ (in this case Eq.~(\ref{eq:energyeval}) applies), it can be evaluated for the regularized potential, and Eq. (\ref{eq:energyeval2}) estimates the mesoscopic energy based on the coarse-grained interaction. Figure \ref{fig:figure10}(b) shows the total field energy versus $\varepsilon$. Unlike the Gaussian core case, the discrepancy between microscopic and mesoscopic energies is rather substantial, as mesoscopic coarse-graining consistently underestimates the repulsive component, highlighting the importance of proper energetic renormalization through GRID CG.

This imbalance between the overestimated attractive and underestimated repulsion becomes more pronounced as $\varepsilon$ decreases. Namely, the energy variance spans six orders of magnitude, clearly indicating that both very small and very large $\varepsilon$ values are unsuitable for the Lennard-Jones system. From this analysis, we identify an optimal $\varepsilon=0.75\,\textrm{\angstrom}$, corresponding to a grid size of $20\,\times\,20$. Notably, this grid size satisfies the high-density condition $N\ge n_g^2$. 

In summary, for interactions with strong short-range repulsion, coarse-graining softens hard-core repulsions, and selecting an optimal resolution is crucial to minimizing the energy mismatch between microscopic and mesoscopic representations. For the Lennard-Jones system, both the energetic optimum and the RIDK grid requirement align at the same grid size. However, this alignment may not hold in general, underscoring the need for careful analysis in applying RIDK to more complex systems. A more systematic investigation of this balance for other interaction forms will be pursued in future work.

\subsection{Mesoscopic RIDK Simulation}
\subsubsection{RIDK Setup}
The GRID CG procedure at the $20\,\times\,20$ grid [Fig. \ref{fig:figure10}(a)] yields a renormalized interaction expressed as a sum of two Gaussian functions: 
\begin{align} \label{eq:ljgrid}
    V^\mathrm{LJ}_\mathrm{GRID}(r) &= 2.3970 \times \exp\left(-\frac{1}{2}\left(\frac{r}{1.1034}\right)^2\right) \nonumber \\ & -0.0034 \times \exp\left(-\frac{1}{2}\left(\frac{r-2.1324}{0.0348}\right)^2\right),
\end{align}
with units in kcal/mol and \angstrom. The corresponding equivalence mapping (SM Sec. VIII B) gives the final RIDK interaction as
\begin{align} \label{eq:gauss2}
    V^\mathrm{LJ}_\mathrm{RIDK}(r) &= 6030.9 \times \exp\left(-\frac{1}{2}\left(\frac{r}{0.2311}\right)^2\right) \nonumber \\ & -8.5545 \times  \exp\left(-\frac{1}{2}\left(\frac{r-0.4466}{7.288\times10^{-3}}\right)^2\right),
\end{align}
where the RIDK simulation was initialized from the coarse-grained final snapshot of the microscopic Lennard-Jones trajectory, using the same numerical settings as in the Gaussian core system.

\begin{figure}
    \includegraphics[width=0.3\textwidth]{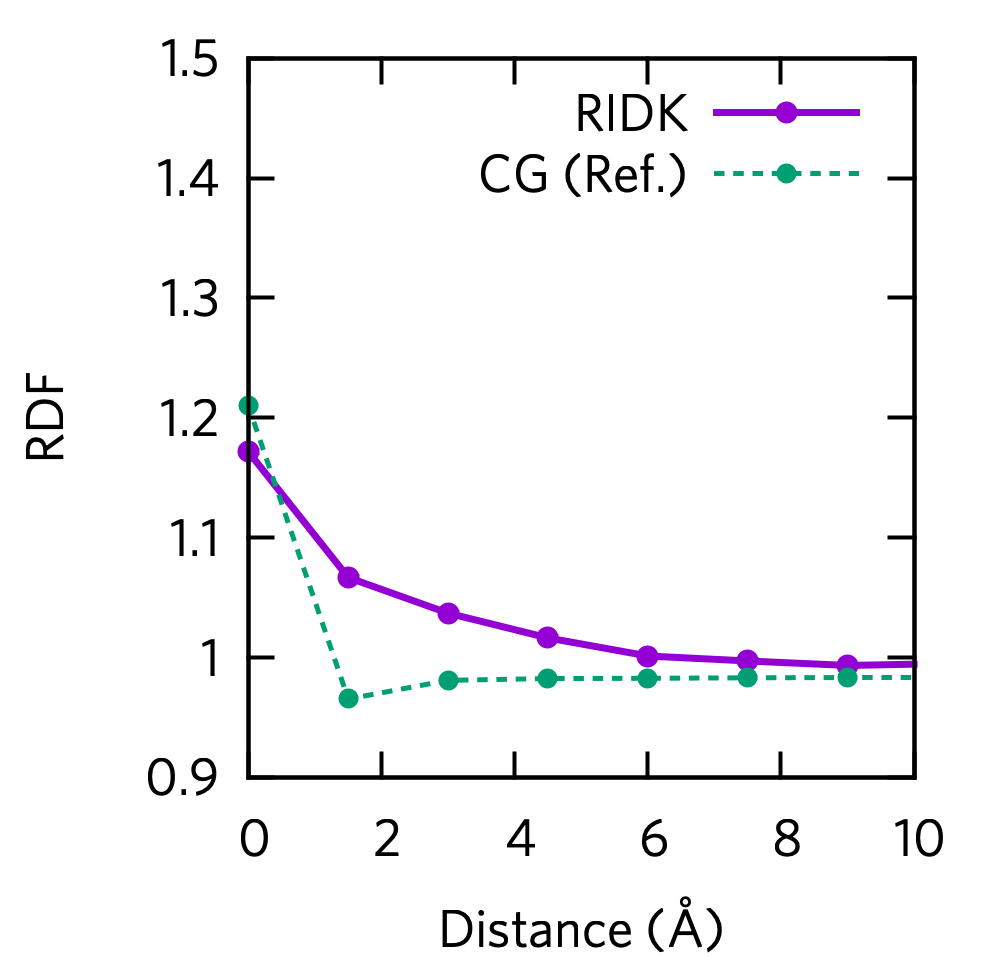}
    \caption{\label{fig:figure11} 
    Mesoscopic correlations estimated from the numerical RIDK simulation using the $20\,\times\,20$ grid setting on the Lennard-Jones model (purple lines), compared to the Gaussian filtered reference correlation (green dots) from Fig. \ref{fig:figure9}(b) with $n_g=20$ (orange).}
\end{figure}

We next analyze the effective correlations obtained from the RIDK simulation. As expected, the RDF computed from the discretized density field appears significantly smoother than the microscopic reference RDF [Fig. \ref{fig:figure9}(b)], with much of the fine structure attenuated. However, Fig. \ref{fig:figure11} shows that the results remain qualitatively consistent with the RDF obtained by manually coarse-graining the microscopic density, with the zero-distance value deviating by less than 5 \%. While structural correlations in the Gaussian core model (Fig. \ref{fig:figure8}) were reproduced almost quantitatively, the Lennard-Jones system shows larger discrepancies. Still, the RIDK results remain qualitatively consistent with the microscopic reference, reproducing key structural signatures such as a change of slope at the location of the correlation hole, the zero-separation value of the RDF, and the overall decay of correlations. We speculate that the stronger deviations for Lennard-Jones arise because its potential is more complex than the Gaussian core case, with steep short-range repulsion and two characteristic length scales, making it more sensitive to approximations introduced during coarse-graining, GRID CG, and RIDK simulation. Thus, even under numerically and physically well-defined settings, key microscopic correlations may already be overly coarse-grained, causing the mesoscopic fields to deviate more noticeably from their microscopic counterparts. Even at moderate resolutions, one should expect different correlation profiles at the mesoscopic level when using RIDK. While our focus in this work has been mainly on static correlations, we note that reduced dynamical correlations can, in principle, be interpreted within the same RIDK framework by rescaling the dynamical correlation function (see SM Sec. IX). Further investigation regarding the dynamical properties of the RIDK simulation will be pursued in follow-up work. 

\section{Outlook and Conclusion}
Rigorously bridging microscopic particle dynamics with a mesoscopic field-theoretic description is a nontrivial challenge, as it requires simplifying complex atomistic degrees of freedom across multiple scales. Dean's equation (or Dean--Kawasaki dynamics) offers a rigorous bottom-up framework for the evolution of the density field \cite{dean1996langevin,kawasaki1994stochastic}. While widely used as a formal tool, its direct application to mesoscopic simulation is hindered by the singular noise structure and the use of microscopic density operators \cite{illien2024dean}. In fact, recent mathematical results have shown that Dean's equation admits no nontrivial solution in its original form, underscoring the need for proper regularization \cite{konarovskyi2019dean,konarovskyi2020dean}. To this end, Cornalba, Shardlow, and Zimmer introduced the RIDK model, which replaces singular Dirac delta distributions with smooth finite kernels under underdamped dynamics \cite{cornalba2019regularized,cornalba2020weakly,cornalba2021well}. Although RIDK yields well-defined, nontrivial solutions, its applications have so far been limited to simple systems such as ideal gases under external fields \cite{cornalba2023dean}. Extending RIDK to realistically interacting liquids has remained unexplored, and this is the frontier addressed by the present work.

We have developed a bottom-up mesoscopic coarse-graining framework that connects the unregularized Dean equation with mesoscopic RIDK dynamics for weakly interacting systems. By coarse-graining particle positions using smooth Gaussian (or von Mises) kernels, the microscopic density is mapped to a mesoscopic density field. Because this mapping alone does not preserve molecular energetics, we introduced the GRID CG method, which systematically determines mesoscopic interactions by matching microscopic energetics. This approach effectively regularizes particle-level pair interactions, enabling accurate and stable simulations at the field level. 

A key technical challenge in RIDK simulations lies in reconciling two coarse-graining parameters: the Gaussian kernel width $\varepsilon$ of the RIDK model and the numerical grid spacing $h$. By analyzing coarse-grained structural features, we have shown that coarse-graining filters density correlations through a Gaussian factor $e^{-\textbf{k}^2 \varepsilon^2}$ and have established the correspondence $h \approx 2\varepsilon$. An optimal $\varepsilon$ was identified by a scaling analysis of the error between the full Dean--Kawasaki dynamics and the approximate RIDK dynamics. We have shown that at high densities, a weaker metric condition suffices, permitting smaller $\varepsilon$ values for RIDK simulations. To control regularization errors, we further derived numerical error bounds (see Appendix B), which provide practical guidelines for future studies aiming to incorporate small $\varepsilon$ while balancing accuracy and stability. 

By establishing a systematic framework to construct mesoscopic field representations directly from microscopic simulations, we extend RIDK to a broad class of interactions, including those with sharply repulsive characteristics. While our results show qualitative agreement with particle-based simulations, the impact of regularized terms in finite element implementations remains and warrants further systematic study. Nonetheless, this work lays the groundwork for advancing multiscale modeling via field-level numerical simulations, enabling the study of systems across extended spatiotemporal scales with improved efficiency and physical fidelity.

\begin{acknowledgments}
J.J. thanks the Arnold O. Beckman Postdoctoral Fellowship for funding and academic support (\url{http://dx.doi.org/10.13039/100000997}). This work was supported by the funding from the Simons Collaboration ``Cracking the glass problem'' (No. 454951). \\
\end{acknowledgments}

\section*{Data Availability}
The data that support the findings of this article are openly available \cite{github}.

\section*{Appendices}
\section*{A. Dynamical Hierarchy in Dean's Equation}
\subsection*{1. Hierarchy of Multi-Particle Densities}
In Sec. II, we derived the exact equation of motion for the one-particle density, Eq. \eqref{eq:dean_general}, which incorporates the two-particle density, $\hat{\rho}^{(2)}$. A similar procedure can be applied to derive the dynamics of the two-particle density by considering an arbitrary, twice continuously differentiable function, $g(\cdot, \cdot)$: $\sum_i \sum_{j(\neq i)} g(q_i(t), q_j(t)) = \int dx dy \hat{\rho}^{(2)}(x,y,t)g(x,y),$ with the time derivative defined as $\int dx dy \partial_t \hat{\rho}^{(2)}(x,y,t)g(x,y) := \sum_i \sum_{j(\neq i)} \frac{d}{dt}g(q_i(t), q_j(t)).$
Using Ito calculus \cite{oksendal2013stochastic}, integration by parts, and the arbitrariness of $g(\cdot,\cdot)$, we obtain the two-particle density evolution equation:
\begin{widetext}
\begin{eqnarray}\label{eq:dym_rho2}
    \partial _t \hat{\rho}^{(2)}(x,y,t) &=& \nabla_x \int dz \hat{\rho}^{(3)}(x,y,z,t)\nabla V(x-z) + \nabla_y \int dz' \hat{\rho}^{(3)}(y,x,z',t)\nabla V(y-z') +\nabla_x\bigg(\hat{\rho}^{(2)}(x,y,t)\nabla V(x-y)\bigg)  \nonumber \\
    &+& \nabla_y\bigg(\hat{\rho}^{(2)}(x,y,t)\nabla V(y-x)\bigg) + T(\nabla_x^2+\nabla_y^2)\hat{\rho}^{(2)}(x,y,t)  + \xi^{(2)}(x,y,t),
\end{eqnarray}
\end{widetext}
where the two-position noise term is given by
\begin{align}
    \xi^{(2)}(x,y,t) = & \nabla_x \bigg(\hat{\rho}^{\frac{1}{2}}(x,t)\bigg[\xi(x,t)\hat{\rho}(y,t) - \gamma(x,y,t)\bigg] \bigg) \nonumber \\ &+ \nabla_y \bigg(\hat{\rho}^{\frac{1}{2}}(y,t)\bigg[\xi(y,t)\hat{\rho}(x,t) - \gamma(y,x,t)\bigg] \bigg).
\end{align}
Here, $\xi(x,t)$ is the stochastic noise defined by Dean \cite{dean1996langevin}, and $\gamma(x,y,t)$ is a white noise field satisfying
\begin{align}
    \langle \gamma(x,y,t)\otimes \gamma(x',y',t')\rangle &= 2T\delta(x-x')\delta(y-y')\delta(t-t') \mathbf{1} \nonumber \\
    \langle \xi(x,t) \otimes \gamma(x', y' ,t')\rangle &= 2T\delta(x-x')\delta(x-y')\delta(t-t')\mathbf{1}.
\end{align}
This procedure can be recursively repeated to derive a full hierarchy of equations involving $\hat{\rho}^{(n)}$.

\subsection*{2. Hierarchy of Correlation Functions}
From the dynamical hierarchy of densities, one can also construct a corresponding hierarchy of correlation functions. Consider the evolution of the one-particle correlation function, obtained via the inverse Fourier transform of the dynamical structure factor:
\begin{align}
    \partial_t \langle \hat{\rho}(x,t)\hat{\rho}(y,0) \rangle = & \nabla_x (\int dy\langle\hat{\rho}(y,0)\rho^{(2)}(x,z,t)\rangle \nabla V(x-z) ) \nonumber \\ &+ T\nabla_x^2 \langle \hat{\rho}(x,t)\hat{\rho}(y,0)\rangle, 
\end{align}
where we used $\langle\xi (x,t) \hat{\rho}^{1/2}(x,t)(y,0)\rangle = \langle\xi (x,t) \rangle \langle \hat{\rho}^{1/2}(x,t)(y,0)\rangle = 0$. The time evolution of the correlation $\langle \hat{\rho}(y,0)\hat{\rho}^{(2)}(x,z,t)\rangle$ can, in turn, be expressed using $\langle \hat{\rho}(z,0)\hat{\rho}^{(3)}(x,y,w,t)\rangle$ , continuing the hierarchy. These hierarchies can be truncated at a certain order by approximating the highest-order correlation as a product of lower-order correlations, similar to the closures in mode coupling theory \cite{reichman2005mode}.

\section*{B. Numerical Estimation of Sobolev Norms in the RIDK Framework}
To quantify the residual terms arising from the convolution approximation of the regularized potential \( V \) in Eq. (\ref{eq:conv_error}), we numerically evaluated the Sobolev norms associated with the bounds of $r_{1,\varepsilon}$ and $r_{2,\varepsilon}$ in Eq. (\ref{eq:sobolev}), i.e., $    r_{1,\varepsilon}\lesssim \left\lVert V\right\rVert_{V^{1,\infty}} \sqrt{\varepsilon}$ and $r_{2,\varepsilon} \lesssim \left\Vert V\right\Vert_{V^{2,\infty}} \sqrt{\varepsilon}$. Here, the Sobolev norms $\left\Vert \cdot \right\Vert_{V^{1,\infty}}$ and $\left\Vert \cdot \right\Vert_{V^{2,\infty}}$ measure the boundedness and smoothness of the potential $V$ and its derivatives, defined as
\begin{align} \label{eq:sobo1}
\left\lVert V\right\rVert_{V^{1,\infty}} &= \left\lVert V\right\rVert_{L^\infty} + \left\lVert \nabla V\right\rVert_{L^\infty}, \\
\left\lVert V\right\rVert_{V^{2,\infty}} &= \left\lVert V\right\rVert_{L^\infty} + \left\lVert \nabla V\right\rVert_{L^\infty} + \left\lVert \nabla^2 V\right\rVert_{L^\infty}. \label{eq:sobo2}
\end{align}

In the RIDK framework, the potential $V$ is regularized through the GRID CG procedure, which renormalizes any singularities and ensures smoothness across the domain. This enables direct computation of the Sobolev norms by evaluating the maximum values ($L^\infty$ norms) of $V,\,\nabla V,$ and $\nabla^2V$, which we performed analytically using the {\tt{SimPy}} module. As an example, here we analyze the Gaussian core model. The RIDK interaction, $V_\mathrm{RIDK}$ [Eq. (\ref{eq:ridk_gauss})], yields $\left\lVert V^\mathrm{GCM}_\mathrm{RIDK}\right\rVert_{L^\infty}=7.1$, $\left\lVert \nabla V^\mathrm{GCM}_\mathrm{RIDK}\right\rVert_{L^\infty}=17.5$, and $\left\lVert \nabla^2 V^\mathrm{GCM}_\mathrm{RIDK}\right\rVert_{L^\infty}=58.7$ (in dimensionless RIDK units). For $n_g=25$, the bounds are
\begin{align}
    \left\lVert V\right\rVert_{V^{1,\infty}} \sqrt{\varepsilon} &\approx (7.1+17.5)\times \sqrt{\frac{\pi}{25}} = 8.71, \\
    \left\lVert V\right\rVert_{V^{2,\infty}} \sqrt{\varepsilon} &\approx (7.1+17.5+58.7)\times \sqrt{\frac{\pi}{25}} = 29.5.
\end{align}
These bounds are further reduced in the final residual estimate [Eq. (\ref{eq:sobolev})], which takes the form $r_{1,\varepsilon}\rho_\varepsilon(x,t)+r_{2,\varepsilon}$, yielding $
r_{1,\varepsilon}\rho_\varepsilon(x,t)+r_{2,\varepsilon}\sqrt{\varepsilon} \lesssim 29.52.$
In summary, this analysis illustrates how numerical estimation of residual bounds via Sobolev norms of the regularized GRID CG interaction can quantify the approximation error introduced by convolution. It further highlights the importance of including higher-order corrections to improve accuracy and support coarse-graining at lower values of $\varepsilon$.
\bibliography{main}
\end{document}

\title{Supplemental Material: Field-Theoretic Simulation of Dean--Kawasaki Dynamics for Interacting Particles}

\author{Jaehyeok Jin}
\email{jj3296@columbia.edu}
\affiliation{Department of Chemistry, Columbia University, 3000 Broadway, New York, NY 10027, USA}
\author{Chen Liu}
\altaffiliation{Current Address: Innovation and Research Division, Ge-Room, Inc., 93160 Noisy le Grand, France}
\affiliation{Department of Chemistry, Columbia University, 3000 Broadway, New York, NY 10027, USA}
\author{David R. Reichman}
\email{drr2103@columbia.edu}
\affiliation{Department of Chemistry, Columbia University, 3000 Broadway, New York, NY 10027, USA}
\date{\today}

\maketitle
\section{Detailed Review of Regularized Inertial Dean-Kawasaki (RIDK) Framework: Without Pair Interactions}
In this section, we provide a concise review of the derivation of the RIDK framework in a physical context, summarizing Refs. \onlinecite{cornalba2019regularized,cornalba2020weakly,cornalba2021well}, while preserving the mathematical rigor of the original work. Following the approach of Cornalba \textit{et al.}, we begin with the RIDK formulation in the absence of pair interactions \cite{cornalba2019regularized} and subsequently extend it to include pairwise interactions \cite{cornalba2020weakly} in Sec. II.

We start from the stochastic evolution equations for $(\rho_\varepsilon, j_\varepsilon)$:
\begin{widetext}
    \begin{align}
    \frac{\partial \rho_\varepsilon}{\partial t}(x,t) &= -\frac{\partial j_\varepsilon}{\partial x} (x,t), \label{eq:rhoepsilon} \\
    \frac{\partial j_\varepsilon}{\partial t}(x,t) &= -\gamma j_\varepsilon(x,t) - j_{2,\varepsilon}(x,t) - \frac{1}{N} \sum_{i=1}^N \left( \frac{1}{N} \sum_{j=1}^N V'(q_i(t)-q_j(t)) \right)\omega_\varepsilon(x-q_i(t))+\frac{\sigma}{N} \sum_{i=1}^N \omega_\varepsilon(x-q_i(t))\dot{\beta}_i.\label{eq:jepsilon}
\end{align}
\end{widetext}

In order to derive a closed formulation for Eqs. (\ref{eq:rhoepsilon}) and (\ref{eq:jepsilon}) in systems without particle pair interactions, two key approximations are introduced. 

\subsection{Kinetic approximation}
First, we close Eq. (\ref{eq:jepsilon}) in terms of higher-order moments of $j_\varepsilon$ by replacing the $j_{2,\varepsilon}$ term. In particular, based on the analysis of the Vlasov-Fokker-Planck equation \cite{duong2013generic}, we assume local equilibrium holds \cite{cornalba2019regularized}, since the position and velocity of particles are separable:
\begin{equation}
    \mathbb{E}\left[ j_{2,\varepsilon}(x,t)\right] =\mathbb{E}\left[p_1^2 (t)\right] \mathbb{E}\left[ \frac{\partial \rho_\varepsilon}{\partial x}(x,t)\right] = \frac{\sigma^2}{2\gamma}\mathbb{E}\left[\frac{\partial \rho_\varepsilon}{\partial x} (x,t) \right].
\end{equation}
Thus, the value of $p_i^2(t)$ can be approximated by the closed form of $\partial \rho_\varepsilon / \partial x$. At the local equilibrium, we can further assume that the particle velocities converge in distribution to a Gaussian with a variance of $\sigma^2/(2\gamma)$. In the limit of low temperature,  $\sigma^2 \ll 2\gamma$, it is reasonable to replace the velocity with the mean value (stationary), since
\begin{equation}
    \mathrm{Var}[p_i^2(t)] \le \frac{C\sigma^4}{(2\gamma)^2} \ll \frac{\sigma^2}{2\gamma} \approx \mathbb{E}[p_i^2 (t)],
\end{equation}
which implies that $\mathbb{E}[j_{2,\varepsilon}]\approx \sigma^2/(2\gamma) \mathbb{E}[\nabla \rho_\varepsilon]$.

Therefore, the local equilibrium approximation closes the kinetic term as follows
\begin{equation} \label{eq:lowTApprox}
    j_{2,\varepsilon} \approx \frac{\sigma^2}{2\gamma} \frac{\partial \rho_\varepsilon}{\partial x}.
\end{equation}

\subsection{Noise replacement}
Next, we replace the stochastic noise $\dot{\mathcal{Z}}_N = \sigma/N \sum_{i=1}^N \omega_\varepsilon(x-q_i(t))\dot{\beta}_i$ with a closed form while retaining a multiplicative structure:
\begin{equation} \label{eq:noise}
    \dot{\mathcal{Y}_N} := \frac{\sigma}{\sqrt{N}} \sqrt{\rho_{\varepsilon/\sqrt{2}}}Q_{\sqrt{2}\varepsilon}^{1/2}\xi,
\end{equation}
where $Q_{\sqrt{2}\varepsilon}$ is a convolution operator involving the regularizing kernel $\omega_{\sqrt{2}\varepsilon}$ acting on $f$ as $Q_{\sqrt{2}\varepsilon}f(\cdot) = \int \omega_{\sqrt{2}\varepsilon}(\cdot -y)f(y)dy$. The particular form of Eq. (\ref{eq:noise}) is derived from the spatial covariance of the exact stochastic noise $\dot{\mathcal{Z}}_N$ \cite{cornalba2019regularized}, i.e., 
\begin{widetext}
\begin{align} \label{eq:gausscov}
    \mathbb{E}[\mathcal{Z}_N(x_1,t)\mathcal{Z}_N(x_2,t)] & = \mathbb{E} \left[ \left(\int_0^t \frac{\sigma}{N} \sum_{i=1}^N \omega_\varepsilon(x_1-q_i(u)) d\beta_i(u) \right) \left(\int_0^t \frac{\sigma}{N} \sum_{i=1}^N \omega_\varepsilon(x_2-q_i(u)) d\beta_i(u) \right)  \right] \nonumber \\
    & = \frac{\sigma^2}{N^2}\mathbb{E}\left[ \sum_{i=1}^N \int_0^t \omega_\varepsilon(x_1-q_i(u)) \omega_\varepsilon(x_2-q_i(u))du \right].
\end{align} 
\end{widetext}
Due to the multiplication rule of Gaussian kernels, we have that $ \omega_\varepsilon(x_1-q_i(u))\omega_\varepsilon(x_2-q_i(u)) = \omega_{\sqrt{2}\varepsilon}(x_1-x_2) \omega_{\varepsilon/\sqrt{2}}((x_1+x_2)/2,u)$ holds for $i \in[1,N]$. Summing over $i$, we obtain $\sum_{i=1}^N \omega_{\varepsilon/\sqrt{2}}(q_i(u)-(x_1+x_2)/2) = \rho_{\varepsilon/\sqrt{2}} ((x_1+x_2)/2,u)$. Combining this equation with Eq. (\ref{eq:gausscov}), $\mathbb{E}[\mathcal{Z}_N(x_1,t)\mathcal{Z}_N(x_2,t)]$ can be reduced to
\begin{align}
    \mathbb{E}[\mathcal{Z}_N(x_1,t)\mathcal{Z}_N(x_2,t)] & = \omega_{\sqrt{2}\varepsilon}(x_1-x_2)  \times \int_0^t \mathbb{E} \left[ \frac{\sigma^2}{N}\rho_{\varepsilon/\sqrt{2}}\left(\frac{x_1+x_2}{2},u \right) \right]du.
\end{align}
The term $\omega_{\sqrt{2}\varepsilon}$ acts as the convolution operator $Q_{\sqrt{2}\varepsilon}$ in $\mathcal{Y}_N$, while the integrand captures the remaining noise, resulting in $\mathcal{Y}_N.$ The convolved noise $\tilde{\xi}_\varepsilon := Q_{\sqrt{2}\varepsilon}^{1/2} \xi$ can be interpreted as a spatially correlated noise that approximates a space-time white noise. 

However, the stochastic noises $\dot{\mathcal{Z}}_N$ and $\dot{\mathcal{Y}}_N$ are not exactly the same. To faithfully apply the final RIDK model to interacting systems, it is essential to identify the difference $\mathcal{R}_N :=\mathcal{Z}_N - \mathcal{Y}_N$ and examine the error bound under a correct scaling. Reference \onlinecite{cornalba2019regularized} established the error bound for the covariance between $\mathcal{Z}_N$ and $\mathcal{Y}_N$: 
\begin{align} \label{eq:stoerr1}
    &|\mathbb{E}[\mathcal{Z}_N(x_1,t)\mathcal{Z}_N(x_2,t)] - \mathbb{E}[\mathcal{Y}_N(x_1,t)-\mathcal{Y}_N(x_2,t)]| \le \frac{C\sigma^2}{N} \omega_{\sqrt{2}\varepsilon}(x_1-x_2)|x_1-x_2|^2,
\end{align}
where the spatial covariance of $\mathcal{Z}_N$ also follows as
\begin{align} \label{eq:stoerr2}
    |\mathbb{E}[\mathcal{Z}_N(x_1,t)\mathcal{Z}_N(x_2,t)]| \le \frac{C\sigma^2}{N}\omega_{\sqrt{2}\varepsilon}(x_1-x_2).
\end{align}
This is valid when $\varepsilon$ is relatively small but not infinitely small. A numerical implementation of the regularized noise using the von Mises kernel and trigonometric basis is provided in Sec. IV for one-dimensional (1D) cases.

\section{Detailed Review of Regularized Inertial Dean-Kawasaki (RIDK) Framework: With Pair Interactions }
\subsection{Mathematical Complexity of Interacting Particle Systems}
Before discussing the replacement of the Dean-Kawasaki interaction term with the convolution form, we would like to remark on the mathematical complexity introduced by weakly interacting particles. While we leave the detailed mathematical derivation and analysis to Ref. \onlinecite{cornalba2020weakly}, formulating this structure is essential for understanding the additional assumptions required in this framework and serves as a foundation for the scaling analysis, which is important for coarse-graining molecular systems.

In Sec. I, we introduced an empirical Gaussian kernel, $\omega_\varepsilon(x)$, to regularize the microscopic delta function (atomistic measure) $\delta(x-q_i(t))$, yielding a well-defined empirical measure. However, when non-trivial interactions are considered, a more complex mathematical framework is needed to regularize these quantities. 

Without interactions between particles, the resulting dynamics is stochastically independent, allowing us to perform Gaussian regularization in the $\mathbb{R}$ domain from a set of particle-level equations:
\begin{align}
  \left\{
    \begin{array}{l}
      \displaystyle \dot{q}_i= p_i,  \vspace{0.2 pc}\\
      \displaystyle \dot{p}_i= -\gamma p_i+\sigma\,\dot{\beta}_i,\qquad i=1,\dots,N. \nonumber
    \end{array}
  \right.
\end{align}
However, when a non-trivial interaction potential between particles, $V$,  the undamped Langevin dynamics can be written as: 
\begin{equation}
  \left\{
    \begin{array}{l}
      \displaystyle \dot{q}_i= p_i,  \vspace{0.2 pc}\\
      \displaystyle \dot{p}_i= -\gamma p_i-\frac{1}{N}\sum_{j=1}^{N}{V'(q_i-q_j)}+\sigma\,\dot{\beta}_i,\qquad i=1,\dots,N,
    \end{array}
  \right.
\end{equation}
Here, the particles $(q_i,\,p_i)_i$ are no longer independent. Therefore, with the pair interaction, an \textit{additional} step is required to introduce an auxiliary Langevin system of particles, where the particles $(\bar{q}_i,\,\bar{p}_i)$ are statistically independent. These variables are governed by the equations
\begin{equation}
\label{eq:vlasov}
  \left\{
    \begin{array}{l}
      \dot{\bar{q}}_i= \bar{p}_i,  \vspace{0.2 pc}\\
          \dot{\bar{p}}_i= -\gamma \bar{p}_i-V'\ast\mu_t(\bar{q}_i)+\sigma\,\dot{\beta}_i,\qquad i=1,\dots,N,
    \end{array}
  \right.
\end{equation}
where $\ast $ denotes the convolution operator and $\mu_t$ is the probability distribution of $\bar{q}_i(t)$. This system is associated with the Vlasov--Fokker--Planck equation \cite{villani2009hypocoercivity,bolley2010trend}. While $(\bar{p}_i,\,\bar{q}_i)_i$ and $(p,q)_i$ evolve differently, the differences between these two configurations upon the propagation of chaos are bound by a constant $C$, as follows: 
\begin{align} \label{eq:chaos}
    \sup_{t \in [0,T]} \mathbb{E} \left[|q_1(t)-\bar{q}_1(t)|^\alpha + |p_1(t)-\bar{p}_1(t)|^\alpha \right]^\frac{1}{\alpha} \leq \frac{C}{\sqrt{N}},
\end{align}
where $\alpha$ is an even natural number. Equation \eqref{eq:chaos} can be proven by considering $\mathbb{E} \left[|q_1(t)-\bar{q}_1(t)|^\alpha + |p_1(t)-\bar{p}_1(t)|^\alpha \right]^\frac{1}{\alpha}$ as a function of time and applying Ito's formula to $|q_i(t)-\bar{q}_i(t)|^\alpha$ and $|p_i(t)-\bar{p}_i(t)|^\alpha$. The detailed proof is provided in Ref. \onlinecite{cornalba2020weakly}.

As the propagation of chaos is bounded under certain limits, the particle coordinates can now be treated independently. In this auxiliary setting, in order to establish the time regularity of Eq. (\ref{eq:vlasov}) \cite{villani2009hypocoercivity}, we need to define our system on a flat torus domain of length one, $\mathbb{T}$, instead of $\mathbb{R}$. Under this toroidal domain, $\mathbb{T}$, a different regularization kernel is required. Reference \onlinecite{cornalba2020weakly} introduced the toroidal equivalent of a Gaussian distribution, known as the periodic von Mises distribution, defined on $\mathbb{T}:=[0,2\pi]$ \cite{forbes2011statistical}
\begin{eqnarray}
    w_{\varepsilon}(x)
:=Z_{\varepsilon}^{-1}e^{-\frac{\sin^2(x/2)}{\varepsilon^2/2}},\qquad Z_{\varepsilon}:=\int_{\mathbb{T}}{e^{-\frac{\sin^2(x/2)}{\varepsilon^2/2}} dx}.
\end{eqnarray}
We note that as $\varepsilon \rightarrow 0$, $\omega_\varepsilon \rightarrow \delta$ (Dirac delta limit), and $\omega_\varepsilon$ follows the same scaling as a Gaussian with variance $\varepsilon^2$. This can be proven from the particular functional form $w(x)=\exp(-V(\sin(x/2)))$, where for small $x$, $V(\sin(x/2))$ approximates to $V(0)+V'(0)\sin(x/2)+0.5V''(0)\sin^2(x/2)$. Thus, $w(x)\approx \exp(-V(0))\exp[-0.5V''(0)\sin^2(x/2)]$ by assuming the symmetry of $V$. By setting $\exp(-V(0))=Z_\varepsilon^{-1}$ and $V''(0)=4\varepsilon^{-2}$, we recover $\omega_\varepsilon(x)$, confirming that the moments of $\omega_\varepsilon$ are consistent with $N(0,\varepsilon^2)$. One key advantage of using the von Mises kernel on the $\mathbb{T}$ domain is its periodicity, making it well-suited for linking to conventional molecular dynamics simulations under periodic interactions and boundary conditions. Given that our main system involves non-trivial interactions between molecules, we will utilize the von Mises kernel for both mathematical analysis and numerical implementation (e.g., scaling analysis). 

\subsection{Interaction Approximation}
\subsubsection{With Pair Potentials}
We aim to replace the interaction term in Eq. (\ref{eq:jepsilon}) containing the pair interactions $V(q_i(t)-q_j(t))$  with the convolution term $\left\{V'\ast \rho_\varepsilon(\cdot,t) \right\}(x)\rho_\varepsilon(x,t)$:
\begin{align} \label{eq:intrx}
    &\frac{1}{N}\sum_{i=1}^N\left(\frac{1}{N}\sum_{j=1}^N V'(q_i(t)-q_j(t)) \right) \omega _\varepsilon (x-q_i(t)) = \left\{V'\ast \rho_\varepsilon(\cdot,t) \right\}(x)\rho_\varepsilon(x,t)+r_{1,\varepsilon}\rho_\varepsilon(x,t)+r_{2,\varepsilon},
\end{align}
where $r_{1,\varepsilon}$ and $r_{2,\varepsilon}$ are stochastic remainders derived from the difference between two terms. 

Next, we further refine $r_{1,\varepsilon}$ and $r_{2,\varepsilon}$ in terms of $\varepsilon$. Following closely the argument presented in Ref. \onlinecite{cornalba2020weakly}, we would like to clarify several approximations that are connected to those derived from the exact form of Dean's equation, as discussed in Sec. II B of the main text. 

We begin by reorganizing the term in the original interaction:
\begin{align} \label{eq:old15}
    &\frac{1}{N}\sum_{i=1}^N\left(\frac{1}{N}\sum_{j=1}^N V'(q_i(t)-q_j(t)) \right) \omega _\varepsilon (x-q_i(t)) \nonumber \\ =&\frac{1}{N} \sum_{i=1}^N \left(\frac{1}{N}\sum_{j=1}^N \left( V'(x-q_j)+V'(q_i-q_j)-V'(x-q_i)\right) \right)  \times\omega_\varepsilon(x-q_i(t)).
\end{align}
We now separate the right-hand side into two contributions: the $V'(x-q_j)$ term (denoted as \textcircled{1}) and the $V'(q_i-q_j)-V'(x-q_j)$ term (denoted as \textcircled{2}).

The first term, \textcircled{1}, can be further simplified as
\begin{align}
    \textcircled{1} & = \rho_\varepsilon(x,t) \left(\frac{1}{N}\sum_{j=1}^N V'(x-q_j(t)) \right),
\end{align}
since $\rho_\varepsilon(x,t) = \left(\sum_{i=1}^N \omega_\varepsilon(x-q_i(t))/N \right)$. From Eq. (\ref{eq:intrx}), we define $r_{1,\varepsilon}$ as the stochastic remainder between \textcircled{1} and the convolution form $\left\{ V' \ast \rho_\varepsilon(\cdot,t)\right\}$:
\begin{align}
    &\rho_\varepsilon(x,t) \left [\left(\frac{1}{N}\sum_{j=1}^N V'(x-q_j(t)) \right) - \left\{ V' \ast \rho_\varepsilon(\cdot,t)\right\}\right]:=  \,r_{1,\varepsilon} \rho_\varepsilon(x,t).
\end{align}
Under this definition, $r_{1,\varepsilon}$ can be expressed as
\begin{widetext}
\begin{align} \label{eq:r1e}
    r_{1,\varepsilon} &= \frac{1}{N}\sum_{j=1}^N V'(x-q_j(t))  - \frac{1}{N} \sum_{j=1}^N \int V'(x-y)\omega_\varepsilon(y-q_j(t))dy = \frac{1}{N}\sum_{j=1}^n \left[ V'(x-q_j(t)) - \int V'(x-y)\omega_\varepsilon(y-q_j(t))dy\right].
\end{align}
\end{widetext}

To further refine Eq. (\ref{eq:r1e}) in terms of $\varepsilon$, we define an effective $\varepsilon$-interval as $A_\varepsilon=(a-\sqrt{\varepsilon},a+\sqrt{\varepsilon})$ and decompose the integral over $\int_\mathbb{T}$ as $\int_\mathbb{T} = \int_{A_\varepsilon} + \int_{\mathbb{T}\backslash A_\varepsilon}$. This separation allows us to evaluate the minimum of the integral over $A_\varepsilon$ under the assumption that the interaction potential $V$ has a finite first-order derivative (i.e., $V'$ is Lipschitz continuous). This additional assumption ensures $|V'(y)-V'(a)|\leq C \sqrt{\varepsilon}$ for all $y\in A_\varepsilon$, given that $|y-a|\leq \sqrt{\varepsilon}.$ Thus, the convolution integral \textcircled{1} is bounded by 
\begin{align}
    \int_\mathbb{T}V'(x-y)\omega_\varepsilon(y-q_j(t))dy &\geq (V'(a)-C\varepsilon) \int_{A_\varepsilon} \omega_\varepsilon(y-a)dy + \min (V') \int_{\mathbb{T}\backslash A_\varepsilon} \omega_\varepsilon(y-a) dy 
\end{align}
Finally, the statistical properties of the von Mises kernel imply that the kernel will rapidly decay outside of the interval $A_\varepsilon$. More quantitatively, this leads to the bound
\begin{equation} \label{eq:bound39}
    \int_{\mathbb{T}\backslash A_\varepsilon} \omega_\varepsilon(y-a)dy \leq C\exp (-C/\varepsilon),
\end{equation}
with a proof outline provided in Subsection B. With Eq. (\ref{eq:bound39}), and the fact that $\int_{A_{\varepsilon}}w_\varepsilon(y-a)dy \leq 1$, we can bound \textcircled{1} by 
\begin{align}
    &\int_\mathbb{T}V'(x-y)\omega_\varepsilon(y-q_j(t))dy - V'(a)  
    \geq \,C (\min(V')-\max(V'))\exp (-C/\varepsilon) - C\sqrt{\varepsilon}, 
\end{align}
which leads to $|r_{1,\varepsilon}| \leq C(V)\sqrt{\varepsilon}$.  

For the second term, \textcircled{2}, a similar argument can be made by the definition of $r_{2,\varepsilon}$ from Eq. (\ref{eq:intrx})
\begin{align}
    r_{2,\varepsilon} = \frac{1}{N} \sum_{i=1}^N & \left(\frac{1}{N} \sum_{j=1}^N \left\{ V'(q_i(t)-q_j(t))-V'(x-q_j(t))\right\} \right) \times \omega_\varepsilon(x-q_i(t)).
\end{align}

The bound for $r_{2,\varepsilon}$ can be established by taking a Taylor expansion 
\begin{align} \label{eq:r2}
    |r_{2,\varepsilon}| & \leq \frac{C}{N} \sum_{i=1}^N |x-q_i(t)|\omega_\varepsilon (x-q_i(t)) \nonumber \\ 
    & =\frac{C}{N} \sum_{i=1}^N |x-\bar{q}_i(t)| \omega_\varepsilon(x-\bar{q}_i(t))+\frac{C}{N} \sum_{i=1}^N\left( |x-{q}_i(t)| \omega_\varepsilon(x-{q}_i(t))-|x-\bar{q}_i(t)| \omega_\varepsilon(x-\bar{q}_i(t)) \right).
\end{align}
The right-hand side of Eq. (\ref{eq:r2}) can be bounded using the propagation of chaos and the regularity condition as:
\begin{align}
    \mathbb{E}\left[ |r_{2,\varepsilon}| \right] \leq C \left\{ \sqrt{\varepsilon} + \varepsilon^\beta \right\},
\end{align}
where $\beta>0$ \cite{cornalba2020weakly}. We note that the complex bounds involving the propagation of chaos and related inequalities arise from the nature of pair interactions, which require auxiliary particle systems to properly account for chaos propagation. 

If we assume that contributions of order $\varepsilon^\beta$ and higher are negligible, we finally arrive at
\begin{equation}
    r_{1,\varepsilon} \rho_\varepsilon(x,t)+r_{2,\varepsilon} \lesssim C_1 \rho_\varepsilon(x,t) \sqrt{\varepsilon} +C_2 \sqrt{\varepsilon},
\end{equation}
where $C_1$ and $C_2$ can be further estimated using Eq. (25) of the main text.

\subsubsection{Interaction Bound for $r_{1,\varepsilon}$}
Here, we will show $\int_{\mathbb{T}\backslash A_\varepsilon}\omega_\varepsilon(y-a)dy \leq C \exp (-C/\varepsilon)$. First, the von Mises kernel is defined as $\omega_\varepsilon(x) = Z_\varepsilon^{-1} \exp \left( - \frac{\sin^2 (x/2)}{\varepsilon^2/2} \right)$ near $x\approx 0$. Hence, outside of $A_\varepsilon$, i.e., $\mathbb{T}\backslash A_\varepsilon$, $\omega_\varepsilon(x)$ decays rapidly and follows this limiting case
\begin{eqnarray}
    \omega_\varepsilon(x)\leq Z_\varepsilon^{-1} \exp \left(-\frac{1}{2\varepsilon}\right)
\end{eqnarray}
where $Z_\varepsilon$ for small $\varepsilon$ can be approximated by integrating the Gaussian-like function over $\mathbb{T}$, giving $Z_\varepsilon \approx \sqrt{2\pi \varepsilon^2}$. Now, consider the integral over $\mathbb{T}\backslash A_\varepsilon$:
\begin{eqnarray}
    \int_{\mathbb{T}\backslash A_\varepsilon}\omega_\varepsilon(y-a)dy \leq \frac{1}{\sqrt{2\pi\varepsilon^2}}\exp (- \frac{1}{2\varepsilon}) \cdot (1 -2\sqrt{\varepsilon}),
\end{eqnarray}
as $1-2\sqrt{\varepsilon}$ is the length of $\mathbb{T}\backslash A_\varepsilon$. For the small $\varepsilon$, we can then bound the integral to $C\exp (-C/\varepsilon)$.

\subsubsection{With External Potentials}
Unlike the pair potentials, bounds from the interaction approximation can be more straightforwardly established for external potentials. In particular, we consider field-induced external interactions, such as those from electric fields, which take the form $V(q)=\nu q^{n}$, where $n\ge2$ and $\nu$ is a constant. Examples include externally coupled harmonic oscillators for $n=2$ and non-linear electronic polarization interaction for $n=3$, etc.

Since external interactions do not involve pair distances, we would like to demonstrate that $N^{-1}\sum_{i=1}^N V'(q_i(t)) \omega_\varepsilon(x-q_i(t))$ can be approximated by $V'(x)\rho_\varepsilon(x,t)$. Based on Ref. \onlinecite{cornalba2019regularized}, this is equivalent to 
\begin{equation}
    \mathbb{E}\left[ |V'(q_1(t))-V'(x)|\omega_\varepsilon (x-q_1(t))\right]\rightarrow 0
\end{equation}
as $\varepsilon \rightarrow 0$. Statistically speaking, evaluating the mean value of this quantity can be formulated as an integral over the probability density function of the phase space ($q_1(t)$), denoted as $f_q$.

To show this, we define the non-negative part of the exponent as $\alpha :=n-2\ge0$ and choose $\tau \in (0,\alpha^{-1})$ to construct the interval $D_\tau(\varepsilon):=[-\varepsilon^{-\tau},\varepsilon^{\tau}]$. Then,
\begin{align} \label{eq:EE}
    & \mathbb{E}\left[ |V'(q_1(t))-V'(x)|\omega_\varepsilon (x-q_1(t))\right] \nonumber \\ = & \int_\mathbb{R} |V'(y)-V'(x)| \omega_\varepsilon(x-y)f_q(y)dy. 
\end{align}

Using the time regularity of the associated Fokker-Planck equation for the probability distribution function $f_q(y)$, the norm is bounded (as shown in Refs. \onlinecite{cornalba2019regularized} and \onlinecite{herau2004isotropic}), which further bounds Eq. (\ref{eq:EE}) as 
\begin{align} \label{eq:EE2}
    & \mathbb{E}\left[ |V'(q_1(t))-V'(x)|\omega_\varepsilon (x-q_1(t))\right] \nonumber \\ \le \,\,&
    C\int_{D_\tau(\varepsilon)}|V'(y)-V'(x)|\omega_\varepsilon(x-y)dy +  C\int_{\mathbb{R}-D_\tau(\varepsilon)}|V'(y)-V'(x)|\omega_\varepsilon(x-y)dy.
\end{align}

In the right-hand side of Eq. (\ref{eq:EE2}), the first term $C\int_{D_\tau(\varepsilon)}|V'(y)-V'(x)|\omega_\varepsilon(x-y)dy$ can be further bounded by applying the mean value theorem to $|V'(y)-V'(x)|$:
\begin{equation}
|V'(y)-V'(x)|\le |y-x|C(\alpha)(1+|y|^\alpha)    
\end{equation}
as $V''(y)$ is bounded within $D_\tau(\varepsilon)$. Since we can bound $(1+|y|^\alpha)\le \varepsilon^{-\tau \alpha}$ in $D_\tau(\varepsilon)$, we arrive at the following inequality:
\begin{align}
    & C\int_{D_\tau(\varepsilon)}|V'(y)-V'(x)|\omega_\varepsilon(x-y)dy \nonumber \\ \le \,\, & C\varepsilon^{-\tau\alpha} \int_{D_\tau(\varepsilon)}|y-x|\omega_\varepsilon(x-y)dy.
\end{align}
Since $\omega_\varepsilon$ is a Gaussian kernel, its first moment is bounded, and $\int_{D_\tau(\varepsilon)}|y-x|\omega_\varepsilon(x-y)dy\le C\varepsilon$ holds. Combining this with the previous result, we find that the first term is bounded by $ C_1\varepsilon^{-\tau\alpha+1}$ for some constant $C_1$. 

For the second term in Eq. (\ref{eq:EE2}), the first derivative $V'(y)$ is also bounded since $|V'(y)|\le C(\alpha)(1+|y|^{\alpha+1})$. Therefore, applying $|V'(y)-V'(x)|\le C_2(\alpha)(1+|y|^{\alpha+1})$ over the $\mathbb{R}-D_\tau(\varepsilon)$ domain, we obtain
\begin{align}
    & C\int_{\mathbb{R}-D_\tau(\varepsilon)}|V'(y)-V'(x)|\omega_\varepsilon(x-y)dy \nonumber \\ \le \,\, & C_2(x,\tau,\alpha) \int_{\mathbb{R}-D_\tau(\varepsilon)} (1+|y|^{\alpha+1})\omega_\varepsilon(x-y)dy.
\end{align}

By combining these results, we achieve a clear bound for the overall terms in Eq. (\ref{eq:EE}):
\begin{align}
     & \mathbb{E}\left[ |V'(q_1(t))-V'(x)|\omega_\varepsilon (x-q_1(t))\right] \nonumber \\ \le & C_1\varepsilon^{-\alpha\tau+1} + C_2(x,\tau,\alpha) \int_{\mathbb{R}-D_\tau(\varepsilon)} (1+|y|^{\alpha+1})\omega_\varepsilon(x-y)dy,
\end{align}
where the right-hand side approaches zero as $\varepsilon\rightarrow0$. Therefore, for external field interactions, one can replace the original equation term with $V'(x)\rho_\varepsilon(x,t)$ from the stochastic analysis. This bound is much clearer and tighter because it does not require introducing an auxiliary system or using the propagation of chaos. 

\subsection{Final RIDK Model}
\subsubsection{Stochastic Equations of Motion}
To summarize this section, we initially began with particle-level dynamics for weakly interacting particles:
\begin{align}
  \left\{
    \begin{array}{l}
      \displaystyle \dot{q}_i= p_i,  \vspace{0.2 pc}\\
      \displaystyle \dot{p}_i= -\gamma p_i+\sigma\,\dot{\beta}_i,\qquad i=1,\dots,N, \nonumber
    \end{array}
  \right.
\end{align}
and, by introducing an auxiliary independent particle system, we derived the closed dynamical equations for the $(\rho_\varepsilon, j_\varepsilon)$ variables. This set of equations is closed by introducing three approximations (kinetic approximation, noise replacement, and interaction convolution approximation; see Sec. IV of the main text for detailed bounds) under a large $\theta$ value following the scaling condition $N\varepsilon^\theta=1$,
\begin{align}
\label{eq:finalridk}
\frac{\partial \tilde{\rho}_{\varepsilon}}{\partial t}(x,t) =& -\frac{\partial \tilde{j}_{\varepsilon}}{\partial x}(x,t), \\
\frac{\partial  \tilde{j}_{\varepsilon}}{\partial t}(x,t)  =& 
  -\gamma  \tilde{j}_{\varepsilon}(x,t)-\left(\frac{\sigma^2}{2\gamma}\right)\frac{\partial  \tilde{\rho}_{\varepsilon}}{\partial x}(x,t) -\{V'\ast \tilde{\rho}_{\varepsilon}(\cdot,t)\} \tilde{\rho}_{\varepsilon}(\cdot,t)
  +\frac{\sigma}{\sqrt{N}}\sqrt{ \tilde{\rho}_{\varepsilon}(x,t)}.
\end{align}

\subsubsection{Numerical RIDK Simulation}
To numerically propagate the RIDK model, the \textit{final approximation} assumes $(\rho,j)\approx (\rho_\varepsilon, j_\varepsilon)$ to avoid additional numerical complexities associated with tracking $\varepsilon$ with additional constraints. The $\varepsilon-$kernel is introduced to construct initial configurations of $\rho_{\varepsilon,0}$ and $j_{\varepsilon,0}$. We then propagate ($\rho,\,j)$ based on the final RIDK equation [Eq. (\ref{eq:finalridk})]. As long as the introduced approximations are valid and bounded, this approach provides a numerically feasible method for conducting field-theoretic simulations.

In practice, we implement the RIDK equation numerically using a discontinuous Galerkin framework \cite{arnold2000discontinuous}. Details of the computational setup for field-theoretic simulation can be found in Sec. V. 

\section{Numerical Implementation of Regularized Noise}
\subsection{Theory}
In this section, we provide a brief description of how to implement stochastic noise in closed form $\dot{\mathcal{Y}}_N:=\sigma/\sqrt{N} \left(\sqrt{\rho_\varepsilon} P_{\sqrt{2}\varepsilon}^{1/2} \xi_1, \cdots, \sqrt{\rho_\varepsilon} Q_{\sqrt{2}\varepsilon}^{1/2} \xi_d  \right)$, where $Q_\varepsilon$ is the convolution operator defined using the von Mises kernel $\omega_\varepsilon$, and $\{\xi_l\}_l$ are independent space-time noises. For simplicity, we outline the numerical implementation in 1D; the extension to two-dimensional (2D) cases follows similarly. Based on Ref. \onlinecite{cornalba2021well}, we first define the $Q-$Wiener representation of the noise $\mathbb{W}_\varepsilon$ as $\dot{\mathbb{W}}_\varepsilon:=\tilde{\xi}_\varepsilon = Q^{1/2}_{\sqrt{2}\varepsilon}\xi$. Then, $\mathbb{W}_\varepsilon$ can be expressed in terms of the spectral properties of $Q_\varepsilon$ \cite{prevot2007concise} using independent Brownian noise $\{\beta_{j} \}_j$, as follows:
\begin{equation}
    \mathbb{W}_\varepsilon := \sum_{j=1}^\infty \sqrt{\alpha_{j,s,\varepsilon}}f_{j,s} \beta_j(t),
\end{equation} 
where $\{f_{j,s}\}_j$ is a set of orthonormal basis of eigenfunctions of $Q_{\sqrt{2}\varepsilon}$, defined as
\begin{equation} \label{eq:fjs}
    f_{j,s}(x) = C(d) \biggl\{ e_{j}(x)\biggr\}\left(1+|j|^2\right)^{-s/2}.
\end{equation}
Here, $s$ is a real number representing the fractional Sovolev space order required for the well-posedness of RIDK on the $\mathbb{T}^d$ domain \cite{cornalba2021well}. Thus, $\{f_{j,s}\}$ is defined on $H^s$ space.  

In Eq. (\ref{eq:fjs}), $e_{j}$ is the trigonometric basis defined as
\begin{align}
 e_{j}(x) \coloneqq 
  \begin{cases}
    \sqrt{1/\pi}\cos(jx), & \mbox{if } j>0, \vspace{0.3 pc}\\
    \sqrt{1/2\pi}, & \mbox{if } j=0,\\
    \sqrt{1/\pi}\sin(jx), & \mbox{if } j<0, \vspace{0.3 pc}
  \end{cases}                   
\end{align}

Note that the eigenvalue of the $P_{\sqrt{2}\varepsilon}$ operator, $\lambda_{j,\varepsilon}$, is related to $\alpha_{j,s,\varepsilon}$ by
\begin{equation}
    \alpha_{j,s,\varepsilon} :=(1+|j|^2)^s \lambda_{j,\varepsilon},
\end{equation}
where $\lambda_{j,\varepsilon}$ for a 1D system is given by
\begin{widetext}
    \begin{align}
  \label{eq:2005}
  \lambda_{j,\varepsilon} 
  & =
    \begin{cases}
      \displaystyle Z^{-1}_{\sqrt{2}\varepsilon}\int_{\mathbb{T}}{e^{-\frac{\sin^2(x/2)}{\varepsilon^2}}\cos(jx)
      \mathrm{d}x}=
   I_{j}\left(\{2\varepsilon^2\}^{-1}\right) / I_0(\{2\varepsilon^2\}^{-1} ), 
      & \mbox{if } j\neq 0, \vspace{0.3 pc}\\
      1, & \mbox{if } j=0,
    \end{cases}
\end{align}
\end{widetext}
where $I_j$ is the modified Bessel function. 

\subsection{Implementation}
We now simplify $\mathbb{W}_\varepsilon$ for the 1D case:
\begin{align}
    \mathbb{W}_\varepsilon(x,t) & = \sum_{j\in\mathbb{Z}} \left( 1+|j|^2\right)^{\frac{s}{2}} \lambda_{j,\varepsilon}^{\frac{1}{2}} \beta_j(t)e_j(x) c_{d=1}\left( 1+|j|^2 \right)^{-\frac{s}{2}}\nonumber \\ & = \sum_{j\in\mathbb{Z}}\left( \lambda_{j,\varepsilon}^{\frac{1}{2}}\beta_j(t)c_{d=1}\right)e_j(x).
\end{align}
Substituting $e_j(x)$ for the 1D case, we obtain
\begin{align} \label{eq:Wenoise}
    \mathbb{W}_\varepsilon(x,t) &= \sum_{j\in\mathbb{Z}^+} \left[ \left( \pi^{-1/2}\lambda_{j,\varepsilon}^{1/2}\beta_j(t)c_{d=1}\right) \cos(jx)\right] \nonumber \\&+ \sum_{j\in\mathbb{Z}^-} \left[ \left( \pi^{-1/2}\lambda_{j,\varepsilon}^{1/2}\beta_j(t)c_{d=1}\right) \sin(jx)\right] \nonumber \\ &+ (2\pi)^{-1/2} \lambda_{0,\varepsilon}^{1/2}\beta_0(t)c_{d=1}.
\end{align}
Next, we define $A_j$ as $A_j := c_{d=1}\pi^{-1/2}\left( \lambda_{j,\varepsilon}\right)^{1/2}\beta_j(t).$ Then, Eq. (\ref{eq:Wenoise}) can be further expressed as
\begin{widetext}
\begin{align}
    \mathbb{W}_\varepsilon(x,t) &= A_0 + \sum_{j\in\mathbb{Z}^+} \frac{1}{2} A_j \left(e^{ijx}+e^{i(-j)x} \right) + \sum_{j\in\mathbb{Z}^-} \frac{1}{2i} A_j \left(e^{ijx}-e^{i(-j)x} \right) \nonumber \\ & = A_0 + \sum_{j\in\mathbb{Z}^+} \left( \frac{1}{2}A_j-\frac{1}{2i}A_{-j}\right)e^{ijx} + \sum_{j\in\mathbb{Z}^-} \left( \frac{1}{2}A_{-j}+\frac{1}{2i}A_j\right)e^{ijx}.
\end{align}
\end{widetext}
We now simplify $\mathbb{W}_{\varepsilon}(x,t)$ as $A_0+\sum_j B_je^{ijx}$, where
    \begin{align} \label{eq:bj}
  B_{j} 
  & =
    \begin{cases}
      \displaystyle \frac{1}{2}\left(A_j+iA_{-j}\right), 
      & \mbox{if } j> 0, \vspace{0.3 pc} \\
      \displaystyle \frac{1}{2}\left(A_{-j}-iA_{j}\right), & \mbox{if } j<0.
    \end{cases}
\end{align}
Plugging $A_j$ and $A_{-j}$ back into Eq. (\ref{eq:bj}) gives
\begin{align} \label{eq:bj2}
  2B_{j} 
  & =
    \begin{cases}
      \displaystyle c_{d=1}\pi^{-\frac{1}{2}}\lambda_{j,\varepsilon}^{\frac{1}{2}}\beta_j+ic_{d=1}\pi^{-\frac{1}{2}}\lambda^{\frac{1}{2}}_{-j,\varepsilon}\beta_{-j}, 
      & \mbox{if } j> 0, \vspace{0.3 pc} \\
      \displaystyle c_{d=1}\pi^{-\frac{1}{2}}\lambda_{-j,\varepsilon}^{\frac{1}{2}}\beta_j-ic_{d=1}\pi^{-\frac{1}{2}}\lambda^{\frac{1}{2}}_{j,\varepsilon}\beta_{j}, & \mbox{if } j<0.
    \end{cases}
\end{align}
Since $\lambda^{1/2}_{j,\varepsilon} = \lambda_{-j,\varepsilon}^{1/2}$, we can further factorize Eq. (\ref{eq:bj2}) into
\begin{align} \label{eq:bj3}
  2B_{j} 
  & =
    \begin{cases}
      \displaystyle c_{d=1}\pi^{-\frac{1}{2}}\lambda_{j,\varepsilon}^{\frac{1}{2}} (\beta_j+i\beta_{-j}), 
      & \mbox{if } j> 0, \vspace{0.3 pc} \\
      \displaystyle c_{d=1}\pi^{-\frac{1}{2}}\lambda_{j,\varepsilon}^{\frac{1}{2}} (\beta_j-i\beta_{-j}), & \mbox{if } j<0.
    \end{cases}
\end{align}
Since $\beta_j$ is real, we observe that $\tilde{\beta}_j := \beta_j+i\beta_{-j}$ satisfies $\tilde{\beta}_{-j}=\tilde{\beta}_j^\ast$. Let $\xi_j \in \mathbb{R}$, then we can define $\tilde{\beta}_j$ as
\begin{equation}
    \tilde{\beta}_j = \frac{\xi_j +\xi_{-j}}{\sqrt{2}}+ i \frac{\xi_j -\xi_{-j}}{\sqrt{2}},
\end{equation}
which satisfies $\tilde{\beta}_{-j}=\tilde{\beta}_j^\ast$. Therefore, by drawing a random variable $\xi_j$, we can construct $\tilde{\beta}_j$, and consequently $A_j$ and $B_j$, providing a complete structure of $\mathbb{W}_\varepsilon$.  

\section{Proof-of-concept Study}
\subsection{Computational Details}
\subsubsection{Numerical Settings}
All simulations were performed on a $50 \times 50$ numerical grid. The friction and diffusion coefficients were set as $\gamma = 1$ and $\sigma = 10^{-6}$ to suppress thermal noise and focus on the effects of pair interactions. The timestep was $dt = 0.01$, and the systems were propagated for $20 dt$, which was sufficient for this proof-of-concept demonstration. Reciprocal $k$-space vectors were constructed on a $40 \times 40$ grid.

\subsubsection{1D Case}
The initial density was set as $\rho_0(x,y) = N_x(\pi,\varsigma_\mathrm{1D})$ with $\varsigma_\mathrm{1D} = 1/4$. The pair interaction was
\[
V(r) = 10 \exp\left[ -0.5 \left( \frac{r}{\varsigma_\mathrm{1D}} \right)^2 \right],
\]
with a cutoff distance of $R_c = 3$. The density evolved into two symmetric peaks along the $x$-axis due to repulsion.

\subsubsection{2D Case}
The initial density was $\rho_0(x,y) = N_x(\pi,\varsigma_\mathrm{2D})N_y(\pi,\varsigma_\mathrm{2D})$ with $\varsigma_\mathrm{2D} = 1$. The interaction was
\[
V(r) = 10 \exp\left[ -0.5 \left( \frac{r}{\varsigma_\mathrm{2D}} \right)^2 \right],
\]
with the same cutoff $R_c = 3$. The density evolved into a ring-like structure, reflecting radial symmetry and repulsive forces.

\subsubsection{Notes on Interaction Potentials}
Note that here we implemented the bare interaction without coarse-graining as a conceptual test; in molecular systems, a coarse-grained potential should replace $V(R)$ to ensure physical realism.

\subsection{Numerical Implementation: Proof-of-Concept}
Building on the mathematical framework and approximations, we implement and test the impact of particle-level interactions on mesoscopic fields. While such interactions are expected to manifest at the field level through the RIDK framework, to the best of our knowledge this study presents the first explicit numerical set of studies. As a proof-of-concept, consider a system with bare pair interactions of a Gaussian repulsive form. The contribution of local interactions up to a cutoff distance $R_c$ can be studied in conjunction with an examination of how the density distribution evolves as $R_c$ is increased. 

\begin{figure}
    \includegraphics[width=0.45\textwidth]{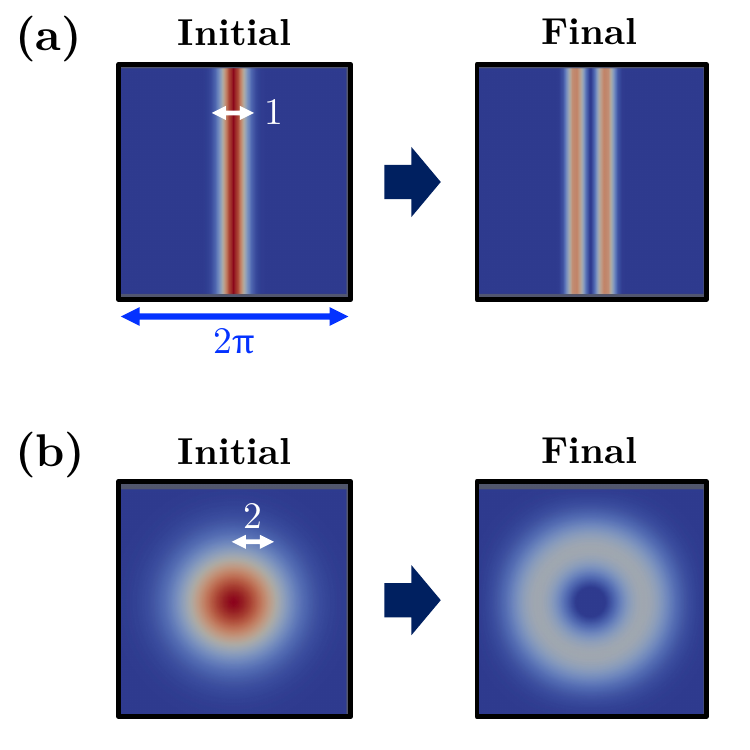}
    \caption{\label{fig:figure2} Role of pair interactions between particles at the mesoscopic field level: Proof-of-concept demonstration of Gaussian interactions in (a) 1D and (b) 2D.}
\end{figure}

Consider a 1D and a 2D system on a $50 \times 50$ grid, respectively. The initial density distribution is taken to be of a Gaussian form: $\rho_0(x,y) = N_x(\pi,\varsigma_\mathrm{1D})$ in 1D [Fig. 2(a)] and $\rho_0(x,y) = N_x(\pi,\varsigma_\mathrm{2D})N_y(\pi,\varsigma_\mathrm{2D})$ in 2D [Fig. 2(b)], with standard deviations $\varsigma_\mathrm{1D} = 1/4$ and $\varsigma_\mathrm{2D} = 1$. We use pair interactions of the Gaussian core form $V(r) = 10 \exp \left[ -0.5 (r/\varsigma)^2 \right]$ with $R_c=3$ and evolve the RIDK equation over time. Figure \ref{fig:figure2} demonstrates how pair interactions drive field-level density redistribution. In 1D, the Gaussian density splits symmetrically along $x$, while in 2D it evolves into a ring due to radial symmetry, confirming that mesoscopic pair interactions can induce nontrivial collective dynamics. We note that introducing interactions requires careful selection of numerical parameters. For example, the interaction range $\varsigma$ should exceed the grid spacing $2\pi/n_g$, where $n_g$ is the number of grid points. In our case, $n_g=50$ with $N=20$ ensures non-vanishing interactions in both dimensions. This constraint becomes more critical in strongly interacting systems and is discussed further in Sec. VII of the main text. Also, we note that $N \varepsilon^\theta = 1$ was not imposed here, as our primary goal is to examine the effect of pair interactions on the evolution of density fields (see Secs. VIII and IX of the main text).

\subsection{Additional Analysis}
\begin{figure}
    \includegraphics[width=0.45\textwidth]{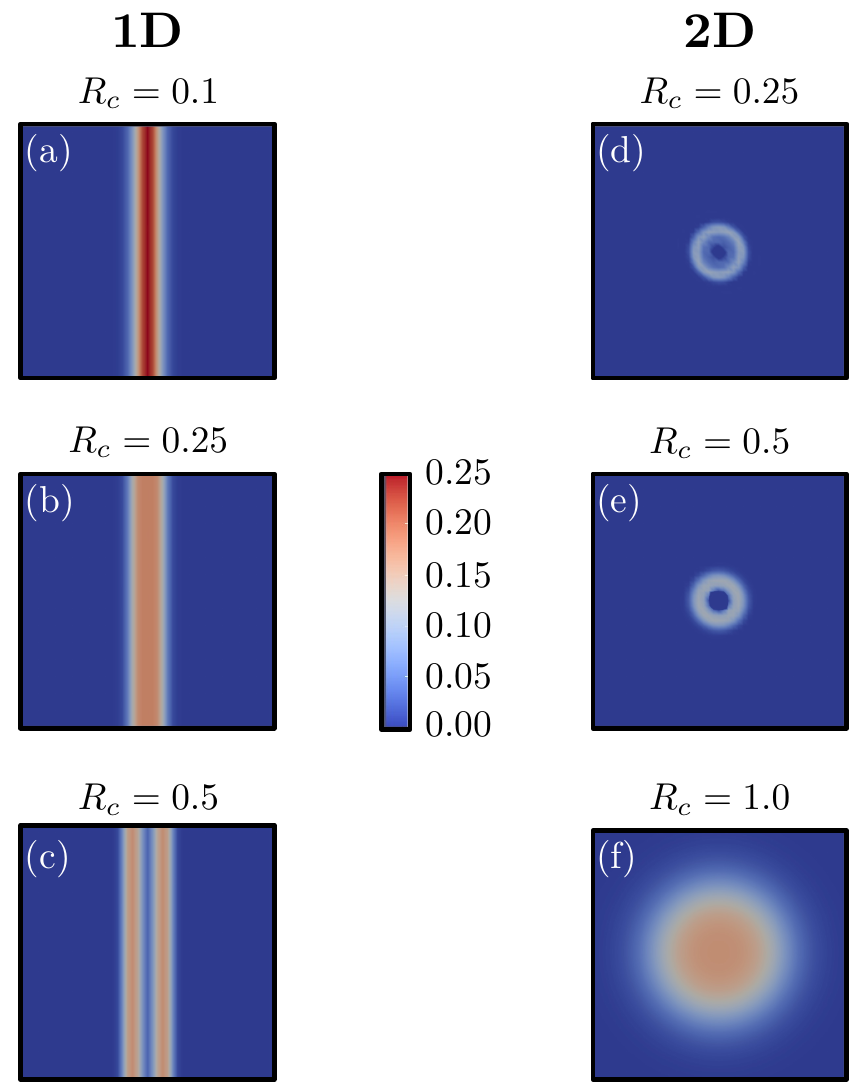}
    \caption{\label{fig:figure12} Systematic analysis of the role of pair interactions at the mesoscopic field level, extending Fig. 2 in the main text with the same initial settings in (a-c) 1D and (d-f) 2D at varying cutoff distances. The final snapshots after field simulations are shown here.}
\end{figure}

Additional computational demonstrations of the proof-of-concept implementation for weak Gaussian interactions are shown in Fig. \ref{fig:figure12}. In particular, Fig. \ref{fig:figure12} depicts the final snapshots of the density field variables after the RIDK simulation (same timestep and duration were used for all cases), discretized into a $50\,\times\,50$ grid.

Compared to the case in Fig. 2 in the main text, where a clear separation of densities due to non-negligible Gaussian repulsions was observed, changing the cutoff distance $R_c$ for pair interactions unambiguously affects the final density profile. For 1D examples [Fig. \ref{fig:figure12}(a)-(c)], decreasing $R_c$ from 3 (Fig. 2) results in a less pronounced separation of the density distribution at the final time of $t_f=0.2$. For even smaller cutoffs, e.g., $R_c=0.25$, the density distribution remains almost unchanged (not separated) or similar to the initial distribution for very small $R_c\ll1$ [Fig. \ref{fig:figure12}(a)]. 

A similar trend is also observed for two-dimensional disks [Fig. \ref{fig:figure12}(d)-(f)]. Decreasing $R_c$ from 3 (as shown in Fig. 2) leads to less repulsion in the final density distribution, as seen in Fig. \ref{fig:figure12}(f) for $R_c=1.0$, and a further narrowing of the distribution at even smaller values, as shown in Figs. \ref{fig:figure12}(d) and \ref{fig:figure12}(e).

\section{Computational Details}
\subsection{Atomistic Simulations: Details}
For microscopic systems, we constructed an effective 2D system with 400 particles in a square box of dimensions $30\,\angstrom\times30\,\angstrom$ by setting up a pseudo-2D periodic box with a \textit{z}-dimension width of $0.2\,\angstrom$ to avoid numerical artifacts. The initial configuration of each system was randomly generated within this pseudo-2D periodic box using the Packmol software \cite{martinez2009packmol}, and the \textit{z}-coordinates were manually set to zero, ensuring an effective 2D configuration for both the Gaussian core model and the Lennard-Jones model. Before running constant \textit{NVT} dynamics to sample the microscopic trajectory, we performed energy minimization using the steepest descent method, followed by the Polak-Ribiere version of the conjugate gradient method \cite{polak1969note} to remove artificial stresses and artifacts within the unit cell. 

At the target thermodynamic state (specified temperature), we carried out constant \textit{NVT} dynamics using the N\'{o}se-Hoover thermostat \cite{nose1984unified,hoover1985canonical} for 10 ns. The last 5 ns of the trajectory was collected and used to compute the correlations and prepare for the field-theoretic simulations. During both the minimization and \textit{NVT} runs, we enforced the 2D nature of the system by applying the {\tt{enforce2d}} fix in LAMMPS \cite{plimpton1995fast,brown2011implementing,brown2012implementing}.

\subsection{Field Simulations: Computational Pipeline}
\begin{figure*}
    \includegraphics[width=0.75\textwidth]{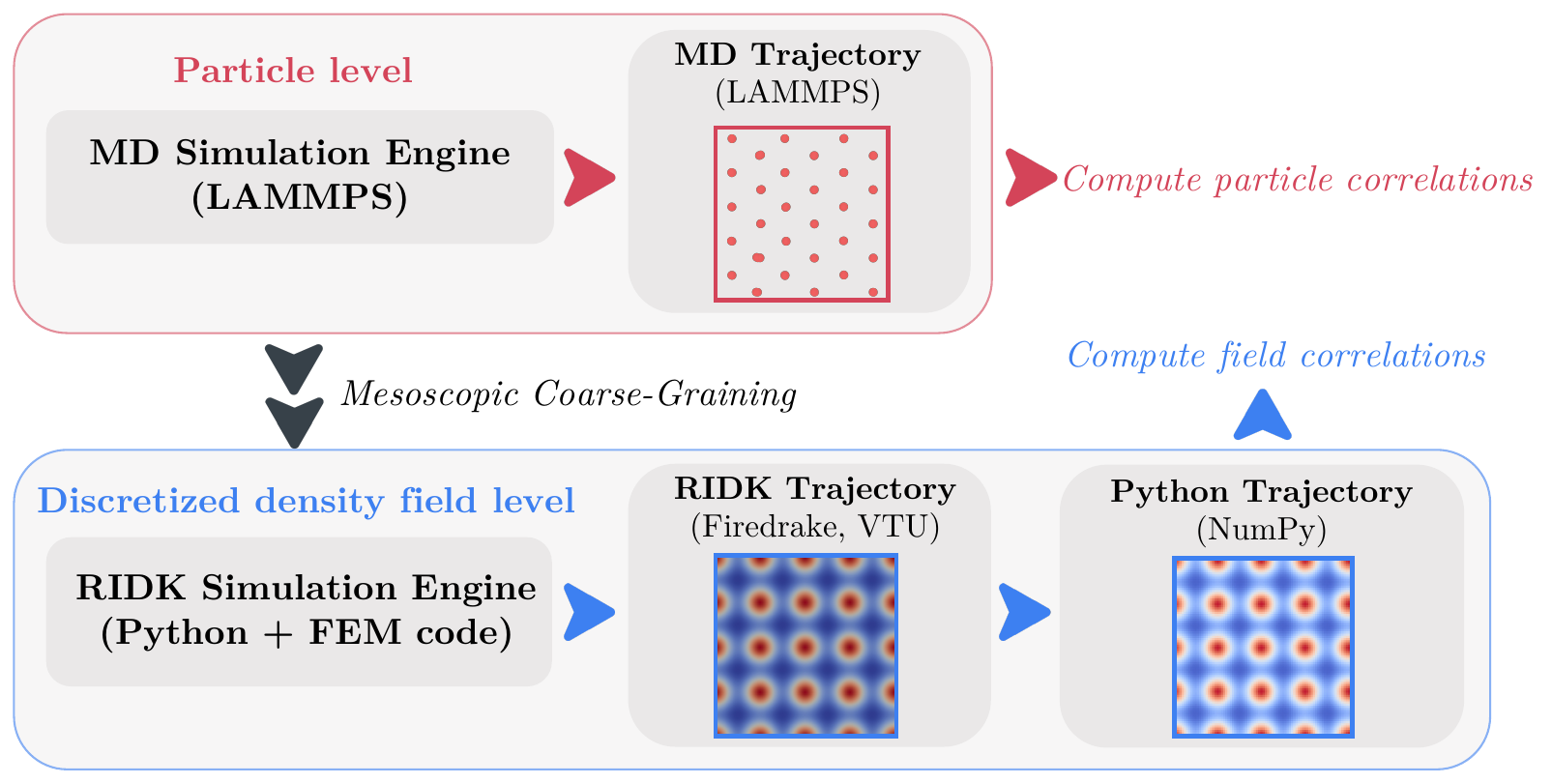}
    \caption{\label{fig:figure13} Computational pipeline developed to bridge the microscopic (particle) and mesoscopic (field) levels. Starting with particle-level MD simulation using the  Large-scale Atomic/Molecular Massively Parallel Simulator (LAMMPS) MD engine \cite{plimpton1995fast,brown2011implementing,brown2012implementing}, particle correlations can be computed from the MD trajectory (top). The microscopic trajectory is then faithfully mapped to a discretized density field for mesoscopic simulation. Using the coarse-grained trajectory and GRID CG interactions, this pipeline proceeds with the RIDK simulation implemented in Python and coupled with {\tt{Firedrake}} for FEM simulation. The resulting grid output files in VTU format are converted into a NumPy array to efficiently estimate field-level properties from density-density correlations.}
\end{figure*}
While our field-level interacting system is built upon the particle-level MD simulation, the particle-level MD and field-level RIDK simulations rely on completely different program bases and numerical methods. To bridge these approaches, we developed a comprehensive computational pipeline that integrates these two distinct methods, see Fig. \ref{fig:figure13}. The first step in our computational pipeline involves conducting a particle-level simulation with the MD simulation engine, as detailed in Subsection A. From this trajectory, we prepare two essential components for the RIDK simulation through mesoscopic coarse-graining.

The first component is the \textit{configuration}. From the last snapshot of the atomistic simulation, we extract the phase space variables and perform mesoscopic coarse-graining by constructing a histogram at the desired coarse-graining level, specified by the numerical grid size $n_g\,\times\,n_g$. For numerical stability, we then fit the mapped initial coordinates, $\rho_\varepsilon$, and momentum, $j_\varepsilon$, to an analytical form using Fourier transforms. This serves as input for our Python code for the field simulation. 

The second component is the \textit{GRID CG interaction}. From half of the \textit{NVT} statistics, we apply the GRID CG method to determine the coarse-grained field-level interaction values at the specified grid spacing. These values are subsequently fitted to an analytical function as input in our Python code. 

With these two inputs prepared, our field-theoretic simulation (see Subsection C) uses the initial density, momentum, and GRID CG interaction in the analytical form to carry out the FEM simulation. This code is open-source and available in Ref. \onlinecite{github}. Finally, we convert the grid trajectories, originally in a Visualization Toolkit for Unstructured grids (VTU), into a NumPy array using the {\tt{meshio}} package. This sequential simulation and conversion process allows for direct analysis of field-level correlations based on density correlations. 

\subsection{Field-Theoretic Simulations: Details}
Using the initial analytical conditions specified by the procedure in Subsection B, we performed field-theoretic simulations with the Python package {\tt{Firedrake}} \cite{rathgeber2016firedrake}. Specifically, we employed the weak form for the Raviart--Thomas mixed finite-element approximation \cite{brezzi2012mixed}, which allows for approximating the numerical fluxes $\rho$ and $j$ across mesh elements in terms of discontinuous functions \cite{arnold2000discontinuous}. Recent mathematical work demonstrated the convergence of this discontinuous Galerkin framework by solving a wave equation form of the RIDK model \cite{cornalba2023regularised}. Additional analytical proofs and discussions for one- and two-dimensional systems are also provided in Ref. \onlinecite{cornalba2023regularised}. 

\section{High-Density Gaussian Core Model}
\subsection{Microscopic Setting}
Similar to the Gaussian core model discussed in the main text (Sec. VIII), we prepared the 400 Gaussian particles in a $30\,\angstrom\,\times\,30\,\angstrom$ box, interacting with parameters $\epsilon = 3.178$ kcal/mol and $\sigma=3.75$ $\angstrom$ at 100 K:
\begin{equation}
    V^\mathrm{hGCM}_\mathrm{MD}(r)=3.178 \times \exp \left( -\frac{1}{2}\frac{r^2}{3.75^2}\right),
\end{equation}
where $\epsilon = 3.178$ kcal/mol at 100 K corresponds to a reduced temperature $T^\ast$ of 16. With $L_\mathrm{MD}=30\,\angstrom$ and 
$\mathbb{L}_\mathrm{MD}=3.75 \,\angstrom$, the system yields a reduced density $\tilde{\rho}=6.25$, indicating a highly dense condition.  

\subsection{Microscopic and Mesoscopic Correlations}
\begin{figure}
    \includegraphics[width=0.5\textwidth]{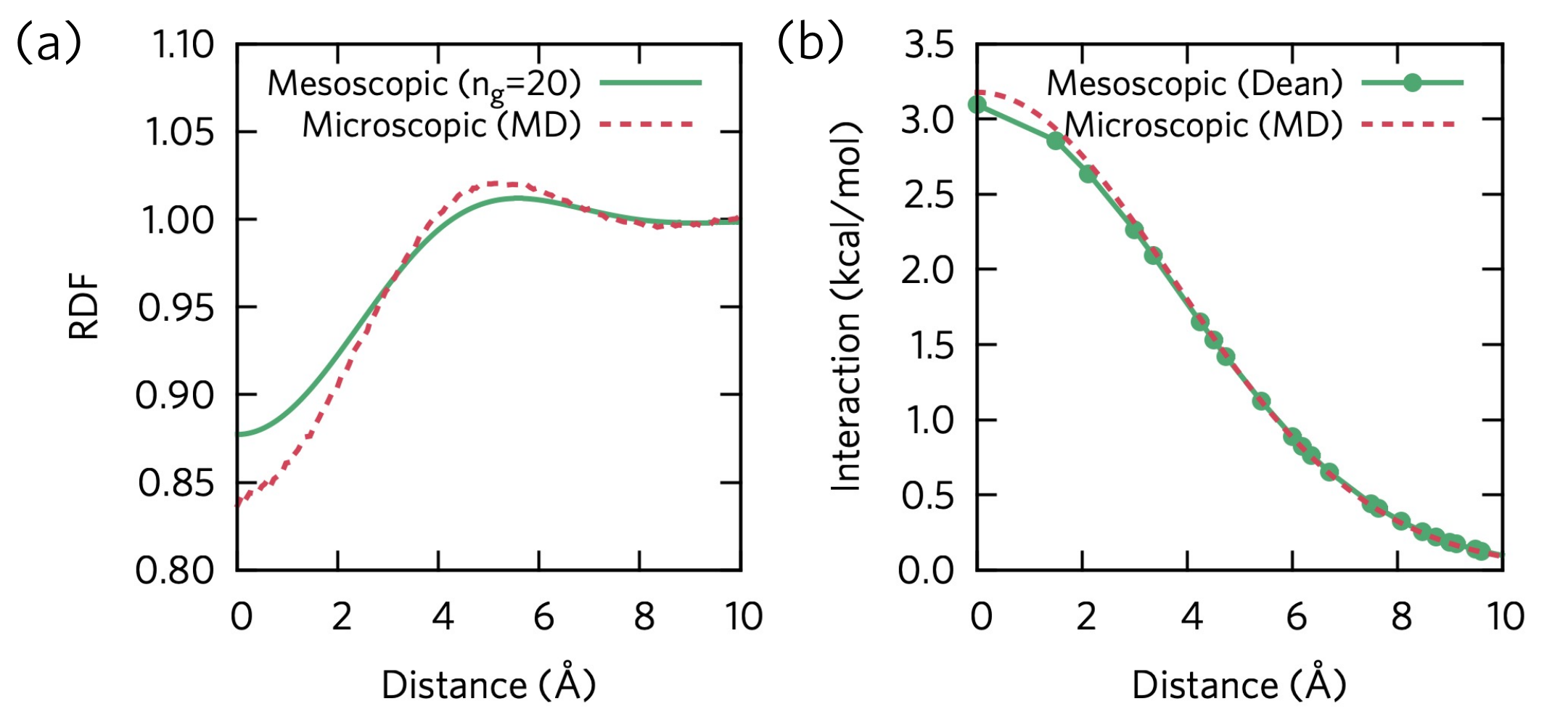}
    \caption{\label{fig:figure15} Microscopic and mesoscopic representations of the Gaussian core model at a high-density setting. (a) Structural correlations (RDF) estimated from the microscopic MD simulation (red dashed) and mesoscopic correlation from the grid setting of $n_g=20$ to satisfy $N\ge n_g^2$ (green line). Note that structural correlations are less pronounced compared to those shown in Fig. 6. (b) Effective interactions at the microscopic reference level (red dashed) and mesoscopic coarse-grained interaction for the RIDK simulation obtained using the GRID CG method (green dashed).}
\end{figure}

While this highly dense condition satisfies the strong metric condition derived from the RIDK framework, we observe that the non-divergent interaction and 2D nature lead to very weak structural correlations. Namely, as shown in Fig. \ref{fig:figure15}(a), the microscopic RDF displays a peak near 5 $\textrm{\AA}$ with an intensity of approximately 1.02. Compared to Fig. 5 in the main text, this peak is already significantly suppressed under the high-density condition at the atomistic level and becomes even weaker after mesoscopic coarse-graining. Figure \ref{fig:figure15}(a) also illustrates the mesoscopic coarse-grained RDF using a grid size of $n_g=20$ (corresponding to $\varepsilon=0.75$) to satisfy the high-density condition. Here, the atomistic RDF peak is even more reduced, with the maximum RDF peak intensity falling below 1.01, resembling an ideal gas peak. This weakened correlation profile indicates that this system may not be suitable for field-theoretic simulations. Hence, we considered a less dense condition in Sec. VII of the main text with $\tilde{\rho}=1.5625$. 

\section{Renormalized GRID CG Interactions}
\subsection{Gaussian Core Model}
The original interaction form 
\begin{align}
        V^\mathrm{GCM}_\mathrm{MD}(r) = 3.178 \times \exp \left( -\frac{1}{2} \left(\frac{r}{1.875} \right)^2 \right)
\end{align}
was coarse-grained to mesoscopic fields discretized using different grid numbers $n_g$ in a 2D setting. The mesoscopic interactions determined through the GRID CG approach were fitted to a Gaussian form, as shown in Table \ref{table:table1}. 
    \begin{figure}
    \includegraphics[width=0.35\textwidth]{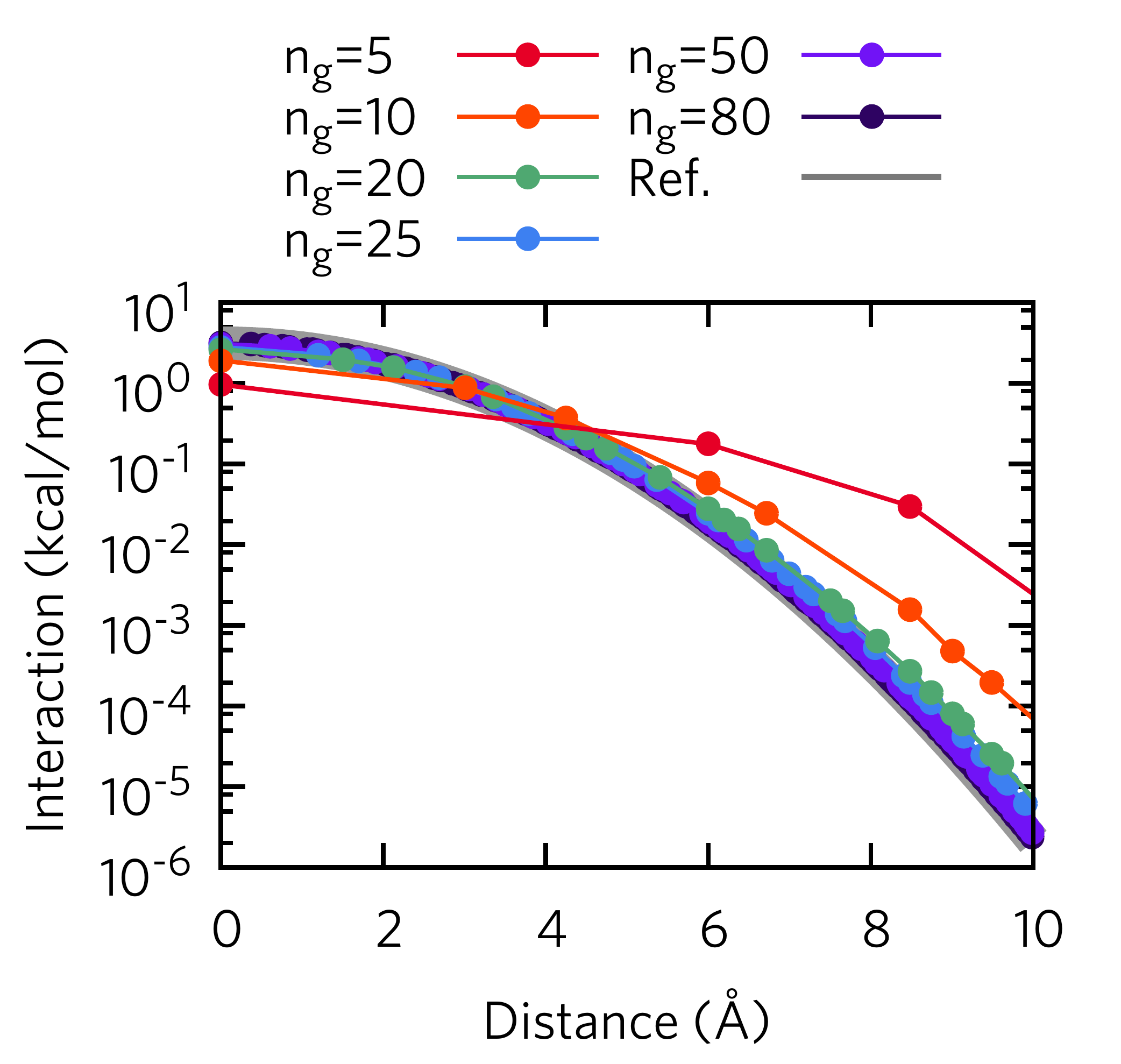}
    \caption{\label{fig:figure16} Crossover between short-range and long-range repulsions upon mesoscopic coarse-graining. The color coding follows that of Fig. 7(a) in the main text. As $n_g$ decreases (indicating increased coarse-grained), a crossover is observed: short-range interactions decrease while long-range interactions increase.}
    \end{figure}
\begin{table}[h] 
\caption{\label{table:table1} Fitted analytical forms using a single Gaussian basis for the parametrized GRID CG interactions of a less dense Gaussian system (as shown in Fig. 7 in the main text) at various $n_g$ values: 100, 90, 80, 70, 60, 50, 40, 20, 10, and 5.\\}
\begin{tabular}{ll}
\hline
$n_g$   & $V^\mathrm{GCM}_\mathrm{GRID}(r)$ (kcal/mol) \\ \hline
100 &  $3.1458 \times \exp\left(-\frac{1}{2} \left( \frac{r}{1.8841}\right)^2\right)$ \\
90  &  $3.1400 \times \exp\left(-\frac{1}{2} \left( \frac{r}{1.8856}\right)^2\right)$ \\
80  &  $3.1295 \times \exp\left(-\frac{1}{2} \left( \frac{r}{1.8887}\right)^2\right)$ \\
70  &  $3.1164 \times \exp\left(-\frac{1}{2} \left( \frac{r}{1.8925}\right)^2\right)$ \\
60  &  $3.0957 \times \exp\left(-\frac{1}{2} \left( \frac{r}{1.8982}\right)^2\right)$ \\
50  &  $3.0654 \times \exp\left(-\frac{1}{2} \left( \frac{r}{1.9071}\right)^2\right)$ \\
40  &  $3.0074 \times \exp\left(-\frac{1}{2} \left( \frac{r}{1.9251}\right)^2\right)$ \\
20  &  $2.6788 \times \exp\left(-\frac{1}{2} \left( \frac{r}{2.0298}\right)^2\right)$ \\
10  &  $1.9487 \times \exp\left(-\frac{1}{2} \left( \frac{r}{2.3606}\right)^2\right)$ \\
5   &  $0.9750 \times \exp\left(-\frac{1}{2} \left( \frac{r}{3.2491}\right)^2\right)$ \\ \hline
\end{tabular}
\end{table}

When comparing mesoscopic coarse-grained interactions at different levels of coarse-graining to the microscopic reference, Fig. \ref{fig:figure16} shows that short-range interactions become less repulsive, while long-range interactions grow more pronounced and extend further in space. This crossover highlights the role of mesoscopic coarse-graining: it smooths out the hard-core short-range repulsion beneath the mesoscopic grid and broadens long-range repulsion across different grid cells.

\begin{figure*}
    \includegraphics[width=0.75\textwidth]{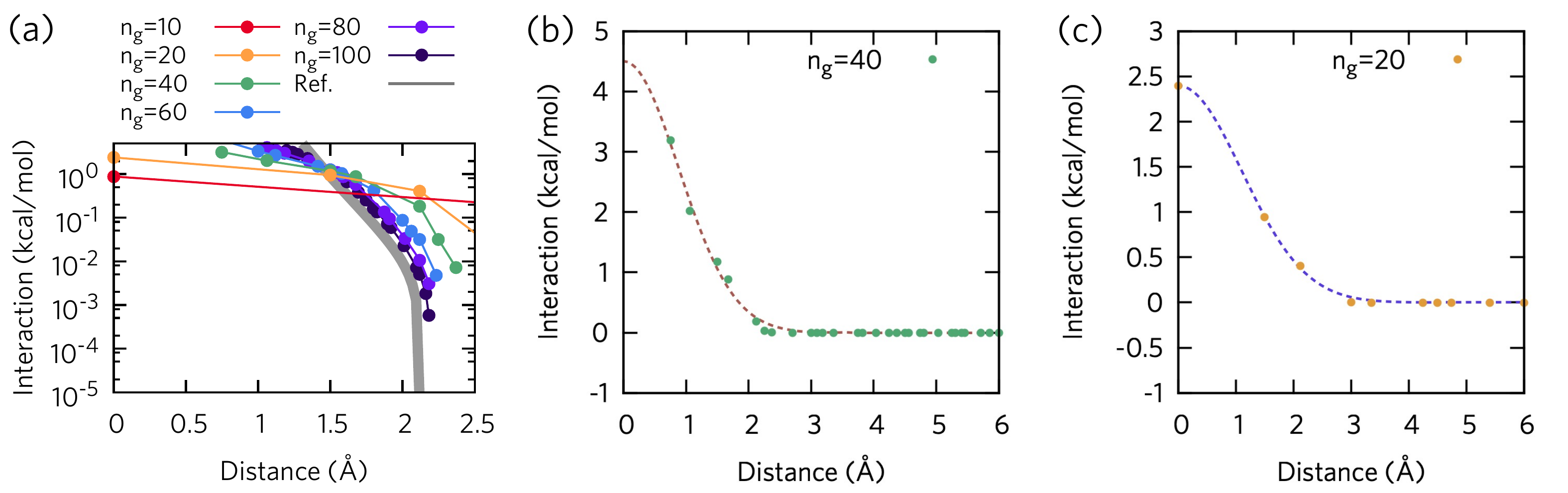}
    \caption{\label{fig:figure17} Mesoscopic coarse-grained interaction for a Lennard-Jones system. (a) Crossover between short-range and long-range repulsions upon mesoscopic coarse-graining. The color coding follows that of Fig. 9(b) in the main text. As $n_g$ decreases (indicating increased coarse-grained), a crossover is observed: short-range interactions decrease while long-range interactions increase, similar to the trend observed in Fig. \ref{fig:figure16} for a Gaussian system. (b-c) Illustration of the fitted GRID CG interactions (lines) as listed in Table \ref{table:table1}, compared with the parametrized GRID CG values (dashed) to demonstrate accuracy: (b) $n_g=40$ case and (c) $n_g=20$ case, which are used in the RIDK simulation in the main text.}
    \end{figure*}
    
\subsection{Lennard-Jones Model}
For the Lennard-Jones interaction of the form:
\begin{align}
        V^\mathrm{LJ}_\mathrm{MD}(r) =& 4\times 3.9745\times10^{-3} \times \left[ \left(\frac{2.121}{r}\right)^{12}-\left(\frac{2.121}{r}\right)^{6}  \right], 
\end{align}
we first applied the GRID CG across varying grid sizes $n_g$. From the parametrized coarse-grained interaction values, we employed a two-step fitting process to capture the divergent nature of the Lennard-Jones interactions, following a parametrization scheme similar to that developed in Ref. \onlinecite{jin2023gaussian}.

Following the approach outlined by Refs. \onlinecite{ma1993approximate} and \onlinecite{jin2023gaussian}, the GRID CG values were fitted to a sum of two Gaussian interactions, as summarized in Table II. Example plots for $n_g=40$ ($\varepsilon=0.375\,\angstrom$) and $n_g=20$ ($\varepsilon=0.75 \,\angstrom$) are illustrated in Figs. \ref{fig:figure17}(b) and (c), respectively. We observe that the microscopic interaction profile due to hard-core repulsion manifests only as a smeared repulsive basin at the mesoscopic level, which is not too different from the coarse-grained Gaussian core interaction. Notably, this resemblance suggests that short-range repulsive characteristics combined with a slowly decaying long-range feature may be representative of mesoscopic interaction at the field level. Further exploration of various interaction profiles and types using the GRID CG approach will be pursued in follow-up work to substantiate this observation.

\begin{table}[h]
\caption{\label{table:table2}Fitted analytical forms using two Gaussian basis sets for the parametrized GRID CG interactions of a Lennard-Jones system (as shown in Fig. 9 in the main text) at various $n_g$ values: 100, 980, 60, 40, 20, and 10.\\}
\begin{tabular}{ll}
\hline
$n_g$   & $V^\mathrm{LJ}_\mathrm{GRID}(r)$ (kcal/mol) \\ \hline
100 &  $55.4575 \exp\left(-\frac{1}{2}\left(\frac{r}{0.5141}\right)^2\right) -0.0112 \exp\left(-\frac{1}{2}\left(\frac{r-0.5815}{1.2956}\right)^2\right)$ \\
80  &  $20.2999 \exp\left(-\frac{1}{2}\left(\frac{r}{0.6089}\right)^2\right) -0.0039 \exp\left(-\frac{1}{2}\left(\frac{r-2.0715}{0.7973}\right)^2\right)$ \\
60  &  $10.8407 \exp\left(-\frac{1}{2}\left(\frac{r}{0.6860}\right)^2\right) -0.0034 \exp\left(-\frac{1}{2}\left(\frac{r- 2.6354}{0.5139}\right)^2\right)$ \\
40  &  $4.7869 \exp\left(-\frac{1}{2}\left(\frac{r}{0.8315}\right)^2\right) -0.0044 \exp\left(-\frac{1}{2}\left(\frac{r- 2.7071}{0.6286}\right)^2\right)$ \\
20  &  $2.3970 \exp\left(-\frac{1}{2}\left(\frac{r}{1.1034}\right)^2\right) -0.0034 \exp\left(-\frac{1}{2}\left(\frac{r-2.1324}{0.0348}\right)^2\right)$ \\
10  &  $0.8783 \exp\left(-\frac{1}{2}\left(\frac{r}{1.6922}\right)^2\right) -0.0099 \exp\left(-\frac{1}{2}\left(\frac{r-3.0861}{0.5603}\right)^2\right)$ \\ \hline
\end{tabular}
\end{table}

Similar to the Gaussian cases (Fig. \ref{fig:figure16}), the crossover between diminishing short-range repulsion and increasing long-range repulsion remains invariant even for the Lennard-Jones interaction, see Fig. \ref{fig:figure17}. 

\section{RIDK Interactions}
\subsection{Gaussian Core Model}
From Sec. VII A, we set up the RIDK simulation as follows. First, from the GRID CG process, we numerically derived the mesoscopic CG interaction potential for this grid setting: $V^\mathrm{GCM}_\mathrm{GRID}(r)=2.820 \times \exp \left[ -\frac{1}{2} \left(\frac{r}{1.981}\right) ^2 \right]$ (in microscopic units), where the interaction length of 0.4149  exceeds the grid spacing ($\pi/25$), ensuring that pair interactions will persist at the mesoscopic level. The convolution cutoff was set to half the system size ($\pi$). 

We then rescaled this interaction for the RIDK simulation domain. Specifically, we adjusted the length scale from $30\,\textrm{\AA}$ to $2\pi$, yielding a rescaled interaction length of $1.981 \,(\textrm{\AA})\times2\pi/30\,(\textrm{\AA})=0.4149$. Next, we rescaled the interaction by matching the reduced temperature $\tilde{\beta}^{-1}=1/16000$ at the microscopic level to the RIDK setting, using $\epsilon^0_\mathrm{RIDK}=8$ and $ T_\mathrm{RIDK}=1/2000$ with $\gamma=1$ and $\sigma=1/\sqrt{1000}=0.0316$. This equivalence mapping gives a rescaled interaction strength of $\epsilon_\mathrm{RIDK}=2.820\times8/3.178=7.100$. Combined together, the final RIDK interaction potential becomes
\begin{equation} \label{eq:ridk_gauss}
    V^\mathrm{GCM}_\mathrm{RIDK}(r) = 7.100 \times \exp \left[ -\frac{1}{2} \left(\frac{r}{0.4149}\right) ^2 \right].
\end{equation}
\subsection{Lennard-Jones Model}
We start from the GRID CG interaction form of
\begin{align} \label{eq:ljgrid}
    V^\mathrm{LJ}_\mathrm{GRID}(r) &= 2.3970 \times \exp\left(-\frac{1}{2}\left(\frac{r}{1.1034}\right)^2\right) \nonumber \\ & -0.0034 \times \exp\left(-\frac{1}{2}\left(\frac{r-2.1324}{0.0348}\right)^2\right),
\end{align}
where units are in microscopic units. The characteristic distances in  $V^\mathrm{LJ}_\mathrm{GRID}(r)$ [1.1034, 2.1324, and 0.0348 in Eq. (\ref{eq:ljgrid})] were rescaled by mapping 30 $\textrm{\AA}$ to 2$\pi$, and the interaction strength was adjusted to match the reduced temperature of $\tilde{T}=500$ by setting $\gamma = 1, \sigma= 100, \epsilon_0 = 10$, i.e., $T/\epsilon=\sigma^2/(2\gamma \epsilon_0) = 500$. The interaction strengths were then rescaled relative to $\epsilon_0$: $2.3970 \times 10/(3.9745\times 10^{-3})=6030.9$ for the repulsive part and $-0.0034\times 10/(3.9745\times 10^{-3})=-8.5545$ for the attractive part. Combining these, we obtained the final RIDK interaction:
\begin{align} \label{eq:gauss2}
    V^\mathrm{LJ}_\mathrm{RIDK}(r) &= 6030.9 \times \exp\left(-\frac{1}{2}\left(\frac{r}{0.2311}\right)^2\right) \nonumber \\ & -8.5545 \times  \exp\left(-\frac{1}{2}\left(\frac{r-0.4466}{7.288\times10^{-3}}\right)^2\right).
\end{align}
\section{Effective Dynamics of RIDK Simulation}
In addition to comparing RDFs, we also analyzed the effective dynamics of the mesoscopic simulation by calculating the intermediate scattering function $F(\textbf{k},t)$. The microscopic (reference) $F(\textbf{k},t)$ is given by \cite{lavcevic2003spatially}:
\begin{equation} \label{eq:fkt}
    F(\textbf{k},t) = \frac{\sum_{ij} \exp{[i\textbf{k}\cdot \textbf{r}_i(0)]}\exp{[i\textbf{k}\cdot \textbf{r}_j(t)]} }{\sum_{ij} \exp{[i\textbf{k}\cdot \textbf{r}_i(0)]}\exp{[i\textbf{k}\cdot \textbf{r}_j(0)]}}.
\end{equation}
In practice, Eq. \eqref{eq:fkt} was evaluated from particle-level correlations in the MD trajectory:
\begin{equation}
    F(\textbf{k},t) = \left\langle\frac{1}{N}\sum_{i=1}^N \sum_{j=1}^N \exp{[i\textbf{k}\cdot\left( \textbf{r}_i(t)-\textbf{r}_j(0)\right)]}\right\rangle.
    \end{equation}
    
\begin{figure} \label{fig:dynamics}
    \includegraphics[width=0.4\textwidth]{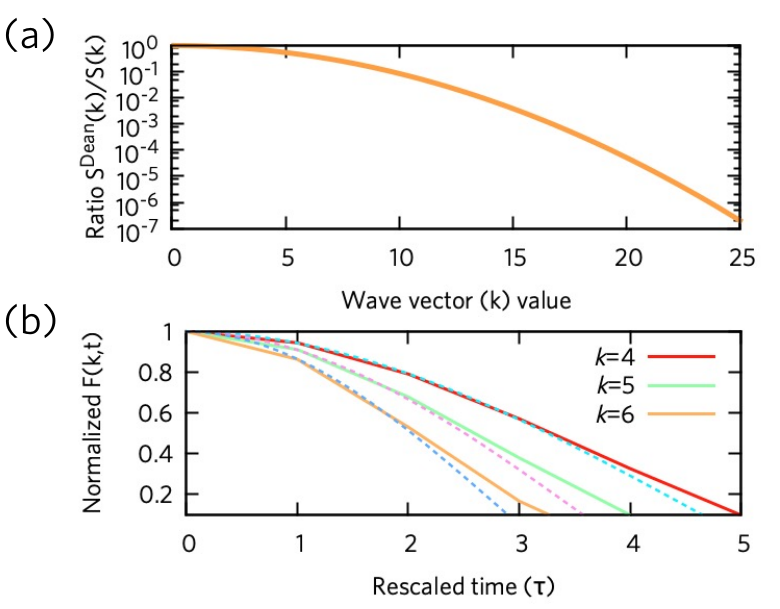}
    \caption{\label{fig:figure18} Effect of mesoscopic coarse-graining on dynamical properties for the Lennard-Jones system: (a) $S(k)$ for $n_g=20$ and (b) $F(k,t)$, normalized by $F(k,0)$ for clarity. The rescaled dynamical correlation functions at selected wave vectors [$k=4$ (red), 5 (green), and 6 (orange)], in dashed lines, quantitatively capture the expected decays (solid lines).}
\end{figure}
As observed for the structure factor $S(\textbf{k})$ [see Fig. \ref{fig:figure18}(a)], mesoscopic coarse-graining directly affects $F(\textbf{k},t)$, where the mesoscopic correlation $F_{\varepsilon}(\textbf{k},t)$ is filtered by Gaussian kernels of width $\varepsilon$: $F_{\varepsilon}(\textbf{k},t)=e^{-\textbf{k}^2\varepsilon^2}F(\textbf{k},t)$. This indicates that much of the microscopic dynamics is smeared out through the Gaussian filtering introduced by coarse-graining. Specifically, when Gaussian (or von Mises) kernels with variance $\varepsilon^2$ are used, the static and dynamic structure factors are effectively rescaled by the factor $\exp\left(-k^2 \varepsilon^2\right)$ at each wave vector $k$, assuming isotropicity. This Gaussian convolution strongly suppresses long-time dynamics, particularly at large wave vectors (e.g., for $k\approx10$ and $n_g=20$, i.e., $\varepsilon=\pi/20$, the ratio drops below $0.1$).

Nevertheless, we show that the reduced dynamical correlations can be interpreted within our framework by rescaling $F_\mathrm{RIDK}(\textbf{k},t)$ against the normalized $F(\textbf{k},t)$, as both should decay at the same rate regardless of $k$. Since the timescale of RIDK simulations is in principle longer than that of atomistic simulations, their time dependence should differ only by a constant factor, corresponding to the timescale ratio, if implemented correctly. Although the RIDK simulation does not explicitly include the $\varepsilon$ parameter, we effectively account for the Gaussian convolution by using $h=2\varepsilon$.

Figure \ref{fig:figure18}(b) demonstrates that rescaling time by a uniform factor $\mathbb{t}_\mathrm{RIDK/MD}=150$ allows the RIDK simulation to qualitatively capture the dynamical correlations of the atomistic reference. For selected wave vectors ($k=4,\,5,$ and 6 on $\mathbb{T}$), this approach successfully recovers short-time dynamics, while long-time behavior exhibits slight deviations. These results suggest that further work is needed to systematically correct coarse-grained dynamics in order to more closely align with the microscopic reference behavior. This direction is akin to recent efforts in molecular coarse-graining, where accelerated CG dynamics are corrected to recover accurate microscopic dynamical properties \cite{jin2023understanding,jin2023understanding2,jin2023understanding3,jin2024understanding,jin2025understanding}. Overall, this agreement indicates that RIDK simulations, despite coarse-graining, preserve the key structural and dynamical features of the underlying microscopic system.

\bibliography{main}